\documentclass[12pt]{article}

\usepackage{setspace}
\usepackage{xr-hyper}
\usepackage{graphicx}
\usepackage{dcolumn}
\usepackage{xcolor}
\usepackage{hyperref}
\usepackage{amsmath} 
\usepackage{amsfonts}
\usepackage{amssymb}
\usepackage{pdflscape}
\usepackage{xltabular}
\usepackage{float}

\def\paperTitle{New News is Bad News}

\def\authorLine{\author{Paul Glasserman\thanks{Columbia Business School, pg20@columbia.edu.}  \and
    Harry Mamaysky\thanks{Columbia Business School, hm2646@columbia.edu.}  \and Jimmy
    Qin\thanks{Columbia Business School, qq2127@columbia.edu. Supported by the W. Edwards Deming
      Center for Quality, Productivity and Competitiveness at Columbia Business School.}}}

\def\re{\mathbb{R}}

\DeclareMathOperator{\sgn}{sgn}
\def\gm{\gamma}

\renewenvironment{thebibliography}[1]
    {\subsection*{\hspace*{0pt}References}%
      \list{\@biblabel{\@arabic\c@enumiv}}%
           {\fs.9.10.\settowidth\labelwidth{\@biblabel{#1}}%
            \leftmargin\labelwidth
            \labelsep5pt
            \advance\leftmargin\labelsep
            \itemsep 2pt plus 2pt minus 0.5pt
            \usecounter{enumiv}%
            \let\p@enumiv\@empty
            \def\newblock{\relax}%
            \renewcommand\theenumiv{\@arabic\c@enumiv}}%
      \sloppy
      \clubpenalty4000
      \@clubpenalty \clubpenalty
      \widowpenalty4000%
      \sfcode`\.\@m}
     {\def\@noitemerr
       {\@latex@warning{Empty `thebibliography' environment}}%
      \endlist}
\def\NatBibNumeric{%
 \renewenvironment{thebibliography}[1]{%
  \bibsection\parindent \z@\bibpreamble\bibfont\list
   {\@biblabel{\arabic{NAT@ctr}}}{%
   \setlength{\labelsep}{5pt}\@bibsetup{##1}\setcounter{NAT@ctr}{0}}%
    \renewcommand\newblock{\hskip .11em \@plus.33em \@minus.07em}%
    \sloppy\clubpenalty4000\widowpenalty4000
    \sfcode`\.=1000\relax
    \let\citeN\cite \let\shortcite\cite
    \let\citeasnoun\cite
 }{\def\@noitemerr{%
  \PackageWarning{natbib}
     {Empty `thebibliography' environment}}%
  \endlist\vskip-\lastskip}%
}

\if@MOOR
\def\NatBibNumeric{%
 \renewenvironment{thebibliography}[1]{%
  \section*{\hspace*{-10pt}References}\parindent \z@\bibpreamble\bibfont\list
   {\@biblabel{\arabic{NAT@ctr}}}{%
   \setlength{\labelsep}{5pt}\@bibsetup{##1}\setcounter{NAT@ctr}{0}}%
    \renewcommand\newblock{\hskip .11em \@plus.33em \@minus.07em}%
    \sloppy\clubpenalty4000\widowpenalty4000
    \sfcode`\.=1000\relax
    \let\citeN\cite \let\shortcite\cite
    \let\citeasnoun\cite
 }{\def\@noitemerr{%
  \PackageWarning{natbib}
     {Empty `thebibliography' environment}}%
  \endlist\vskip-\lastskip}%
}%
\fi

\def\scaledCoeffDesc{The risk premia coefficients are scaled by the standard deviations of betas in
  the first stage of Fama-MacBeth.}

\usepackage{verbatim}
\usepackage{adjustbox}
\usepackage{longtable}
\usepackage{tabularx}
\usepackage{subcaption}
\usepackage{filecontents}

\usepackage{natbib}
 \bibpunct[, ]{(}{)}{,}{a}{}{,}%
 \def\bibfont{\small}%
 \def\newblock{\ }%
\def\numParams{177,488}

\begin{document}

\title{\paperTitle}
\authorLine

\date{\today}
\maketitle

\begin{abstract}
  An increase in the novelty of news predicts negative stock market returns and negative
  macroeconomic outcomes over the next year.  We quantify news novelty -- changes in the
  distribution of news text -- through an entropy measure, calculated using a recurrent neural
  network applied to a large news corpus. Entropy is a better out-of-sample predictor of market
  returns than a collection of standard measures. Cross-sectional entropy exposure carries a
  negative risk premium, suggesting that assets that positively covary with entropy hedge the
  aggregate risk associated with shifting news language. Entropy risk cannot be explained by
  existing long-short factors.
\end{abstract}

Online Appendix: \url{https://sites.google.com/view/hmamaysky}

\thispagestyle{empty}
\newpage
\setcounter{page}{1}

\section{Introduction} \label{s: intro}

Several studies have documented that the sentiment of news text forecasts short-term changes in
asset prices, with negative news forecasting negative returns. We find that a change in the
distribution of news text --- a measure of the novelty or unusualness of the news rather than its
sentiment --- forecasts negative market returns and macroeconomic outcomes over the subsequent
year. Consistent with this pattern, we find that assets that positively covary with our measure
carry a negative risk premium: investors accept lower compensation to hold assets that hedge the
risk associated with a shift in the language of news.

To motivate the idea, consider the string of words ``the securities and exchange commission issued
an.'' What is the likelihood that the next word in the sentence is ``agreement,'' ``order,''
``emergency,'' or any other English language word? Using a recurrent neural network re-estimated in
rolling windows over 1.6 million Reuters news articles spanning a period of 27 years, we can
quantify exactly how this distribution evolves over time. For example, the likelihood that the next
word is ``agreement'' has a decreasing time trend, while the likelihood that the next word is
``order'' has an increasing time trend. The likelihood that the subsequent word is ``emergency''
spikes during the financial crisis and again during the COVID-19 pandemic.%
\footnote{The methodology to calculate these conditional probabilities is explained in Section
  \ref{s:data-overview}.}
Although in principle such shifts in language structure could be neutral, we find that they
generally contain negative news about markets and the economy. This finding leads us to explore and
subsequently uncover a rich structure of risk pricing associated with exposures to the changing
distribution of news language.

\begin{figure}[h]
\begin{center}
  \mbox{
    \includegraphics[width=0.33\textwidth]{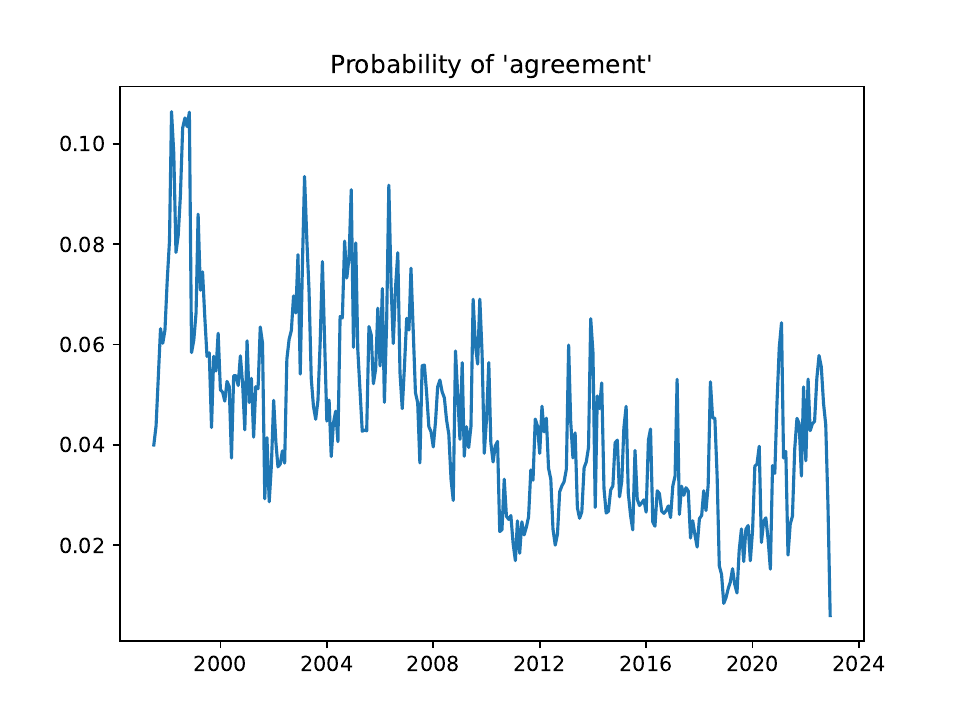}
    \includegraphics[width=0.33\textwidth]{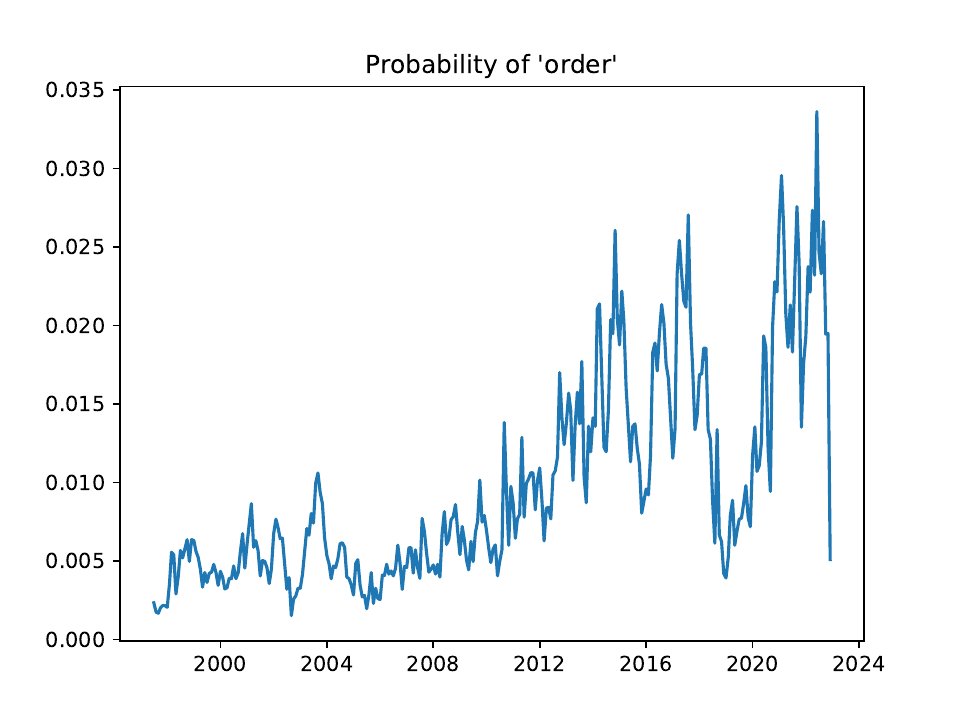}
    \includegraphics[width=0.33\textwidth]{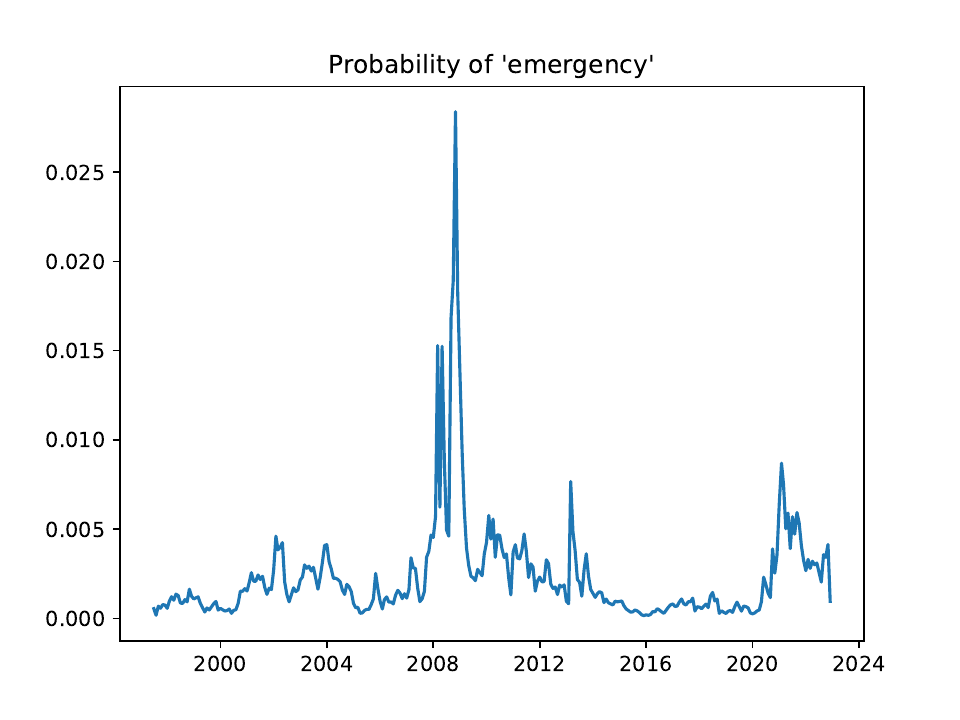}}
\end{center}
\vskip -20pt
\caption*{\footnotesize The probability of ``agreement'', ``order'', and ``emergency'' following the
  string of words ``the securities and exchange commission issued an.'' The distributional shift
  over time is estimated by a recurrent neural network.}
\end{figure}

We refer to our measure as entropy; it can also be seen as a measure of the cross entropy (or
dissimilarity) between the current and past distributions of news text. \cite{glasserman2019does}
used an entropy measure to forecast volatility, rather than the direction of returns. To estimate
conditional word probabilities, they used the empirical frequencies of word sequences in what is
known as an $n$-gram method in the natural language processing (NLP) literature; see
\cite{jurafskyspeech}. Exact phrases, like in the above example, occur rarely if they are more than
a few words long, limiting the accuracy of the $n$-gram method.

Here we use a far more powerful recurrent neural network (RNN) model with a long short-term memory
(LSTM) architecture (\citealt{mikolov2010recurrent, hochreiter1997long}) to estimate conditional
word probabilities and thus to calculate entropy. Whereas in practice the $n$-gram approach
conditions on only three or four words prior to the predicted word, an RNN model captures far more
contextual information in assigning probabilities to words. We apply the RNN model with word
embeddings that encode rich information about the relationships between words, adding to the power
of the method.  Unlike the latest and most complex large language models (like GPT-4, BERT, and
their relatives), our RNN can be trained relatively quickly. This allows us to retrain our RNN
monthly in rolling windows over our news data, and then use the most recent model to assign an
entropy score to the next month's news articles, which reflects the degree to which the text in
these articles deviates from the RNN model's conditional word distributions. We can then assess the
in- and out-of-sample forecasting performance of our entropy measure.

Each month we calculate \textit{ENT}, the change in our raw entropy measure over the previous 12
months. For simplicity, we often refer to \textit{ENT} simply as \textit{entropy} in the rest of the
paper.  In in-sample regressions, we find that a one standard deviation increase in \textit{ENT}
forecasts a 3\% decline in cumulative S\&P 500 returns over the next 12 months, even after
controlling for a large number of alternative predictors. The effect is highly statistically
significant.

To check out-of-sample predictability, we compare univariate forecasts for a large number of
predictors; the out-of-sample R-squared for \textit{ENT} is positive at multiple horizons but
overwhelmingly negative for the other predictors. In particular, entropy is the best out-of-sample
predictor of market returns when compared to a long list of variables proposed in prior studies,
including the inverse of the cyclically adjusted price-to-earnings ratio of \cite{campbell1988stock}
\textit{CAPE}, the variables tested in \cite{welch2008comprehensive}, \cite{campbell2008predicting},
and \cite{david2022survey}, as well as economic policy uncertainty (\textit{EPU},
\citealt{baker2016measuring}), the \textit{VIX} (\citealt{martin2017expected}), and the aggregate
consumption to wealth ratio of \cite{lettauludv2001}. We are careful to estimate our RNN model in
rolling windows to ensure the test is truly out-of-sample. Given the extensive literature devoted to
identifying successful out-of-sample risk premium forecasters, the out-of-sample forecasting
performance of {\it ENT} is remarkable.

If an increase in \textit{ENT} forecasts a decline in the stock market, then an asset that
positively covaries with \textit{ENT} provides a hedge against this deterioration in the investment
opportunity set. We, therefore, expect \textit{ENT} to carry a negative risk premium: investors will
accept lower expected returns on assets that covary positively with \textit{ENT} because of the
hedge they provide. This ICAPM-style (\citealt{merton1973}) argument provides the intuition
underlying the sign consistency property proposed by \cite{maio2012}. Through cross-sectional
regressions that control for standard factors, we estimate a monthly risk premium for \textit{ENT}
that is indeed negative, ranging from $-0.06\%$ to $-0.09\%$ per unit standard deviation of
cross-sectional entropy loading, depending on the choice of test assets. While the magnitude of this
risk premium is in line with that of the other factors, none of the other cross-sectional pricing
factors satisfy the \cite{maio2012} consistency property that a factor's market return forecasting
direction and cross-sectional risk premium have the same sign.

The structure of our \textit{ENT} measure allows its decomposition into two components, which we
refer to as \textit{news updates} and \textit{model updates}. The first component measures the
change in entropy, holding the distributional model of text fixed, and the second measures the
change in the distributional model. The first component is thus more focused on the novelty of the
text in the current month. We find that the predictive power of entropy stems primarily from the
first component, which negatively forecasts future returns out-of-sample and carries a negative
cross-sectional risk premium.

Entropy and its news innovation component forecast year-ahead macroeconomic outcomes. Increased
entropy is associated with future increases in unemployment and the \textit{VIX}, and with future
declines in industrial production, inflation, interest rates, and corporate earnings. Consistent
with an ICAPM-style hedging argument, market participants are willing to accept lower expected
returns for securities that hedge entropy risk. However, entropy also negatively forecasts aggregate
market returns suggesting that markets do not fully reflect the information content of entropy
either because of investor informational constraints as in \cite{sims2003} or because of slow-moving
institutional capital and limits to arbitrage \citep{gk2021,gv2010}. While improving data and
analytics may alleviate the former channel, the latter institutional-constraint channel is likely to
persist.

Lastly, we show that the replicating portfolio for our entropy measure cannot be explained by the
multitude of existing long-short factors
\citep{harvey2016and,feng2020taming}. \cite{jensen2021there} argue that most factor-based asset
pricing results can be replicated, and in doing so construct a database of over 150 long-short
factor portfolios.%
\footnote{\cite{chenzimm2021} make a similar point and also provide factor-replication code and
  data.}
Using this database of over 150 long-short factors we show that our entropy-change replicating
portfolio is among the long-short portfolios that are most poorly explained by existing
factors. Furthermore, entropy appears largely unrelated to several measures of economic uncertainty
\citep{baker2016measuring,bekaert2022time,jurado2015measuring,azzimonti2018partisan,david2022survey}
and contains information that is distinct from news sentiment. Entropy, with its sign-consistent
price of risk and out-of-sample forecasting performance, is also largely unspanned by existing
long-short factors and measures of uncertainty.

\subsection{Literature Review}

Our paper is related to the literature examining the impact of text similarity in financial
markets. \cite{tetlock2011all} measures news staleness as the similarity of news stories to prior
news stories about the same firm, and finds that firms' stock returns are less responsive to stale
news but that retail investors overreact to stale news. \cite{cohen2020lazy} show that firms with
quarter-over-quarter changes in the language of their 10-Ks and 10-Qs experience significantly lower
future stock return relative to non-changers, which they interpret as market underreaction to the
information content of corporate reports. \cite{hoberg2016text, hoberg2018text} construct firm peer
groups using textual similarity of product descriptions contained in 10-K reports and show that
there are cross-momentum effects within text-based peer groups.

The traditional definition of dissimilarity has focused on differences in word counts between small
groups of related documents. \cite{glasserman2019does} extend this concept to measure how unusual an
article's language is relative to the text of {\it all} prior articles. They show that, at the
firm-level and in aggregate, unusual news forecasts increases in future realized and implied
volatility. Prior measures of dissimilarity or informativeness suffer from several problems: word
counts cannot account for article context or synonyms, while the $n$-gram approach of
\cite{glasserman2019does} suffers from the sparsity issue that many feasible $n$-grams are never
observed, especially for large $n$.  Our entropy methodology offers a dramatic improvement over
prior efforts by using word embeddings inside a deep-learning framework.

Our paper is related to a long literature in computer science and linguistics which aims to create
probabilistic models for text using deep learning. \cite{hochreiter1997long} introduce the long
short-term memory (LSTM) architecture, which revolutionized the ability of recurrent neural networks
to successfully model large text corpora (\citealt{charniak2019}). \cite{pennington2014glove}
introduces the GloVE model of word embeddings which map words into a high-dimensional space where
vector operations reflect the semantic content of words. While the transformer underlying ChatGPT
\citep{radford2019language} is a far larger model than the one we use, it has been trained on an
extensive corpus obtained from the internet, and therefore cannot be used for backtesting trading
strategies because the model contains future information when applied to historical data. Our
approach, while restricted to a smaller model and data set, is trained in rolling windows, and as
such, it makes forecasts using only data that would have been available to market participants in
real-time, which can be used for out-of-sample testing.

Several papers are concerned with aggregate measures of text or news flow. \cite{jiang2019manager}
show that an index of aggregate manager sentiment, based on the text of firms' 10-Ks, 10-Qs, and
conference calls, negatively forecasts stock returns, suggesting stock investors overreact to the
information content of management communication. Using a dictionary-based approach,
\cite{shapiro2022measuring} construct an aggregate economic sentiment measure from articles in 16
major U.S. newspapers and show that positive economic news is associated with growth in future
consumption, output, and real rates. \cite{baker2016measuring} introduce the economic policy
uncertainty (\textit{EPU}) measures which count the number of times terms like {\it uncertainty},
{\it economic}, and {\it legislation} (and other similar word combinations) appear in close
proximity to each other in articles from 10 major U.S. newspapers. They then use a weighted average
of the frequency of major news discussing economic policy uncertainty, expiring tax provisions, and
forecaster disagreement about government purchases to construct their measure. They show that
\textit{EPU} negatively forecasts investment, output, and employment, and at the firm-level
\textit{EPU} predicts ``greater stock price volatility and reduced investment and employment'' for
firms most heavily exposed to government policy.

\cite{brogaard2015asset} test the asset pricing implications of \textit{EPU} and show that it
positively forecasts market returns but carries a negative cross-sectional price of risk.%
\footnote{Lin (2022) is a related paper that shows that another economic uncertainty index
  positively forecasts stock market volatility, but does not earn a significant price of covariance
  risk in a cross-section of 25 Fama-French size and book-to-market sorted portfolios, which is also
  a violation of the ICAPM sign property proposed in Maio and Santa-Clara (2012).}
Based on these findings, \textit{EPU} does not satisfy the \cite{maio2012} sign property which
suggests that if \textit{EPU} positively forecasts future returns, it should have a positive
cross-sectional price of risk. When controlling for other forecasting variables and for
\textit{ENT}, we also find that \textit{EPU} forecasts returns positively and that it carries a
negative risk premium, but the forecasting results are not consistent and don't hold out-of-sample,
and the negative risk premium is not significant in all specifications. In contrast, \textit{ENT} is
a significant return forecaster in- and out-of-sample, satisfies the ICAPM-style sign property, and
has a significant price of risk across different specifications.

It is natural to contrast our entropy measure with measures of model uncertainty. Among others,
\cite{hansensargent200814}, \cite{ait2021uncertainty}, and \cite{brenner2018asset} link risk premia
to uncertainty, distinguishing uncertainty about the true model and ambiguity about the true outcome
probabilities from the market or economic volatility.  An increase in \textit{ENT} might suggest heightened
uncertainty, but theory commonly associates uncertainty with a positive risk premium, and we find a
negative risk premium for \textit{ENT}.
 
The remainder of this paper proceeds as follows. Section \ref{s:data} discusses the data we use.  In
Section \ref{s:method}, we define the entropy measure and compare several estimation
methods. Section \ref{s:ts} provides an empirical assessment of the forecasting power of entropy for
future market returns. Section \ref{s:cs} investigates the cross-sectional association between
aggregate entropy and returns. Section \ref{s:channels} decomposes entropy into news and model
innovation components, investigates the macroeconomic forecasting properties of entropy, and
discusses potential channels for why entropy forecasts market returns. Section \ref{s:robust} offers
some robustness checks and Section \ref{s:conclusion} concludes. Technical details and supplementary
results are included in an Online Appendix.

\section{Data and Variable Construction} \label{s:data}

We construct our entropy measure using the Thomson Reuters News Feed Direct archive from January
1996 to December 2022, which is the time frame of our analysis. We removed articles that
represent multiple rewrites of the same initial story, retaining only the first article in a given
chain.%
\footnote{Thomson Reuters tracks articles by assigning a Primary News Access Code (PNAC). Articles
  that share the same PNAC are duplicates. This generally happens when there is an update to the
  coverage of the same event.}
We only kept news articles that were written in English and discussed S\&P500 companies.%
\footnote{A detailed description of the article selection methodology is in \cite{glm2023}. The
  methodology involves mapping Reuters company names to CRSP company names using fuzzy matching.}
We excluded articles with headlines containing certain keywords such as ``shh margin trading'',
``nyse'', ``imbalance'', ``machine generated'', and ``research alert;'' these keywords typically
signal messages that are not really news articles or exhibit irregularities. For example, ``machine
generated'' articles were produced using algorithms or automated processes, rather than being
created manually by human writers and editors. As such, their content may not have the same level of
context, nuance, or analysis as content created by human writers. ``[R]esearch alert'' articles
present hundreds of duplications. We further discarded articles with fewer than 30 words. This
filtering process leaves us 1,642,517 articles. Figure~\ref{f:numArt} depicts the number of articles
and average article length per month, respectively. The monthly article count increased from 1996
and peaked in the early 2000s, remaining steady until 2010 before gradually decreasing. The average
number of words per article steadily increased from 1996, reaching an approximate steady-state level
of 250 words per article in the early 2000s, with fluctuations observed after 2013.

We retrieve the Chicago Board Options Exchange's CBOE Volatility Index (\textit{VIX}) from the
Federal Reserve Economic Data (FRED) server. We write \textit{VIX2} for the square of
\textit{VIX}. We include additional uncertainty measures from \cite{bekaert2022time} (\textit{BEX})
and \cite{jurado2015measuring} (\textit{JLN}), and the \cite{azzimonti2018partisan} Partisan
Conflict Index (\textit{PCI}), all of which can be retrieved from the \cite{david2022survey} review
article data set. As a proxy for the risk-free rate, we use the market yield on U.S. Treasury
securities at ten-year (\textit{DGS10}) and two-year (\textit{DGS2}) constant maturity, also
obtained from FRED. We construct the 2s-10s spread, \textit{DGS10-2}, as \textit{DGS10} minus
\textit{DGS2}. As conditioning variables for stock returns, we include the monthly S\&P 500 dividend
yield (\textit{DY}), monthly cyclically adjusted PE ratio for the S\&P 500 (\textit{CAPE}), and the
consumption-wealth ratio (\textit{CAY}). We also use daily and monthly returns of
\cite{fama2015five} 5 factor model (\textit{MKT}, \textit{SMB}, \textit{HML}, \textit{RMW},
\textit{CMA}) and the momentum factor (\textit{UMD}).%
\footnote{They are obtained from the Nasdaq Data Link \url{https://data.nasdaq.com/}, Robert
  Shiller's website \url{http://www.econ.yale.edu/~shiller/data.htm}, Martin Lettau's website
  \url{https://sites.google.com/view/martinlettau/data}, and Ken French's website
  \url{http://mba.tuck.dartmouth.edu/pages/faculty/ken.french/biography.html} respectively.}
\textit{MKT} refers to the CRSP value-weighted index return net of the risk-free rate. We write
\textit{Return1} for the market return of the previous month, \textit{Return12} for the cumulative
market return of the previous 12 months excluding the most recent month, and \textit{Return60} for
the cumulative market return of the previous 60 months including the most recent month. Similarly,
we construct \textit{SMB60} (\textit{HML60}, \textit{RMW60}, \textit{CMA60}, \textit{UMD60}) as the
\textit{SMB} (\textit{HML}, \textit{RMW}, \textit{CMA}, \textit{UMD}) factor return of the previous
60 months. We use the last 60-month lagged factor returns in the forecasting regressions in Section
\ref{s:ts} to convert factor returns to state variables as suggested by \cite{maio2012}.%
\footnote{\cite{maio2012} use 60-month lagged returns for two factors, but use a different
  construction for converting \textit{SMB} and \textit{HML} to state variables. For consistency, we
  use the 60-month lagged returns for all factors.}

\section{The Entropy Measure} \label{s:method}

We build upon and extend the idea from \cite{glasserman2019does} to construct an entropy score for
the unusualness of news text as a market signal.  A text is considered unusual if it has a low
likelihood relative to a model of language probability. This problem has been studied in the natural
language processing literature on word prediction; see, in particular, Chapter 3 of
\cite{jurafskyspeech}.

\subsection{The Entropy of Text}

Our goal is to estimate the probability of a new set of articles (an \textit{evaluation text}) under
a probability model $\mathbb{P}$ estimated from past articles (referred to as the \textit{reference
  text} or \textit{training corpus}).  We can represent an evaluation text as a sequence of $N$
words $w_1 w_2 \cdots w_N$. Its probability is given by the product of conditional probabilities
\begin{equation}
  \mathbb{P}(w_1 \cdots w_N)=\prod_{k=1}^{N}\mathbb{P}(w_k | w_1\cdots w_{k-1}),
  \label{eq: prob of text}
\end{equation}
in which the first factor is $\mathbb{P}(w_1)$, the unconditional marginal probability of word
$w_1$. The average negative log probability per word is then given by
\begin{equation}
  -\frac{1}{N}\ln \mathbb{P}(w_1 \cdots w_N)=-\frac{1}{N}  \sum_{k=1}^{N} \ln \mathbb{P}(w_k | w_1\cdots w_{k-1}). 
  \label{eq: log prob of text}
\end{equation} 
If $\mathbb{P}$ correctly models the word-generating process, and if this process is stationary and
ergodic, then (\ref{eq: log prob of text}) converges to the entropy of the process as $N$ increases
(see Section 3.8 of \citealt{jurafskyspeech}). For fixed $N$, (\ref{eq: log prob of text}) measures
the cross-entropy between the empirical distribution of the evaluation text and the model
$\mathbb{P}$. For brevity, we will refer to (\ref{eq: log prob of text}) as the entropy of the
evaluation text $w_1\cdots w_N$. High entropy means low probability and thus signals an unusual
text, relative to $\mathbb{P}$.

To calculate entropy, we need a model or estimate for $\mathbb{P}$. The $n$-gram approach (often
with $n$ equal to four or five) truncates the conditioning on the right side of (\ref{eq: log prob
  of text}) to the $n-1$ words immediately preceding $w_k$ and thus misses additional context
provided by words that came earlier. The $n$-gram approach estimates conditional probabilities of
the form $\mathbb{P}(w_k|w_{k-n+1}\cdots w_{k-1})$ as the ratio of the number of occurrences of the
sequences $w_{k-n+1}\cdots w_{k-1}w_k$ and $w_{k-n+1}\cdots w_{k-1}$, with adjustments to the
numerator and denominator for strings that were never previously observed, which we refer to as the
\textit{sparsity} issue. Choosing a larger $n$ captures more conditioning information but makes it
more likely that the conditioning sequence has rarely or never been observed before. The inability
to condition on a large number of prior words leads to {\it approximation} errors in \eqref{eq: log
  prob of text}.

\subsection{A Neural Network Approach}

A neural network approach to modeling the probability $\mathbb{P}$ addresses the two shortcomings
(approximation and sparsity) of the $n$-gram method. Recurrent Neural Networks (RNNs) are a type of
neural network specifically designed for modeling sequential data, such as language. RNNs process
input sequences one element at a time while maintaining an internal state that captures the context
of the preceding elements in the sequence. This enables the network to capture complex dependencies
between elements in the sequence, making them particularly useful for tasks like language modeling,
where understanding the context of a word is critical for predicting the subsequent word in the
sequence. We use an RNN with an LSTM (long short-term memory) architecture to incorporate additional
conditioning information. \cite{charniak2019} provides an excellent introduction.

\begin{figure}[ht]
  \begin{center}
    \includegraphics[width=16cm]{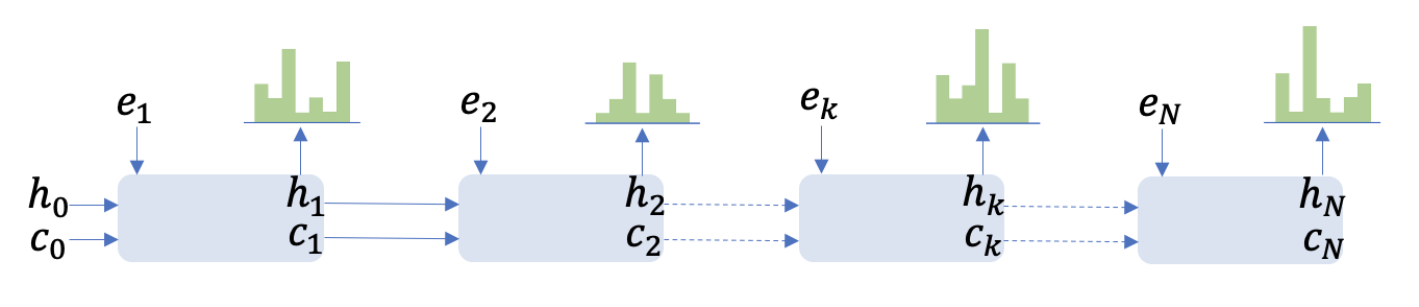}
  \end{center}
  \vskip -20pt
  \caption*{\footnotesize Illustration of RNN model with LSTM architecture. $h$ contains the model's
    hidden state and $c$ contains the long-term memory.}
\end{figure}

The above figure illustrates the evolution of the RNN model.  The state of the RNN has two
components, $h$ and $c$. The $h$ component is a vector of numerical values summarizing the context
of the input sequence processed up to that point. It serves as a summary of the past inputs seen by
the RNN up to the current time step and is updated iteratively as the RNN processes each word in the
input sequence. The cell state, denoted by $c$, is an internal memory that stores and propagates
information across time steps to capture long-term dependencies; this is the distinguishing feature
of the LSTM structure. It allows the RNN to learn and remember long-term dependencies by regulating
the flow of information through a mechanism called the ``gates.'' The cell state is updated in
parallel with the hidden state and is used to decide what information to retain, forget, or update.

The state of the RNN is updated after each word is read. Each word is represented through a word
embedding (shown in the figure as the $e$), which is a vector of numerical values that encodes
information about the meaning and use of the word.  We use the well-known GloVe word embeddings
(\citealp{pennington2014glove}), which map each word to a 100-dimensional vector.%
\footnote{We obtain these from \url{https://nlp.stanford.edu/projects/glove/}.}

At each step, the current state $(h,c)$ determines a probability distribution over the model's
vocabulary, which assigns a probability to each possible next word. The RNN thus embodies a model of
$\mathbb{P}$. Once the RNN is trained, we evaluate the probabilities of the form
$\mathbb{P}(w_k|w_1\cdots w_{k-1})$ by feeding the embeddings of $w_1,\dots,w_{k-1}$ into the RNN
and then evaluating the RNN's conditional distribution $\mathbb{P}(\cdot|w_1\cdots w_{k-1})$ at each
observed word $w_k$. Unlike the $n$-gram approach, the conditional probabilities returned from the
RNN reflect information from a very large number of prior words in the document.

Each of the arrows in the LSTM figure corresponds to a combination of linear and nonlinear
transformations involving a large number of parameters. These are described in more detail in
Section \ref{s:lstm} of the Online Appendix. Training the RNN means calibrating these parameters to
a reference or training text, a process we turn to next.

\subsection{Model Training and Updating} \label{s:rnn-train}

Our neural network consists of an embedding (input) layer, an LSTM layer, and a fully connected
(output) layer. The embedding layer maps words to their vector representations. The LSTM layer
models temporal dependencies between words in a document. We set the dimension of the $h$ and $c$
vectors at 16. The LSTM layer, whose architecture is detailed in Section \ref{s:lstm} of the Online
Appendix, contains four $W$ matrixes which transform word embeddings into an internal state and have
dimension $16 \times 100$, four $U$ matrixes which transform the internal state and have dimension
$16 \times 16$, and four bias vectors of dimension 16 each. The total number of parameters in the
LSTM layer is $4 \times (1600 + 256 + 16) = \text{7,488}$. The output of the LSTM consists of a
fully connected layer which maps the final hidden state $h$ into a probability distribution over the
vocabulary of 10,000 words. This layer contains a $10,000 \times 16$ matrix $U_s$ which transforms
the state $h$ to 10,000 outputs and a 10,000-length bias vector for a total of 170,000
parameters. Thus the network has \numParams{} parameters which must be trained.

All articles in the corpus are preprocessed by removing all punctuation strings and converting text
to lowercase. We retain only the 10,000 most frequent words in the whole corpus, representing all
other words as ``UNK'' for unknown words. We chunk each article into segments of 100 words.%
\footnote{For segments containing fewer than 100 words (at the end of articles, for example), we pad
  zero vectors to the actual words. When a batch of sequences is passed through the model,
  fixed-size length sequences allow the model to use vectorized operations and take advantage of
  hardware optimizations (i.e., parallel processing on GPUs). This results in faster training times
  and more efficient use of available computational resources.}
Training the RNN uses mini-batches, which are subsets of the entire dataset. The model's
  likelihood function is evaluated on the mini-batch, and parameter updates occur using the average
  gradient from each mini-batch item. This process requires specifying a batch size and the number
of epochs. The batch size is the number of 100-word segments in a mini-batch. A larger batch size
typically results in more stable updates to the weights but requires more memory and may lead to
longer training times. Conversely, a smaller batch size allows for faster training times but may
cause more fluctuations in weight updates. We choose a batch size of 128, i.e., the model updates
its parameters after processing each mini-batch of 128 samples, each of length 100 words. An epoch
is a full pass through the training set, during which the network processes every sequence of words
in the training set once. Increasing the number of epochs can lead to better convergence of the
network weights and improved performance on the validation set, but it also increases the risk of
overfitting. We choose the number of epochs to be 50. These values for the batch size and number of
epochs are standard (see \citealt{goodfellow2016deep,keskar2016large}).

At the beginning of each epoch, the entire training dataset is randomly shuffled and divided into
mini-batches without replacement.%
\footnote{If the dataset is not perfectly divisible by the batch size, only the last batch in an
  epoch will contain fewer samples than the specified batch size.}
During each epoch, the model processes one mini-batch at a time and updates its weights using
backpropagation for each length-100 element of the mini-batch.%
\footnote{We train the model using the categorical cross-entropy loss function and the Adam
  optimizer \citep{kingma2014adam}. The optimization uses gradient descent with backpropagation: the
  loss function is computed for each step in the sequence of words in the training text, and
  gradients are backpropagated through the entire sequence, updating the weights of the model at
  each time step to reduce the loss function \cite{werbos1990backpropagation}.}
This process is repeated for all the mini-batches in the dataset until the entire dataset has been
used for training in that epoch. Since the dataset is shuffled at the beginning of each epoch, the
composition of the mini-batches changes from epoch to epoch, providing the model with a diverse set
of samples to learn from. This shuffling means that, during training, the network state from
100-word sequence $A$ that precedes 100-word sequence $B$ in the same article will not carry over
from $A$ to $B$.

We initialize the training using the first six months of data (January 1996 -- June 1996) to compute
the entropy for all articles in the following month (July 1996). Write $m_{[t-6,t-1]}$ for the RNN
model trained using articles in the six months prior to $t$, and write $m_{[t-6,t-1]}(t)$ for the
equal-weighted average entropy score of articles in month $t$ using the RNN model trained over the
prior six months. We do this to detect month $t$ articles that may seem unusual based on information
known only prior to month $t$.

Suppose in month $t$ we have $n_t$ articles in the archive, and the $i$-th article, represented by
$w_{ti1}w_{ti2}\cdots w_{tiN_{ti}}$ of length $N_{ti}$.  By forming the sample counterpart of
\eqref{eq: log prob of text}, the entropy of this article calculated using model $m_{[t-6,t-1]}$ is
\begin{equation} \label{eq:ent-art}
H_{ti} = -\frac{1}{N_{ti}}\sum_{k=1}^{N_{ti}} \ln(p_{tik}^{x_{tik}})
\end{equation} 
where $p_{tik}$ is the output of a vector of length 10,000 when we feed
$w_{ti1}w_{ti2}\cdots w_{ti(k-1)}$ into $m_{[t-6,t-1]}$, $x_{tik}$ is the position of word $w_{tik}$
in the vector, and $p_{tik}^{x_{tik}}$ is the estimated probability of the $k$-th word being
$w_{tik}$ given the preceding words. Thus,
\begin{equation}
  m_{[t-6,t-1]}(t)
  = \frac{1}{n_t}\sum_{j=1}^{n_t} H_{tj}
  = -\frac{1}{n_t}\sum_{j=1}^{n_t} \frac{1}{N_{tj}}\sum_{k=1}^{N_{tj}} \ln(p_{tjk}^{x_{tjk}})
\end{equation}
Unlike in model training, we keep track of the internal network state ($h,c$) in calculating all
$p_{tjk} \in \re^{10,000}$ vectors when assigning probabilities to the words in article $j$ in month
$t$. The state of the RNN is then reset to zero before calculating the entropy of article $j+1$.

From August 1996 onward, in order to calculate the entropies for all articles in month $t$, we first
retrain the model $m_{[t-7,t-2]}$ to obtain an updated model $m_{[t-6,t-1]}$. The retraining starts
from the parameter values obtained in $m_{[t-7,t-2]}$ and updates the values by running the
parameter optimization on new text in month $t-1$ together with randomly sampled text from month
$t-2$ to month $t-6$. In particular, we take all articles from month $t-1$, half of the articles
randomly sampled from month $t-2$, one-quarter of the articles randomly sampled from month $t-3$,
and so on until $\frac{1}{2^5}$ of the articles are randomly sampled from month $t-6$. Each
retraining happens with batches of 128 length-100 word sequences with 50 epochs. Our main variable
of interest, monthly entropy or $\textit{ENT}_t$, is the difference between the average entropy in
month $t$ and month $t-12$, using the most recently available model at each time, or
\begin{equation} \label{eq:ent}
  \textit{ENT}_t \equiv m_{[t-6,t-1]}(t) – m_{[t-18,t-13]}(t-12)
\end{equation}
Our first observation is for July 1997.

The major benefit of using a relatively small language model, with ``only'' \numParams{} parameters,
is that we are able to estimate it in rolling windows, which means that $\textit{ENT}_t$ would have
been available to market participants in real-time.%
\footnote{Running on a GPU server with four NVIDIA T4 cards requires 21 hours to estimate the model
  in all rolling windows.}
Furthermore, $\textit{ENT}_t$ can be decomposed into a new information component and a model update
component, a feature of the model we exploit in Section \ref{s:channels}.

\subsection{Data Overview} \label{s:data-overview}

Table \ref{tab: summary-stats-month} shows the summary statistics of variables used in subsequent
analysis. All our analysis takes place at a monthly frequency. \textit{ENT} ranges from -0.140 to
0.188, with a mean close to 0 and a standard deviation of 0.061. The table also shows statistics for
the Economic Policy Uncertainty (\textit{EPU}) index of \cite{baker2016measuring} and the San
Francisco Fed's News Sentiment (\textit{SEN}) index introduced by
\cite{shapiro2022measuring}. \textit{EPU} is a well-known 
measure that is plausibly related to \textit{ENT},
and
thus we pay special attention to \textit{EPU} in our analysis. \textit{SEN} is also a related
news-based measure. The construction of the \textit{EPU} and \textit{SEN} indexes is described in
Online Appendix Section \ref{s:epu-and-sen}.

The time series of \textit{ENT} and \textit{EPU} are plotted in the top-left panel of Figure
\ref{f:ent-ts} as solid dark red and dashed light red lines, respectively. While \textit{ENT} and
\textit{EPU} sometimes exhibit similar patterns (e.g., during the 2007--08 financial crisis), they
often diverge. The time series of \textit{ENT} and \textit{SEN} are plotted in the top-right panel
of Figure \ref{f:ent-ts} as solid dark red and dashed light red lines, respectively. The peaks in
\textit{ENT} mostly coincide with either peaks or troughs in the sentiment measure. For example,
during the 2007--08 financial crisis period, \textit{SEN} is very negative while \textit{ENT} is
high. The time series of \textit{ENT} and \textit{VIX} are plotted in the bottom-left panel of
Figure \ref{f:ent-ts} as solid dark red and dashed light red lines, respectively. The peaks of
\textit{VIX} often coincide with peaks of \textit{ENT}, particularly during the financial crisis
period.

Figure \ref{f:corrHeatmap} reports the average contemporaneous correlations between \textit{ENT} and
other control variables. \textit{ENT} is not very correlated with any other control variables; its
strongest correlation (0.394) is with \textit{DY}. \textit{ENT} has little correlation with
\textit{EPU} (0.115) and \textit{SEN} (-0.280). The bottom-right panel of Figure \ref{f:ent-ts}
plots \textit{ENT} and 12-month ahead cumulative returns. There appears to be some tendency for
large positive next-12-month returns to occur at times of low entropy, and for low future returns to
occur at times of moderate to high entropy. We investigate this tendency more formally in Section
\ref{s:ts}.

Figure \ref{f:ent-ts} shows that January 2007, November 2010, and October 2014 were all months with
large \textit{ENT} peaks (and, interestingly, all of these were also low {\it EPU}
months). Table~\ref{t:ent-ex} shows sample articles with high entropies from each of these
months. We saw in the example of Section~\ref{s: intro} how the conditional probabilities of words
can change over time. We now give examples of how these changes contribute to the peaks in
\textit{ENT} through the entropy calculated in \eqref{eq:ent-art}:

\begin{itemize}
\item In January 2007, there were several high-entropy news articles related to Fannie Mae and
  Freddie Mac. In one such article, the phrase ``lack of progress in reining in mortgage lenders
  Fannie Mae and Freddie Mac leaves the economy at'' was followed by the word ``risk.'' But in
  January 2007, the word ``risk'' was perceived by the model to be unlikely in the context of this
  phrase, giving the sentence high entropy.%
  \footnote{The conditional probability is calculated by feeding the embeddings of the preceding
    string of words (e.g., ``lack of progress in reining in mortgage lenders Fannie Mae and Freddie
    Mac leaves the economy at'') into the RNN model from the prior month (e.g., December 2006) and
    then evaluating the probability at the target word (e.g., ``risk'').}
  As illustrated in the middle panel of Figure~\ref{f:anecdotal-example}, the conditional
  probability of ``risk'' increased dramatically through the mortgage crisis in late 2007 and the
  conservatorship of Fannie Mae and Freddie Mac by the U.S. government in 2008.  The initial high
  entropy score thus signaled an important shift in the news.

  This distributional shift can also be illustrated through the string of words ``Fannie Mae and
  Freddie Mac growth will be'' from another article in January 2007. The actual next word was
  ``muted.'' The word clouds in Figure~\ref{f:anecdotal-cloud} compare the conditional distributions
  of the next word as calculated in December 2006 (left panel) and December 2008 (right panel). The
  onset of the global financial crisis in between these dates is clearly reflected in the word
  clouds, where the sizes of words reflect their relative likelihoods under the models estimated at
  each date.
\item In late November 2010, there were high entropy articles discussing Ireland's financial
  strains.  In one such article, the phrase ``the Euro falls to a 5 week low on growing concerns
  Ireland will be forced to'' was followed by the word ``default.''  But ``default'' had low
  conditional probability under the October 2010 model, thus giving the sentence high entropy. The
  right panel of Figure~\ref{f:anecdotal-example} shows how the conditional probability of
  ``default'' changed over time, and became much higher by 2012 and again in 2014. In this example,
  as in the previous examples, the initial increase in entropy signaled a change in the distribution
  of text associated with economic developments.
\item Lastly, the \textit{ENT} peak in October 2014 we observed in Figure \ref{f:ent-ts} coincides
  with the Fed's ending of QE3.  Many moderate to high entropy articles from that month mentioned
  QE.
\end{itemize}

\section{Time Series of Returns} \label{s:ts}

In this section, we present evidence that entropy forecasts future market returns, and we compare
entropy with other predictors from the literature.

\subsection{In-Sample Regressions} \label{s:insample}

We estimate a variety of time-series forecasting regressions of the form
\begin{equation} \label{eq: reg-is}
  R_{t+1, t+12} = \beta_{0} + \beta_{\textit{ENT}} \textit{ENT}_{t} + \gamma^{\top} Control_{t} + \epsilon_{t}
\end{equation}
where $R_{t+1, t+12}$ is the cumulative market return from month $t+1$ to month $t+12$ and
$\textit{ENT}_t$ is the entropy measure in month $t$. $Control_{t}$ contains sentiment
(\textit{SEN}), squared implied volatility (\textit{VIX2}), interest rates (\textit{DGS10} and
\textit{DGS10-2}), the dividend yield (\textit{DY}), the difference between actual consumption and
the consumption level predicted by wealth and income (\textit{CAY}), the market return of the
previous month (\textit{Return1}), the cumulative market return of the previous 12 months excluding
the most recent month (\textit{Return12}), the cumulative returns of the Fama-French five factors
(\citealt{fama2015five}) and momentum over the previous 60 months (\textit{Return60},
\textit{SMB60}, \textit{HML60}, \textit{RMW60}, \textit{CMA60}, and \textit{UMD60}). As explained in
Section \ref{s:data}, using the 60-month lagged returns converts factors to state variables, as
suggested in \cite{maio2012}. We also use \textit{1/CAPE}, the inverse of the cyclically adjusted
price-to-earnings ratio, as a forecasting variable because, together with the rate variables, these
span the excess earnings yield forecasting variable from the Fed model proposed in \cite{maio2013}.

Column (1) of Table~\ref{tab:IS-std} reports the coefficient estimates (and $t$-statistics) for
Equation~\eqref{eq: reg-is}, after scaling each coefficient estimate by the standard deviation of
the independent variable.%
\footnote{Table~\ref{tab: IS} in the Online Appendix shows the unscaled coefficient estimates for
  Equation~\eqref{eq: reg-is}. Table~\ref{tab:IS-std} shows that \textit{CAY} forecasts 12-month
  ahead market return negatively, whereas \citealt{lettauludv2001} find that \textit{CAY} forecasts
  returns positively. We verified that, as a standalone forecaster, \textit{CAY} forecasts return
  positively when restricted to the pre-2002 period but negatively in the full sample.}
A one standard deviation increase in entropy predicts a 2.821\% decrease in the 12-month ahead
cumulative market return. Column (2) replaces \textit{ENT} in Equation~\eqref{eq: reg-is} with
\textit{EPU}. The results show that \textit{EPU} positively predicts future market returns, which is
consistent with \cite{brogaard2015asset}.  Column (3) includes both \textit{ENT} and \textit{EPU} in
a single regression model. After controlling for these other measures \textit{ENT} still forecasts
future market returns, and the standardized coefficient for \textit{ENT} (-2.607\%) is nearly
unchanged from column (1).

The full-sample analysis incorporates data from the COVID-19 period, which raises the concern that
this time period may disproportionately impact our results. To address this concern, we replicate
the analyses detailed in the first three columns using pre-COVID-19 data, which are presented in
columns (4)--(6). In the pre-COVID-19 analysis, entropy is calculated up to the end of 2018 and used
to predict through the end of 2019. We find that \textit{ENT} consistently and negatively forecasts
12-month ahead cumulative returns: a one standard deviation increase in entropy is associated with a
2.7\% year-ahead negative market return. Conversely, \textit{EPU} does not significantly predict
market returns once the COVID-19 period is dropped from the analysis.

\subsection{Out-of-Sample Regressions} \label{s:outofsample}

We saw in the previous section that an increase in entropy negatively forecasts future market
returns in in-sample regressions. We now examine whether entropy provides out-of-sample
predictability.

Let $R_{t+1,t+12}$ denote the cumulative return over months $t+1, \cdots, t+12$. For a fixed
training window of length $h$, for each $t$, we run the univariate regression
\begin{equation} \label{eq:oos}
  R_{t-i-11,t-i} = \alpha_t + \beta_t \textit{ENT}_{t-i-12} + \epsilon_{t-i-12}, ~\mbox{  }~ i=1,2,\dots,h,  
\end{equation} 
estimating $\hat{\alpha}_t$ and $\hat{\beta}_t$ from the $h$ monthly observations in the training
window (we discuss the choice of $h$ momentarily). Then we predict the future return $R_{t+1,t+12}$
as $\hat{R}_{t+1,t+12} = \hat{\alpha}_t + \hat{\beta}_t \textit{ENT}_t$. The out-of-sample R-squared
is calculated as
\begin{equation} \label{eq:oos-R2}
  1-\frac{\textit{MSE} (R_{t+1,t+12} - \hat{R}_{t+1,t+12})}{\textit{MSE} (R_{t+1,t+12} - \frac{1}{h}\sum_{i=1}^{h} R_{t-i-11,t-i})}
\end{equation}
where the numerator is the mean square error of predicting with the model in \eqref{eq:oos}, and the
denominator is the mean square error based on using past historical means to forecast future
returns. In other words, we are comparing the prediction accuracy of a model which uses
$\textit{ENT}_t$ to make a conditional return forecast versus a model which just uses the rolling
mean over the past $h$ months. If predictions using entropy are better than the rolling mean
predictions, we will get a positive out-of-sample R-squared. If predicting with entropy is worse
than using historical means, we will get a negative out-of-sample R-squared.

Table~\ref{tab: OOS} shows the out-of-sample R-squared of entropy as well as all other control
variables we used in the in-sample analysis. Each variable is used separately in univariate
regressions of the form in \eqref{eq:oos}. Each column represents a different choice of window
length $h \in \{12, 15, 18, 21, 24\}$ months, for estimation of the forecasting model in
\eqref{eq:oos}. In the first row, we see that entropy produces a positive out-of-sample R-squareds
in four out of the five estimation windows, and the out-of-sample R-squareds range from 0.024 (with
$h=15$ months) to 0.051 (with $h=21$ months). In the second row, we see that \textit{ENT\_NEWS},
which captures the part of entropy due to news innovation while holding the text model fixed (we
formally define \textit{ENT\_NEWS} in Section~\ref{s:channels}) produces a positive out-of-sample
R-squareds in all five estimation windows, with the out-of-sample R-squareds ranging from 0.060
(with $h=24$ months) to 0.096 (with $h=18$ months). In fact, this news innovations measure
outperforms {\it all} other predictors, most of which have negative R-squareds in all estimation
windows. Only two other measures have non-negative out-of-sample R-squareds: \textit{Return60}
produces positive R-squareds in all estimation windows and \textit{RMW60} produces positive
R-squareds for estimation window lengths of 18, 21, and 24 months.

Table~\ref{tab: OOS-coeff} in the Online Appendix shows the average of $\beta_t$ from \eqref{eq:oos}
for each of the training horizons used in the out-of-sample forecasting. The average forecasting
coefficients associated with entropy are all negative and are highly statistically significant for
horizons of 15 months and above. Though the $\beta_t$'s vary over time, the average rolling entropy
forecasting coefficient for future returns is consistent with the full sample results in
Table~\ref{tab:IS-std}.

Given the difficulty of out-of-sample market return forecasting
\citep{welch2008comprehensive,campbell2008predicting}, the performance of \textit{ENT} and
\textit{ENT\_NEWS} in our out-of-sample tests is remarkable.

\section{Cross-Section of Returns} \label{s:cs}

According to the argument in \cite{maio2012}, the sign with which an ICAPM state variable forecasts
aggregate market returns should match the cross-sectional risk price associated with that
variable. If entropy negatively forecasts the return of the aggregate stock market, then high
entropy is associated with an unfavorable investment opportunity set for investors. In this case,
securities that do well during high entropy times provide a useful hedge, and should therefore earn
a negative risk premium. In this section, we investigate how exposure to entropy is priced in the
cross-section of stock returns.

In examining how entropy is priced in the cross-section of expected stock returns, we want to test
whether entropy is a priced risk factor and estimate the price of aggregate entropy risk. According
to the \cite{merton1973} ICAPM framework, in equilibrium, an asset's risk premium is determined by
its conditional variances with ex-post market returns and innovations of state variables of the form
\begin{equation} \label{eq:theoreticMotive}
  \mathbb{E}_{t}[r_{i, t+1}-r_{f, t+1}] = \beta Cov_{t}(r_{i, t+1}, r_{m, t+1}) + \gamma Cov_{t}(r_{i, t+1}, \Delta x_{t+1}) 
\end{equation} 
where $r_{i, t+1}$ is the ex-post return on asset $i$, $r_{f, t+1}$ is the risk-free rate,
$r_{m, t+1}$ is the ex-post market return, and $x_t$ is state variable that affect the investment
opportunity set. The coefficients $\beta$ and $\gamma$ are risk
premia. Equation~\eqref{eq:theoreticMotive} -- which reproduces equation (8) of \cite{maio2012} --
states that investors receive compensation for the covariance of stock returns with variables that
impact future market returns. Based on the argument of \cite{maio2012}, if $\Delta x_t$ forecasts
stock returns positively, then an ICAPM-type argument suggests that stocks that positively covary
with $\Delta x_t$ should earn positive risk premia because they add to investor risk. On the other
hand, if, like a change in entropy, $\Delta x_t$ forecasts future negative stock returns, then
stocks that covary positively with $\Delta x_t$ hedge against unfavorable changes in the investment
opportunity set and should earn negative risk premia. The ICAPM sign property of \cite{maio2012} can
be summarized as:
\begin{equation} \label{eq:sgn}
  \sgn(\gm) = \left\{
    \begin{array}{cc}
      + & \text{  if $\Delta x_t$ predicts market returns positively,} \\
      - & \text{  if $\Delta x_t$ predicts market returns negatively.} \\
    \end{array}
  \right.
\end{equation}
We have seen that entropy forecasts a decline in market returns, so we expect it to carry a negative
risk premium.

\subsection{Factor-Mimicking Portfolios} \label{s:mimick}

Because \textit{ENT} and \textit{EPU} are not directly tradeable, we approximate them with
factor-mimicking portfolios, as follows. In the first step, we extract innovations from the
\textit{ENT} and \textit{EPU} time series by fitting an $AR(p)$ model.  The number of lags in the
$AR$ model is determined by minimizing the Bayesian information criterion (BIC). The final models
are $AR(12)$ for \textit{ENT} and $AR(1)$ for \textit{EPU}:
\begin{equation} \label{eq:TS}
  \begin{aligned}
    \textit{ENT}_t &= \beta_{\textit{ENT},0} + \sum_{i=1}^{12} \beta_{\textit{ENT},i} \textit{ENT}_{t-i} + \gamma_{\textit{ENT}} \textit{MKT}_{t-1} +
    \epsilon_{\textit{ENT},t} \\
    \textit{EPU}_t &= \beta_{\textit{EPU},0} + \beta_{\textit{EPU},1} \textit{EPU}_{t-1} + \gamma_{\textit{EPU}} \textit{MKT}_{t-1}  + \epsilon_{\textit{EPU},t}
  \end{aligned} 
\end{equation}
Following \cite{breeden1989empirical,lamont2001financial,ang2006cross}, in the second step, we
create the mimicking factor $F_{\textit{ENT}}$ ($F_{\textit{EPU}}$ ) to track innovations in
\textit{ENT} (\textit{EPU}) by estimating the coefficient $b_{ENT}$ ($b_{EPU}$) in the following
regressions
\begin{equation} \label{eq:mimic}
  \begin{aligned}
    \hat{\epsilon}_{\textit{ENT}, t} &= a_{\textit{ENT}} + b^{\top}_{\textit{ENT}} X_{t} + u_{\textit{ENT}, t} \\ 
    \hat{\epsilon}_{\textit{EPU}, t} &= a_{\textit{EPU}} + b^{\top}_{\textit{EPU}} X_{t} + u_{\textit{EPU}, t}
  \end{aligned}
\end{equation}
where $\hat{\epsilon}_{\textit{ENT},t}$ and $\hat{\epsilon}_{\textit{EPU},t}$ denote the innovations
of \textit{ENT} and \textit{EPU} from \eqref{eq:TS}, and $X_t$ denotes the excess returns on a set
of base assets. We estimate \eqref{eq:mimic} over the full sample and construct factor mimicking
portfolios as follows
\begin{equation} \label{eq:fs}
  \begin{aligned}
    F_{\textit{ENT},t} &= \hat{b}^{\top}_{\textit{ENT}} X_t \\
    F_{\textit{EPU},t} &= \hat{b}^{\top}_{\textit{EPU}} X_t
  \end{aligned}
\end{equation}
where $\hat{b}_{\textit{ENT}}$ and $\hat{b}_{\textit{EPU}}$ are the coefficient estimates from
\eqref{eq:mimic}. Since $X_t$ represents the excess returns of base assets portfolios, the
coefficient vectors $\{b_{\it ENT},b_{\it EPU}\}$ can be interpreted as weights of a zero-cost
portfolio with returns given by \eqref{eq:fs}.

We consider four sets of base assets: the 25 Fama-French portfolios sorted by size and momentum,
size and book to market, size and investment, and size and profitability, respectively. In the risk
premia tests of the next section, the base assets are selected to correspond to the test assets in
the \cite{fm1973} analysis. Table~\ref{tab: summary-stats-day} shows summaries of the
\cite{fama2015five} factor and momentum returns, as well as those of the \textit{ENT} and
\textit{EPU} factor replicating portfolios using different base assets.

\subsection{Factor Risk Premia} \label{s:fmrp}

To estimate the risk premium associated with the entropy factor mimicking portfolio, we run a
\cite{fm1973} analysis using \textit{MKT} (market minus risk-free), \textit{SMB}, \textit{HML},
\textit{RMW}, \textit{CMA}, \textit{UMD}, as well as $F_{\textit{EPU}}$ and $F_{\textit{ENT}}$ as
the factors. To properly reflect estimation risk from the first stage regressions, we use the
following general method of moments (GMM) system in our estimation
\begin{align}
    \mathbb{E}[r_{i,t}-\alpha_i-\beta_i^{\top}f_t] &= 0 ~\mbox{  }~ i=1,2,\cdots,N \nonumber \\
    \mathbb{E}[(r_{i,t}-\alpha_i-\beta_i^{\top}f_t)f_t] &= 0_{K} ~\mbox{  }~ i=1,2,\cdots,N \label{eq:gmm} \\
    \mathbb{E}[\beta(r-\beta^{\top}\lambda)] &= 0_{K} \nonumber
\end{align}
where $r_{i,t}$ is the excess return of test asset $i$ on day $t$, $\beta_i$ is a $K$ by 1 vector of
factor loadings for test asset $i$, $f_t$ is a $K$ by 1 vector of factors, and $0_K$ is a $K$ by 1
vector of zeros. In addition, $\beta = [\beta_1 \ \beta_2 \ \cdots \ \beta_N]$ is the $K \times N$
matrix of the $N$ test asset betas, $r$ is a vector of the $N$ test asset average excess returns,
and $\lambda$ is a $K$ by 1 vector of factor risk premia.%
\footnote{The risk premia $\lambda$ in (\ref{eq:gmm}) differ from $\gamma$ in
  \eqref{eq:theoreticMotive} in that they are premia on regression coefficients rather than
  covariances.}
The parameters of the system are $\{\alpha_1,\cdots,\alpha_N,\beta_1,\cdots,\beta_N,\lambda\}$. The
first two moment conditions, which correspond to the first-stage Fama-MacBeth regressions, exactly
identify $\alpha$ and $\beta$. The last moment condition represents the second stage regressions,
with no constant, which pin down the prices of risk. See \citealt{cochrane2009} page 241 for
details.

Table~\ref{t:fm-same-ent-std} shows the risk premia from \eqref{eq:gmm}, but scaled by the
standard deviations of the first-stage betas to highlight the magnitudes of the effects.%
\footnote{Table~\ref{t:fm-same-ent} in the Online Appendix shows the raw risk premia.}
Columns (1) -- (4) show monthly risk premia using different sets of 25 Fama-French portfolios as the
base and test assets: size-momentum, size-book/market, size-investment, and size-profitability. We
see that of the four sets of test assets, $F_{\textit{ENT}}$ carries a negative and significant risk
premium in all cases, with the monthly risk premium for a standard deviation increases in entropy
beta ranging from -0.060\% to -0.091\%. The magnitude of the entropy risk premium is on par with
that of the other factors, suggesting the risk compensation associated with hedging entropy risk is
economically large.

In comparison, $F_{\textit{EPU}}$ carries a negative and significant risk premium in two out of the
four tests as well, with a risk premium ranging from -0.063\% to -0.083\% per unit of standard
deviation of \textit{EPU} betas. The sign and the magnitude of the $F_{\textit{EPU}}$ risk premium
are consistent with the results in \cite{brogaard2015asset} (compare Panel B of their Table 5
against the unscaled risk premia for $F_{\textit{EPU}}$ in Table~\ref{t:fm-same-ent} of the Online
Appendix). Among the five factors of \cite{fama2015five}, only \textit{MKT} has a significant risk
premium for more than two out of four test assets, with a risk premium ranging from 0.048\% to
0.083\%. The momentum factor carries a positive and significant risk premium in two tests: 0.083\%
for size and book-to-market portfolios and 0.064\% for size and investment portfolios.

Column (5) of Table~\ref{t:fm-same-ent-std} shows the sign of the ICAPM property from
\eqref{eq:sgn}. The determination of whether a factor forecasts year-ahead market returns positively
or negatively is based on the forecasting regression in \eqref{eq: reg-is}, shown in column (3) of
Table~\ref{tab:IS-std}. A ``+'' or ``--'' in column (5) of Table~\ref{t:fm-same-ent-std} corresponds
to significant positive and negative coefficients at 5\% level, respectively; a blank indicates a
lack of significance in the forecasting regression. The ICAPM sign property posits that any
significant coefficients in columns (1) -- (4) of Table \ref{t:fm-same-ent-std} should have the same
signs as the corresponding entry in column (5).%
\footnote{Recall that \textit{UMD60} in Table \ref{tab:IS-std} corresponds to \textit{UMD} in Table
  \ref{t:fm-same-ent-std}, \textit{Return60} corresponds to \textit{MKT}, and so on. \textit{ENT}
  and \textit{EPU} are the state variable equivalents of $F_{\textit{ENT}}$ and $F_{\textit{EPU}}$.}
\textit{ENT} is the only factor for which this property holds. \textit{CMA} has positive signs in
column (5) but only one out of the four tests shows a significant risk premium.

\subsection{Which Securities Hedge Entropy Risk} \label{s:which-hedge}

In light of the price of entropy risk results from the prior section, we now investigate which types
of securities are useful for entropy risk hedging. We use decile portfolios from univariate sorts on
size, book/market, operating profitability, and investment obtained from Ken French's website as
test assets, and estimate their loadings on $F_{\textit{ENT}}$.  In particular, we estimate each
asset $i$'s factor exposures using a full-sample time-series regression:
\[
  r_{i, t} = \beta_{i,0} + \beta_{i,F_{1}} F_{1,t} + \cdots + \beta_{i,F_{8}} F_{8,t} + \epsilon_{i,t}
\]
where $r_{i, t}$ is the excess return of asset $i$ on day $t$, $F_{i,t}$ is the day-$t$ return on
factor $i$, and the indices $j=1,\dots,8$ refer to $\textit{MKT}$, $\textit{SMB}$, $\textit{HML}$,
$\textit{RMW}$, $\textit{CMA}$, $\textit{UMD}$, $F_{\textit{EPU}}$, and $F_{\textit{ENT}}$,
respectively. For this analysis, $F_{EPU}$ and $F_{ENT}$ were constructed using 100 base assets,
corresponding to the 25 size-$X$ sorted portfolios for $X \in$ \{momentum, book/market, investment,
profitability\}. Each $\beta_{i,F_{j}}$ is asset $i$'s loading on factor $j$. From this set of
equations, we compute estimates $\hat{\beta}_{i,F_{1}}$, $\dots$, $\hat{\beta}_{i,F_{8}}$ of the
factor loadings for each asset $i$.

In Figure~\ref{f:FM-univariate-average}, we show the $F_{\textit{ENT}}$ loadings
($\hat{\beta}_{i,F_8}$) corresponding to each test asset. The first panel shows these loadings for
size-sorted portfolios. The U-shaped pattern indicates that small and large companies hedge entropy
risk, while medium-sized firms are negatively correlated with entropy innovations. Companies
  sorted on momentum (left panel, middle row) show that past losers hedge entropy risk while past
  winners do poorly during high entropy times, though the effect is non-monotonic. Companies sorted
on book/market (right panel, middle row) show a similar pattern of exposures to the size-sorted
portfolios, with growth firms (low book/market) and value firms (second highest book/market) having
positive entropy betas, and all other portfolios having negative entropy betas.%
\footnote{We interpret the highest book/market decile as distressed, rather than value, stocks with
  extreme high book values relative to market values.}
In contrast, sorting on investment (left panel, bottom row) yields an inverted U-shape, indicating
that very low and very high investment companies do poorly during times of high entropy, while
medium-investment firms do well. The bottom right panel shows that companies with high operating
profitability offer a hedge to entropy risk, while less profitable companies are
entropy-risky. {Given the typical high minus low long-short portfolio approaches, our finding of
  generally non-monotonic entropy exposure is notable.} As a robustness check,
Figure~\ref{f:FM-univariate} in the Online Appendix repeats the analysis in each panel of
Figure~\ref{f:FM-univariate-average} but using $F_{\textit{ENT}}$ constructed from different base
assets. The results are similar.


\section{Channels} \label{s:channels}

We've shown that entropy negatively and robustly forecasts aggregate market returns both in- and
out-of-sample. Furthermore, entropy is a priced cross-sectional risk and is the only factor in our
analysis that satisfies the \cite{maio2012} ICAPM sign property between its forecasting direction
for the market (negative) and its cross-sectional price of risk (also negative). In this section, we
investigate some potential mechanisms underlying these findings. First, we show that entropy can be
decomposed into a news innovation component and a model innovation component, and that the news
innovation part plays a large role in our findings. Second, we study how entropy and its different
components forecast macroeconomic outcomes, and speculate on how this translates to market return
predictability.

\subsection{Entropy Decomposition} \label{s:decomp}

We decompose the entropy measure from Equation~\eqref{eq:ent} into two parts
\begin{equation} \label{eq:ent-decomp}
  \textit{ENT}_t = \textit{ENT\_NEWS}_t + \textit{ENT\_MODEL}_t.
\end{equation}
The first part is
\[
  \textit{ENT\_NEWS}_t = m_{[t-18,t-13]}(t) – m_{[t-18,t-13]}(t-12)
\]
which uses the text model from one year prior to month $t$, but applies it to month $t$ news
flow. This captures the change in the information content of news. The second part is
\[
  \textit{ENT\_MODEL}_t = m_{[t-6,t-1]}(t) – m_{[t-18,t-13]}(t)
\]
which looks at the difference between month $t$ entropy as seen through the lens of today's and the
year prior's text models. In order to understand what drives the predictive power of \textit{ENT},
we replace $\textit{ENT}_t$ by $\textit{ENT\_NEWS}_t$ and $\textit{ENT\_MODEL}_t$, and repeat the
time series prediction of 12-month ahead cumulative returns in-sample as in Section \ref{s:insample}
and out-of-sample as in Section \ref{s:outofsample}.

Online Appendix Table \ref{tab:IS-std-news-model} columns (1) and (2) show the in-sample results,
where the coefficient estimates are normalized by the standard deviation of the independent
variables.%
\footnote{Online Appendix Table~\ref{tab:IS-news-model} shows the unscaled results.}
Column (1) indicates that a standard deviation increase in \textit{ENT\_NEWS} is associated with a
-1.433\% market return over the next 12 months. A standard-deviation increase in
\textit{ENT\_MODEL}, column (2), is associated with a -1.393\% next 12-month return. However,
neither effect is significant.

Table \ref{tab: OOS} presents the out-of-sample R-squareds for \textit{ENT\_NEWS} and
\textit{ENT\_MODEL}. \textit{ENT\_NEWS} produces a positive out-of-sample R-squared irrespective of
the model estimation window length, with the largest R-squared of 0.096 occurring at an 
estimation window length of 18 months. On the other hand, \textit{ENT\_MODEL} never produces a
positive out-of-sample R-squared. The predictive power of \textit{ENT} for future market returns
comes from the change in the information content of news rather than from the year-over-year change
in the model. We return to this finding in Section~\ref{s:fundamentals}.

While \textit{ENT\_NEWS} is a very robust forecaster of year-ahead market returns, the direction of
forecastability fluctuates over time. This can be seen by the negative, but insignificant average
$\beta_t$ coefficient from \eqref{eq:oos} for \textit{ENT\_NEWS} shown in Table \ref{tab:
  OOS-coeff} of the Online Appendix. While the average direction of forecastability from
\textit{ENT\_NEWS} to year-ahead market returns is negative, there are also times when this sign is
positive, which renders the average impact of the effect only marginally significant.

Next, we explore whether \textit{ENT\_NEWS} and \textit{ENT\_MODEL} are priced cross-sectionally. We
construct factor-mimicking portfolios for these two components of \textit{ENT} using the procedure
from Section~\ref{s:mimick}. The base assets for the factor mimicking portfolios and the test assets
are the 25 Fama-French portfolios: size-momentum, size-book-to-market, size-investment, and
size-profitability. For all tests, the base assets are the same as the test assets.

Table \ref{t:fm-same-part1-std} shows that $F_{\textit{ENT\_NEWS}}$ has a negative risk premium in
three out of the four cases, ranging from -0.055\% to -0.060\% per month.%
\footnote{Tables \ref{t:fm-same-part1} of the Online Appendix shows the unscaled results.}
Though it is negative in all cases, the risk premium is significant for the book-to-market,
investment, and profitability sorts, but not for momentum. The magnitudes of the risk premia are
again large in comparison to those associated with the other factors, and are consistent with the
results in Table \ref{t:fm-same-ent-std}, where we use $F_{\textit{ENT}}$ instead of
$F_{\textit{ENT\_NEWS}}$. Column (5) shows that none of the factors satisfy the ICAPM property in
\eqref{eq:sgn} when the signs of the coefficients in the forecasting regression in \eqref{eq:
  reg-is} are obtained from column (1) of Table \ref{tab:IS-news-model}. However, as Tables
\ref{tab:IS-news-model} and \ref{tab: OOS-coeff} show, the general predictive sign of {\it
  ENT\_NEWS} for future market returns is negative, though marginally significant.

Tables \ref{t:fm-same-part2-std} (scaled coefficients) and \ref{t:fm-same-part2} (unscaled) of the
Online Appendix show that \textit{ENT\_MODEL} is not associated with a significant risk premium. And
this part of entropy does not forecast future market returns in- or out-of-sample. The evidence
points to the market demanding a negative risk premium for securities that help hedge the part of
entropy -- \textit{ENT\_NEWS}, or news innovation -- which is useful for forecasting market
returns in our out-of-sample tests.

\subsection{Entropy and Macro Outcomes} \label{s:fundamentals}

Having established that the predictive power of \textit{ENT} stems largely from year-over-year news
innovation, we now explore whether such changes in news flow predict economic fundamentals.  We
estimate time-series forecasting regressions of the form
\begin{equation} \label{eq:fund}
  Y_{t+12} = \beta_0 + \beta_{\textit{ENT}} \textit{ENT}_t + \gamma^{\top} Control_t + \beta_{Y} Y_t + \epsilon_t.
\end{equation} 
The dependent variable\footnote{12-month real earnings per share inflation-adjusted, constant May
  2023 dollars.} in our analysis is one of:
\vspace*{-10pt}
\begin{center}
  \begin{tabular}{p{2.75cm}p{9cm}p{3cm}}
    Variable & Description & Units \\ \hline
    \textit{EPU} & economic policy uncertainty & standard units \\
    \textit{UNRATE} & unemployment rate & percent \\
    \textit{INDPRO{\_}YOY} & year-over-year change in industrial production & percent \\
    \textit{CPI{\_}YOY} & year-over-year change in the Consumer Price Index & percent \\
    \textit{DGS10}, \textit{DGS2} & 10-year and 2-year Treasury yield & percent \\
    \textit{DGS10-2} & 10-year minus 2-year slope & percent \\
    \textit{VIX} & CBOE Volatility Index & percent \\
    \textit{EPS} & S\&P500 last-twelve-month earnings per share & dollars \\ 
    \hline
  \end{tabular}
\end{center}
\vskip 5pt
\noindent
For each dependent variable, we use the lagged value of entropy ($\textit{ENT}_t$), the lagged value of the
variable in question, and the lagged values of all other dependent variables as controls.

Table~\ref{tab: macro-std-short} summarizes the results. Each column of Table~\ref{tab: macro-std-short} corresponds to one of the above dependent variables. The first set of rows of the table shows the $\beta_{\textit{ENT}}$ coefficient from \eqref{eq:fund}, scaled by the full-sample standard deviation of entropy, and the associated t-statistic. The second set of rows shows the results of \eqref{eq:fund} where $\textit{ENT}_t$ is replaced with the news innovation part of entropy, $\textit{ENT\_NEWS}_t$. The third set of rows shows the results for $\textit{ENT\_MODEL}_t$.%
\footnote{The full results reporting the scaled coefficients are in Online Appendix Tables~\ref{tab:
    macro-std}--\ref{tab: macro-part2-std}. The full results reporting the raw coefficients are in
  Online Appendix Tables~\ref{tab: macro}--\ref{tab: macro-part2}.}

The table shows that the forecasting results for \textit{ENT} and \textit{ENT\_NEWS} are very
similar. An increase in either measure positively forecasts the unemployment rate and the
\textit{VIX} volatility index, and negatively forecasts industrial production, CPI, the level of
interest rates, and the S\&P 500 earnings per share. There is a weak positive forecasting
relationship from these two entropy measures for \textit{EPU}, and no relationship for the 2s-10s
curve. \textit{ENT{\_}MODEL} does not significantly forecast any of the macro outcomes variables,
with the exception of the \textit{VIX}, where the relationship is negative, not positive as with the
other entropy measures. Overall, entropy and its news innovation component robustly forecast
negative economic outcomes one year ahead.

\subsection{Discussion} \label{s:disc}

As in the return forecasting regressions, high \textit{ENT} and high \textit{ENT\_NEWS} significantly
forecast negative macroeconomic outcomes. On the other hand, the model innovation part of entropy
(\textit{ENT\_MODEL}) does not forecast macroeconomic outcomes. Furthermore, exposure to both
\textit{ENT} and \textit{ENT\_NEWS} receives a significant and negative cross-sectional risk premium,
while \textit{ENT\_MODEL} is not a priced risk. The evidence points to a somewhat puzzling
finding. Market participants understand that there is predictability from entropy (newness of news)
to future market and macroeconomic outcomes because they are willing to give up expected returns to
hold securities that allow them to hedge this risk. However, the aggregate market does not react to
this information instantaneously, which is why entropy and its news innovation component forecast
year-ahead market returns. One may have expected entropy to forecast aggregate market returns
positively, as market participants demand compensation to hold risky assets heading into difficult
economic periods. However, we do not find this result. Instead, our finding of aggregate
underreaction is consistent with recent evidence of asset-class-level macro momentum
\citep{aqrecontrend2023}.

The non-instantaneous reaction of markets to entropy may have two causes. First, consistent with the
rational inattention theory of \cite{sims2003}, market participants may not have been aware of times
of high news entropy. Since our entropy measure is difficult to observe -- depending on tens of
thousands of articles today and one year ago, as well as sophisticated NLP techniques -- it is
possible that, historically, market participants were not immediately aware of high entropy
times. Even if such times were only identifiable after the fact, securities that did well in these
ex-post high entropy periods (see discussion in Section \ref{s:which-hedge}), may have been bid up
by investors in anticipation of similar hedging properties in the future. Thus it is possible to
observe cross-sectional entropy risk premia even if market participants could not observe entropy in
real-time. With advances in computational linguistics and data availability, this channel may or may
not be as relevant in the future.

The second, arguably more enduring, reason for such underreaction, is a combination of slow-moving
institutional capital \citep{gk2021,glm2023} with limits to arbitrage \citep{gv2010}. Institutions
may be constrained by mandate to not deviate excessively from a particular set of portfolio targets,
for example, a risk level consistent with a 60/40 stock/bond portfolio. Even if such institutions
begin to believe that markets will not do well over the next year, the amount of uncertainty around
that belief is still large (recall that even the highest out-of-sample R-squareds for year-ahead
market returns in Table \ref{tab: OOS} are less than 10\%). Such institutions may need board or
investment consultant approval before they can change their portfolio allocations, and these
approvals may be slow in coming. Arbitrage capital, like actively managed mutual funds or hedge
funds, may not be big enough relative to the amount of capital controlled by constrained
institutions to provide a sufficient offset. As such, macro-level information may take time to work its
way into stock prices. This channel is unlikely to be impacted by the availability of better or
faster information, suggesting that market return forecastability by entropy may be a persistent
market phenomenon.

\section{Robustness} \label{s:robust}

In this section, we check whether entropy is spanned by existing risk factors, whether the entropy
risk premium can be explained by a news sentiment risk premium, and whether other measures of
uncertainty span entropy or can explain the forecasting power of entropy for market returns.

\subsection{Existing Pricing Factors}

Entropy is distinctive because of its impressive in- and out-of-sample forecasting power for market
returns, the fact that it is the only factor that satisfies the ICAPM sign property of
\eqref{eq:sgn}, and its macroeconomic forecasting ability. However, it is possible that the
information content of entropy is spanned by one of the multitude of existing factors in the
literature. To check for this, we use the 153 value-weighted factors discussed in
\cite{JensenKellyPedersen2022}, whose daily returns are available at
\url{https://jkpfactors.com/}. We augment this set of factors with the \textit{ENT} and \textit{EPU}
factor replicating portfolios, $F_{\textit{ENT}}$ and $F_{\textit{EPU}}$, using size-momentum as the
base assets.%
\footnote{The results for the other base assets are similar.}
We regress each of the 155 factor return series on the remaining 154 factors. 

In Figure \ref{f:channel_r2} we show these R-squareds for each factor, ranked from low to
high. Factors close to the lower left corner are not well explained by the existing factor zoo,
while factors close to the upper right corner are largely spanned by existing, alternative
factors. We label the top and bottom five factors based on this spanning R-squared measure and also
indicate \textit{ENT} by red text. $F_{\textit{ENT}}$, the entropy mimicking portfolio with
size-momentum as the base assets, has the seventh lowest R-squared among the 155 factors. This shows
that \textit{ENT} conveys novel information relative to existing factors.

\subsection{Economic Sentiment}

Another well-known news-based variable is \textit{SEN}, the San Francisco Fed's news sentiment
index. The main return forecasting results in Table \ref{tab:IS-std} show that entropy is a strong
forecaster of year-ahead market returns even after controlling for the information content of
\textit{SEN}, while \textit{SEN} is not a significant return forecaster when \textit{ENT} is
included in the set of forecasting variables. Similarly, Table \ref{tab: OOS} shows that
\textit{SEN} is a poor out-of-sample forecaster of market returns. Online Appendix
Tables \ref{t:fm-same-std-incSEN} (scaled coefficients) and \ref{t:fm-same-incSEN} (unscaled)
show the Fama-MacBeth risk premium analysis which includes $F_{\textit{SEN}}$, the \textit{SEN}
replicating portfolio constructed using the method described in Section~\ref{s:mimick}, in addition
to the other factor exposures. Exposure to \textit{SEN} carries a positive and significant risk
premium in two out of the four tests but does not satisfy the ICAPM property because it is not a
significant in-sample market return forecaster. Even in the presence of $F_{\textit{SEN}}$, exposure
to $F_{\textit{ENT}}$ continues to carry a negative and significant risk premium, consistent with
the results of Section \ref{s:fmrp}.

\subsection{Other Uncertainty Measures}

Another concern is that entropy may be spanned by other uncertainty measures studied in the
literature. To check for this, we utilize measures of macroeconomic uncertainty discussed in the
survey article \cite{david2022survey}.%
\footnote{All the data used in this part of the analysis are available from the survey article.}
In particular, we consider the \cite{bekaert2022time} (\textit{BEX}) uncertainty measure of
time-varying risk aversion, the \cite{jurado2015measuring} (\textit{JLN}) variable which provides
econometric estimates of the conditional volatility of the purely unforecastable component of the
future values of ``hundreds of macroeconomic and financial indicators,'' the
\cite{azzimonti2018partisan} Partisan Conflict Index (\textit{PCI}) which tracks the degree of
disagreement among U.S. politicians at the federal level, as well as \textit{ENT}, \textit{EPU},
\textit{SEN}, and \textit{VIX}.

We regress each of the seven uncertainty measures on the other six uncertainty measures and
calculate the resultant R-squared. In Figure \ref{f:macro_r2} we show these R-squareds for each
uncertainty measure, ranked from lowest to highest. Variables close to the lower left corner are not
well explained by the remaining ones, while variables close to the upper right corner are largely
spanned by the remaining measures. \textit{ENT}, indicated by a red dot, has the lowest R-squared
(0.13) among all considered uncertainty measures, while \textit{EPU} has the highest R-squared
(0.70). The second smallest is \textit{PCI}, whose R-squared of 0.47 is still substantially greater
than that of \textit{ENT}. Our entropy measure is the one that is furthest removed from the
information content of standard measures of uncertainty.

We expand the in-sample analysis of Section~\ref{s:insample} to include multiple uncertainty
measures along with the more standard controls used previously in Table~\ref{tab:IS-std}.  We
estimate a variety of time-series forecasting regressions of the form
\begin{equation} \label{eq:reg-is-robust}
  R_{t+1, t+12} = \beta_{0} + \beta_{\textit{ENT}} \textit{ENT}_t + \eta^{\top} Uncertain_{t} + \gamma^{\top}
  Control_{t} + \epsilon_{t}
\end{equation}
where $R_{t+1, t+12}$ is the cumulative market return from month $t+1$ to month $t+12$ and
$\textit{ENT}_t$ is the entropy measure in month $t$. $Uncertain_{t}$ contains non-text-based
uncertainty measures: \textit{BEX}, \textit{JLN}, \textit{PCI}, and squared implied volatility
(\textit{VIX2}). $Control_{t}$ contains interest rates (\textit{DGS10} and \textit{DGS10-2}), the
dividend yield (\textit{DY}), the difference between actual consumption and the consumption level
predicted by wealth and income (\textit{CAY}), the inverse of the cyclically adjusted
price-to-earnings ratio (\textit{1/CAPE}), the market return of the previous month
(\textit{Return1}), the cumulative market return of the previous 12 months excluding the most recent
month (\textit{Return12}), the cumulative returns of the Fama-French five factors
(\citealt{fama2015five}) and momentum over the previous 60 months (\textit{Return60},
\textit{SMB60}, \textit{HML60}, \textit{RMW60}, \textit{CMA60}, and \textit{UMD60}). These are
  the forecasting variables used in the analysis in Table \ref{tab:IS-std}.

Columns (1) and (2) of Table~\ref{tab:IS-std-short} report the scaled coefficient estimates (and
t-statistics) for \eqref{eq:reg-is-robust}, without and with the uncertainty controls.%
\footnote{The full regression results are shown in Table~\ref{tab:IS-std-robust} (scaled coefficients) and Table~\ref{tab:IS-robust} (raw coefficients) of the Online Appendix.}
A one standard deviation increase in entropy predicts a 2.266\%-2.530\% decrease in the 12-month
ahead market return. Columns (3) and (4) replace \textit{ENT} in \eqref{eq:reg-is-robust}
with two text-based measures: economic policy uncertainty (\textit{EPU}) and sentiment
(\textit{SEN}). The results show that \textit{EPU} positively predicts future market returns but
\textit{SEN} does not have a significant impact. Columns (5) and (6) include all text-based
uncertainty measures \textit{ENT}, \textit{EPU}, and \textit{SEN} in a single regression model,
again without and with uncertainty controls. After controlling for these other measures \textit{ENT}
still forecasts future market returns, and the standardized coefficients for \textit{ENT} (from
-2.306\% to -2.413\%) are nearly unchanged from columns (1) and (2). In all cases, introducing
uncertainty controls (Columns 2, 4, 6) does not meaningfully change the results without the
uncertainty controls (Columns 1, 3, 5).

We also replicate the analysis in Section~\ref{s:outofsample} to test if any of the uncertainty
measures are robust out-of-sample forecasters of aggregate market returns. The results for
\textit{BEX}, \textit{JLN}, \textit{PCI} and \textit{VIX2} in Table~\ref{tab: OOS} show that none of
the uncertainty measures forecast market returns out-of-sample.

\section{Conclusion} \label{s:conclusion}

This paper combines natural language processing with the tools of empirical asset pricing to
investigate a novel aspect of how news affects prices at the aggregate level. We use a recurrent
neural network to derive entropy, a measure of the novelty or unusualness of aggregate news. This measure
extends the prior literature \citep{glasserman2019does} by applying modern NLP tools to solve the
sparsity and context problems with prior entropy measures.

We show that entropy negatively forecasts next twelve-month market returns, even after controlling
for a multitude of known in-sample return forecasters. In particular, entropy does better than
either economic policy uncertainty or news sentiment in our in-sample tests. In an out-of-sample
market forecasting horse race, we find that, remarkably, entropy is the best forecaster of
year-ahead market returns out of a large set of candidate variables.

Using a Fama-MacBeth GMM framework, we show that entropy has a negative risk price in the
cross-section of portfolio returns. Together with the finding that entropy forecasts market returns
negatively, it turns out entropy is the {\it only} factor in our study that is consistent with the
\cite{maio2012} ICAPM sign property. Entropy thus proxies for a factor that negatively impacts
investors' opportunity set and investors are willing to give up expected return to hedge against
entropy innovations.

We show that entropy can be decomposed into one part that reflects the change in news flow and the
other part that reflects the change in the text model, and that it is the former, change-in-news,
that accounts for entropy's time series forecasting properties and its cross-sectional risk
pricing. We further show that the factor mimicking portfolio for entropy is among the least
well-spanned out of 155 long-short factors we obtain from \cite{JensenKellyPedersen2022}.

The forecasting power of entropy for future market returns likely stems from its ability to forecast
future fundamental variables. Both entropy and its news innovation component positively forecast
12-month ahead unemployment and volatility, and negatively forecast industrial production,
inflation, interest rates, and S\&P500 corporate earnings. While entropy risk is priced, markets
appear to not fully react to the information content of entropy either because of informational
constraints or because of slow-moving institutional capital.

Our paper opens up several interesting directions for future work. First, the field of natural
language processing is witnessing rapid advances which may produce more successful network
architectures than recurrent neural networks for measuring news unusualness
\citep[e.g.,][]{mikolov2010recurrent,lee2017end,devlin2018bert,xiong2019pretrained}. Second, our
analysis focuses on the impact of entropy at the aggregate level, while there may also be
interesting applications at the level of individual stocks \citep{glasserman2019does}. Finally, our
entropy factor replicating portfolio should be a
useful factor for asset pricing models.

\appendix
\bibliography{ref.bib}


\begin{figure}[h]
    \centering
    \begin{subfigure}[b]{0.8\textwidth}
        \includegraphics[width=\textwidth]{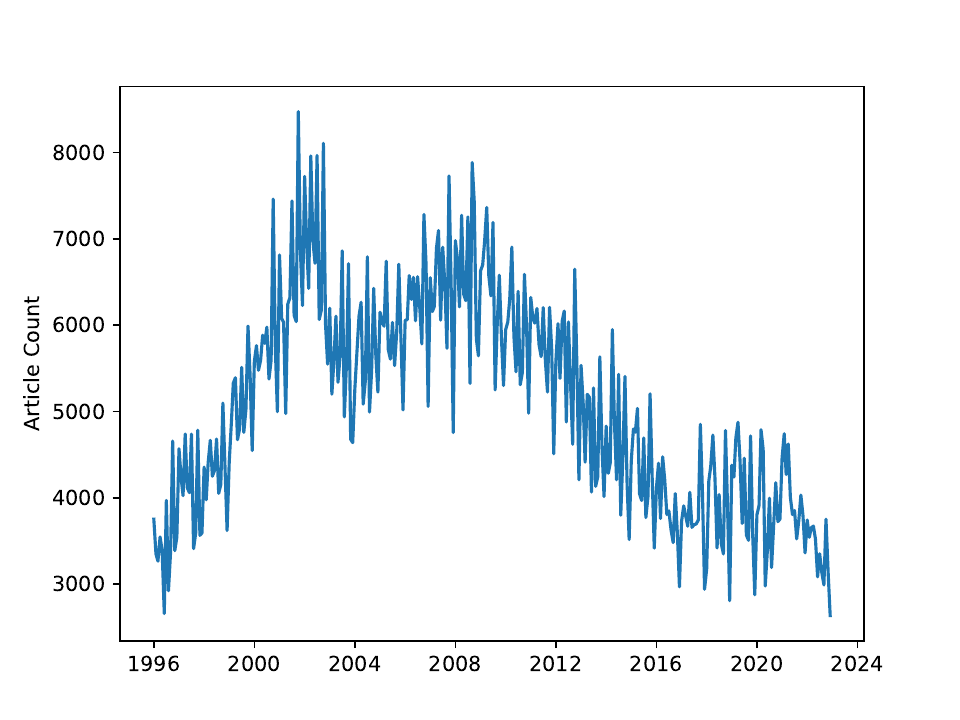}
        \caption{Time series of number of articles in each month.}
    \end{subfigure}
    \\
    \begin{subfigure}[b]{0.8\textwidth}
        \includegraphics[width=\textwidth]{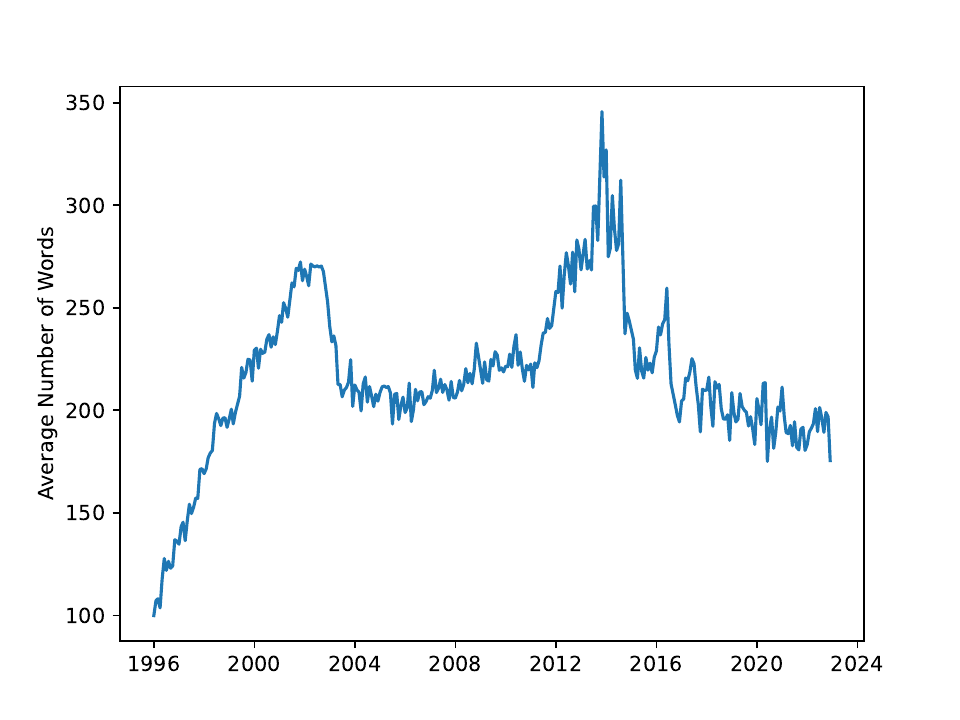}
        \caption{Time series of average article length in each month.}
    \end{subfigure}
    \caption{Thomson Reuters News Feed Direct archive articles characteristics from January 1996 to December 2022.}
    \label{f:numArt}
\end{figure}

\begin{figure}[ht]
  \centering
  \mbox{\includegraphics[width=0.5\textwidth]{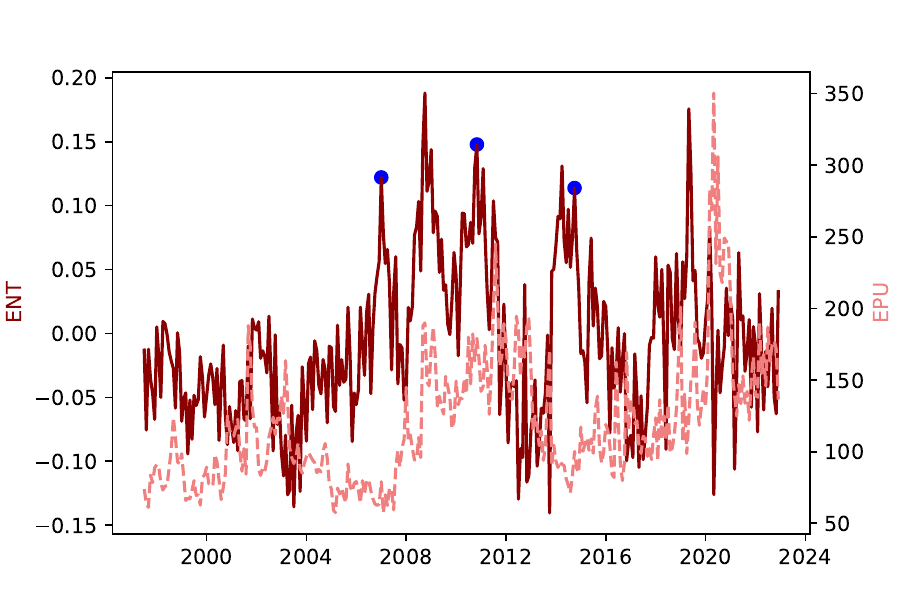}
    \includegraphics[width=0.5\textwidth]{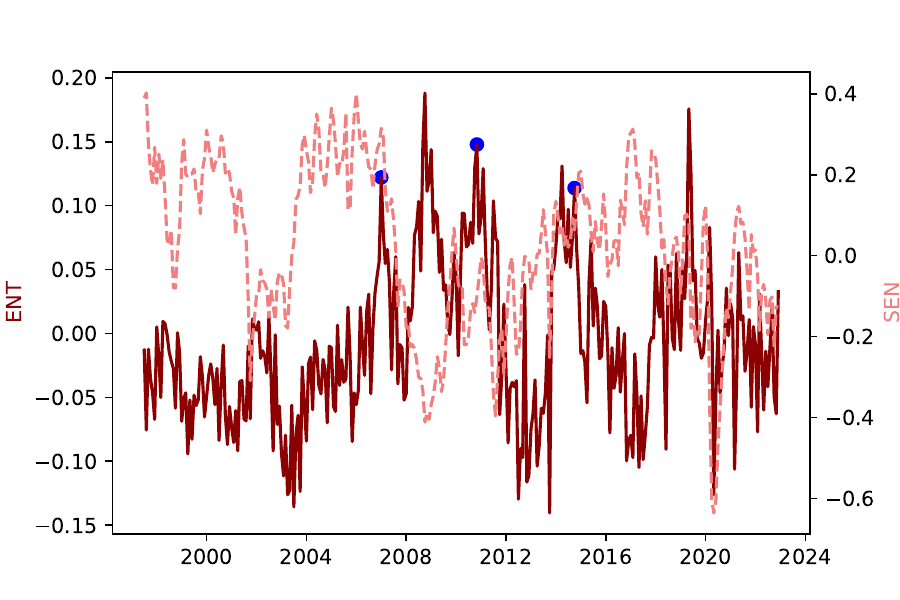}}
  \mbox{\includegraphics[width=0.5\textwidth]{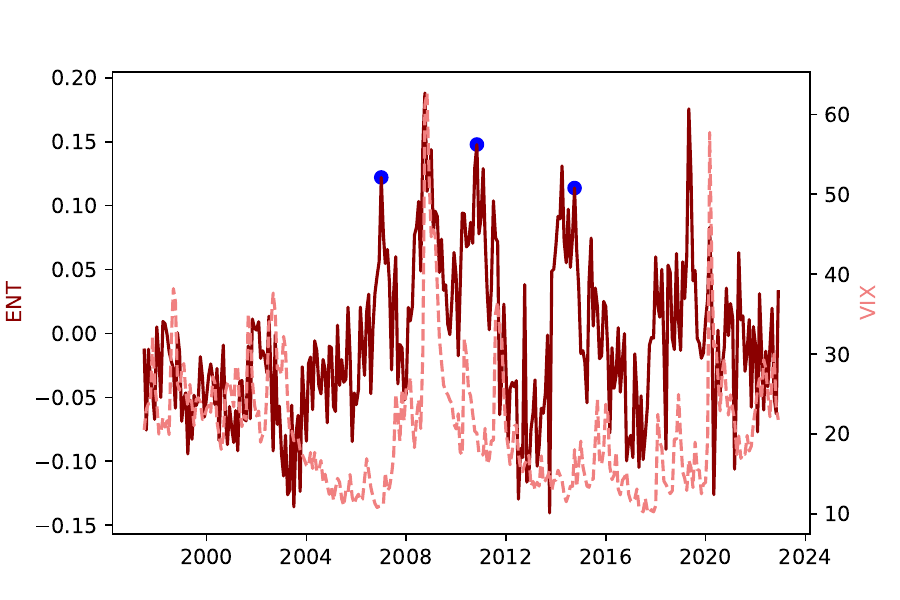}
    \includegraphics[width=0.5\textwidth]{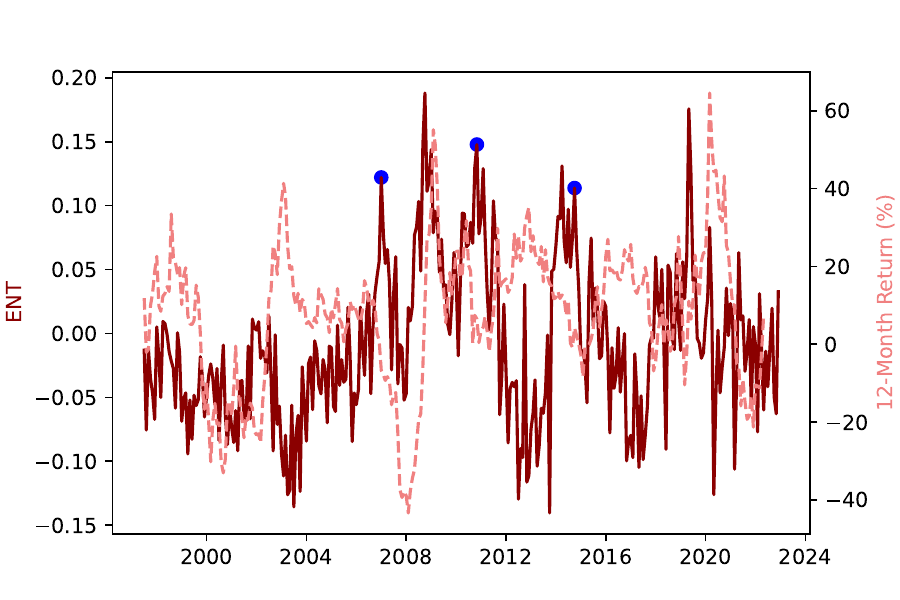}}
  \caption{The top-left panel plots \textit{ENT} and \textit{EPU}. The top-right panel plots
    \textit{ENT} and \textit{SEN}. The bottom-left panel plots \textit{ENT} and the \textit{VIX}
    index. The bottom-right panel plots \textit{ENT} and 12-month ahead cumulative returns on the
    CRSP value-weighted index (ends on December 2021). In all cases, \textit{ENT} is plotted as the
    solid line, with the other series plotted as the dashed line. Data are shown on a monthly
    frequency. The blue dots on each series correspond to the months January 2007, November 2010,
    and October 2014, which are discussed in Section \ref{s:data-overview}.}
  \label{f:ent-ts}
\end{figure}

\begin{landscape}
\begin{figure}[ht]
  \centering
  \includegraphics[width=1.75\textwidth]{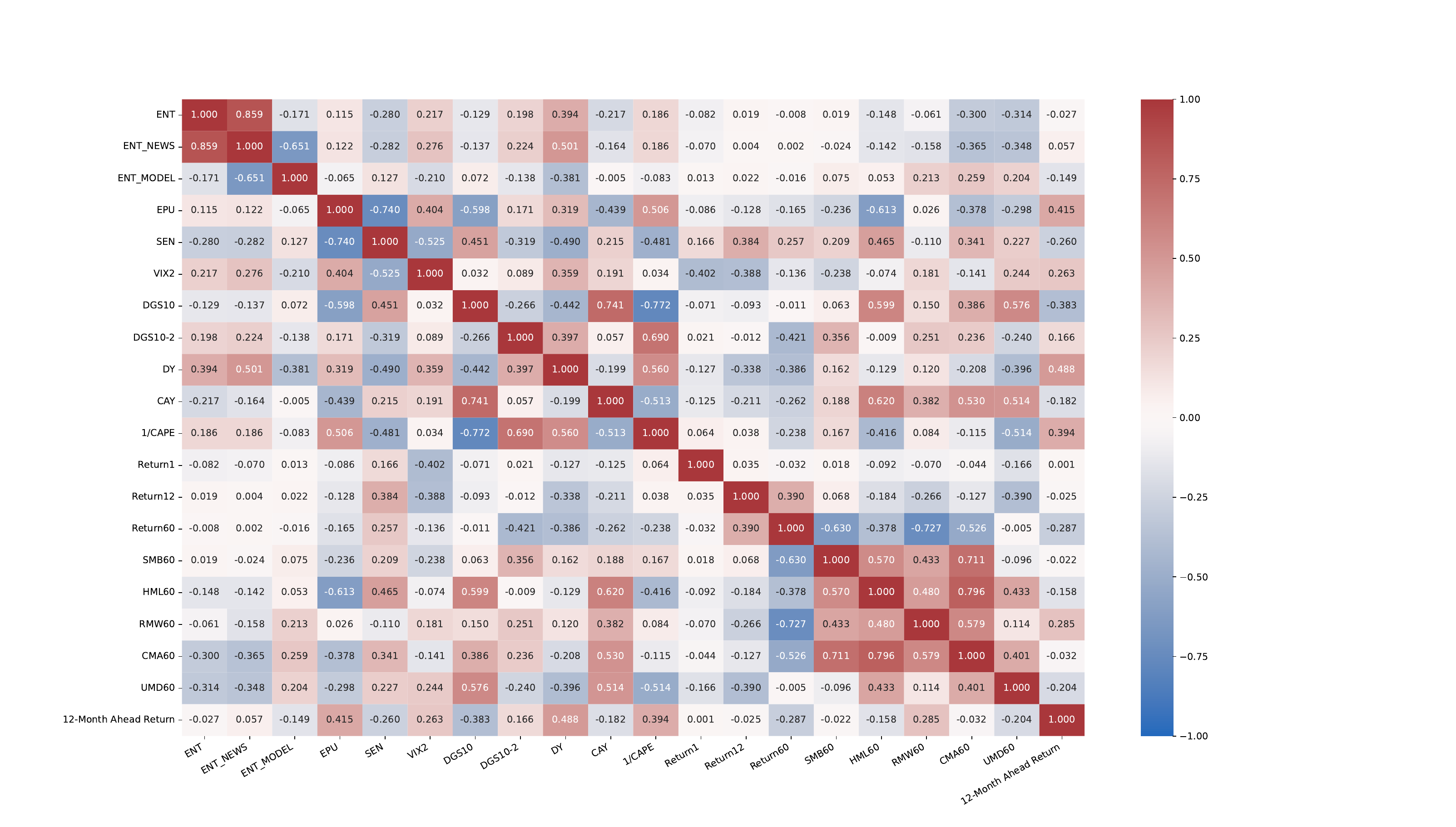}
  \caption{Correlations at monthly observation using data from July 1997 to December 2022, inclusive.}
  \label{f:corrHeatmap}
\end{figure}
\end{landscape}


\begin{figure}[ht]
  \centering
  \mbox{
    \includegraphics[width=0.33\textwidth]{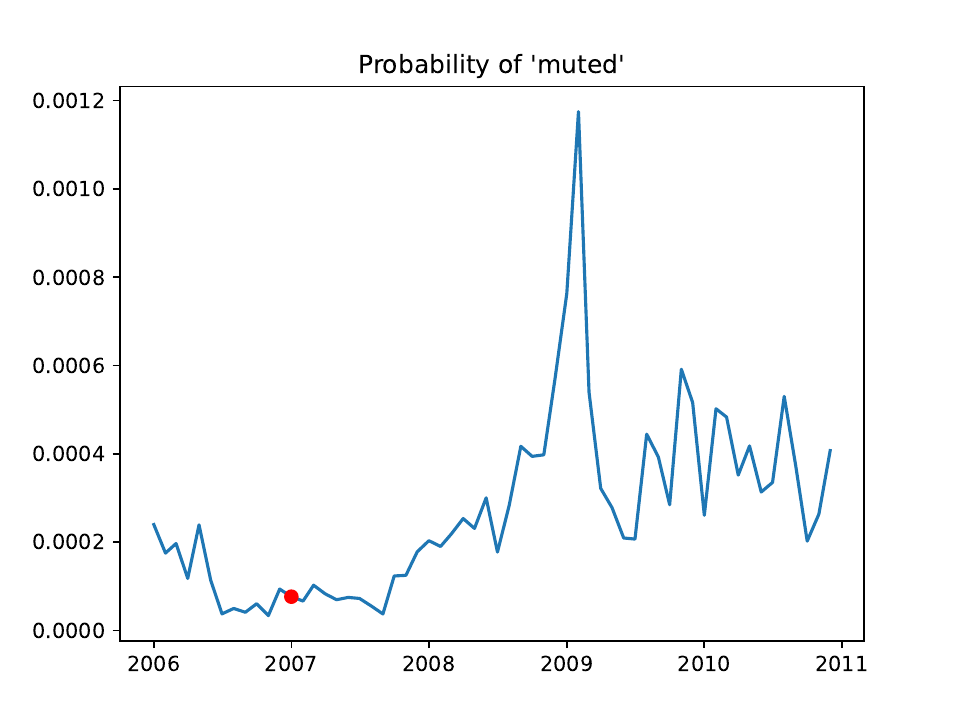}
    \includegraphics[width=0.33\textwidth]{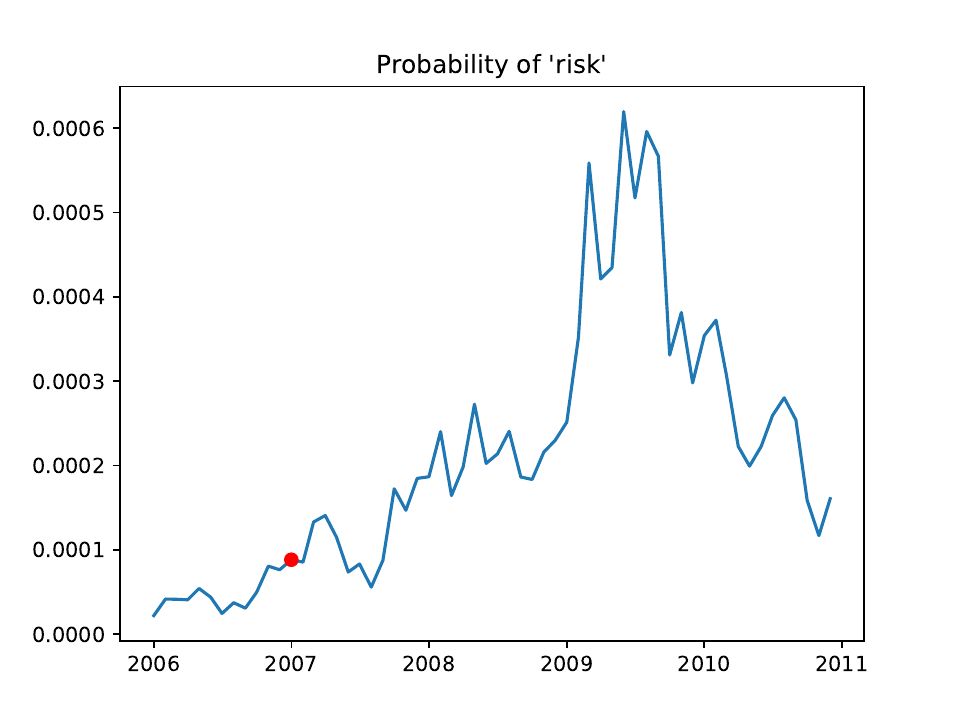}
    \includegraphics[width=0.33\textwidth]{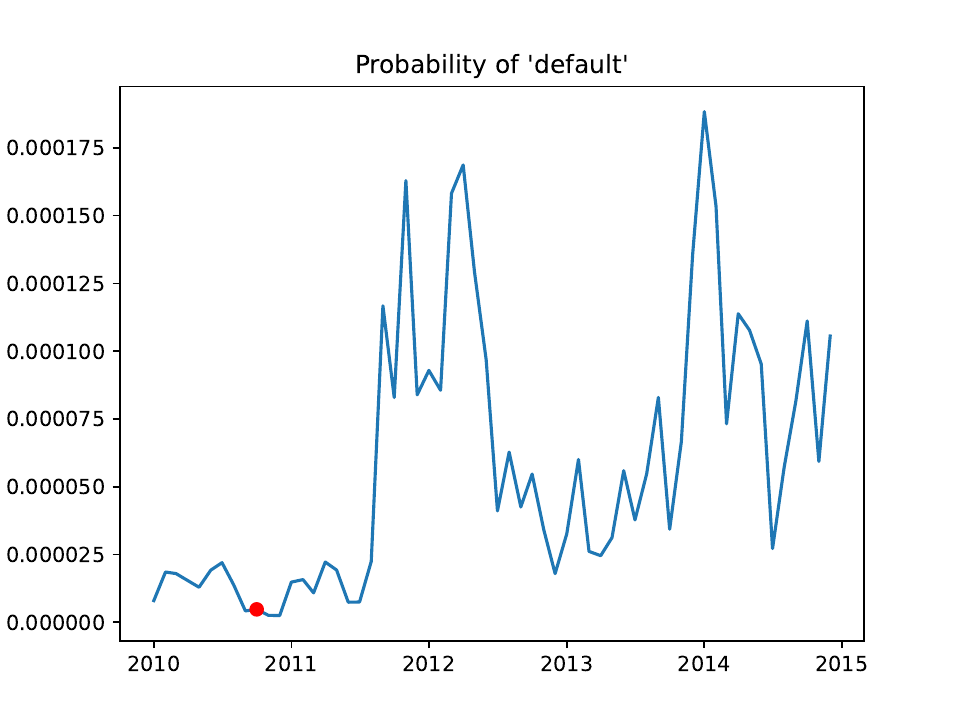}}
  \caption{The left panel plots the probability of ``muted'' following ``fannie mae and freddie mac
    growth will be.'' The middle panel plots the probability of ``risk'' following ``lack of
    progress in reining in mortgage lenders fannie mae and freddie mac leaves the economy at.'' The
    right panel plots the probability of ``default'' following ``growing concerns Ireland will be
    forced to.'' The red dot in each panel marks the date the corresponding sentence appears in the
    database.}
  \label{f:anecdotal-example}
\end{figure}

\begin{figure}[ht]
  \centering
  \mbox{
    \includegraphics[width=0.5\textwidth,trim={2.5cm 1cm 2.5cm 1cm},clip]{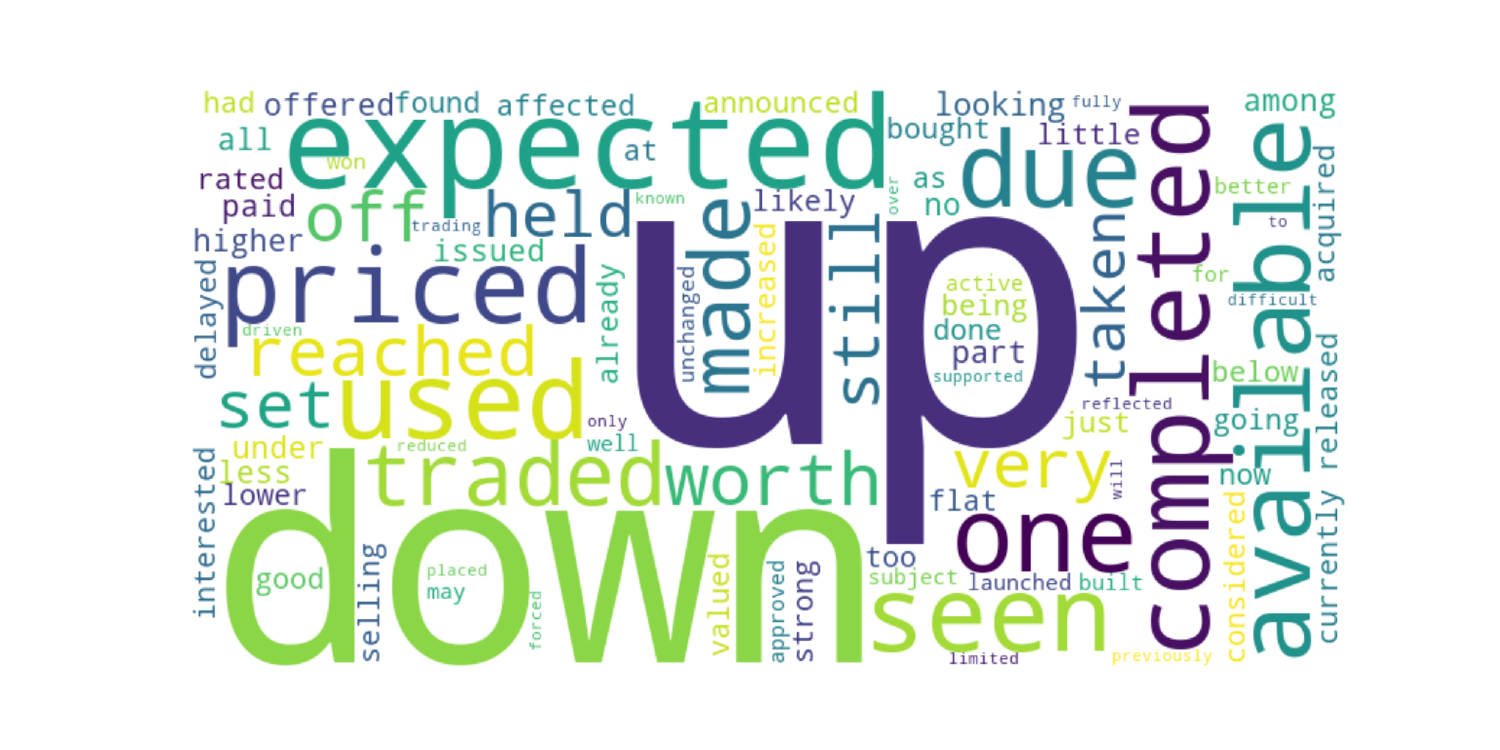}
    \includegraphics[width=0.5\textwidth,trim={2.5cm 1cm 2.5cm 1cm},clip]{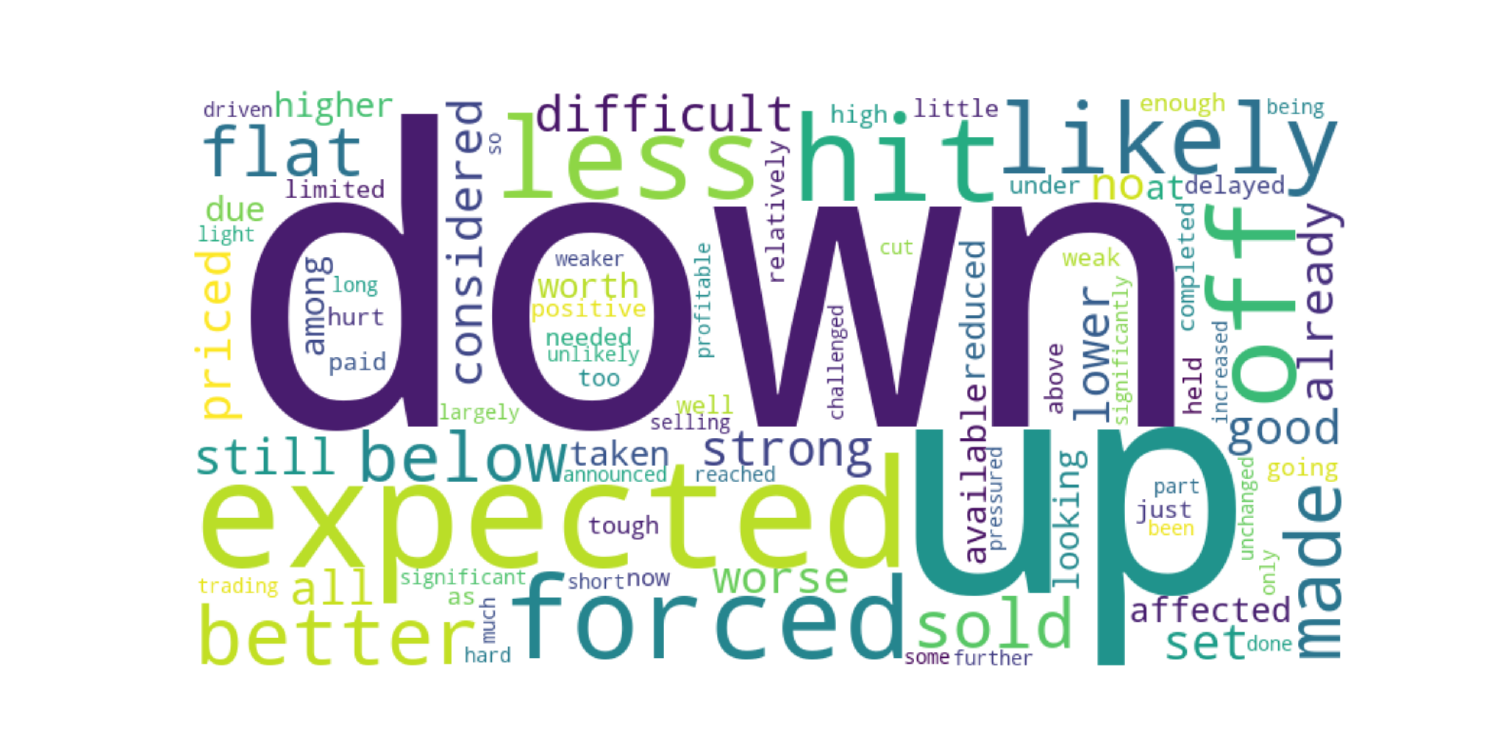}
  }
  \caption{The most probable 100 words following the string of words ``fannie mae and freddie mac
    growth will be'' under the December 2006 model (left panel) and under the December 2008 model
    (right panel). The size of each word is proportional to its probability.}
  \label{f:anecdotal-cloud}
\end{figure}

\begin{figure}[h]
    \centering
    \mbox{
    \includegraphics[width=0.5\textwidth]{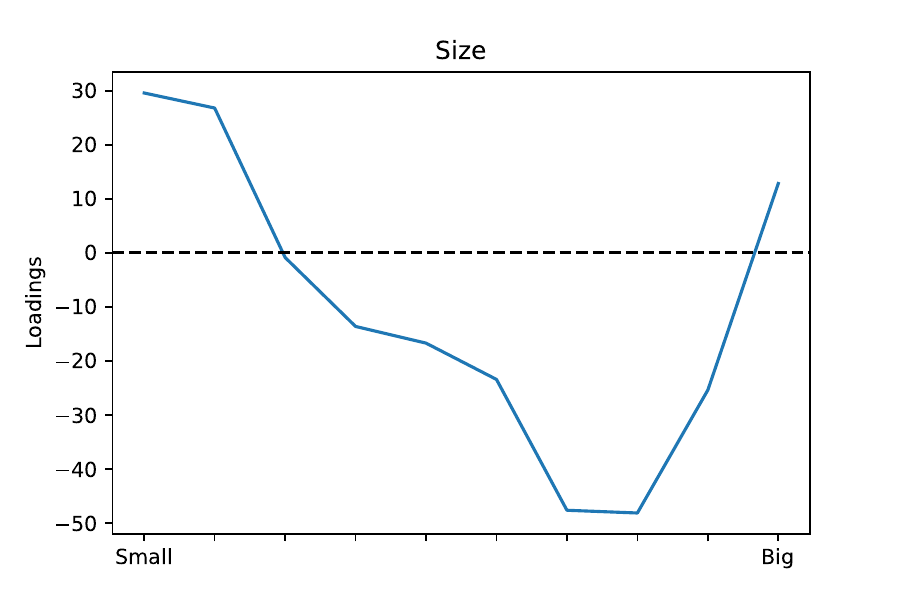}}
    
    \mbox{
    \includegraphics[width=0.5\textwidth]{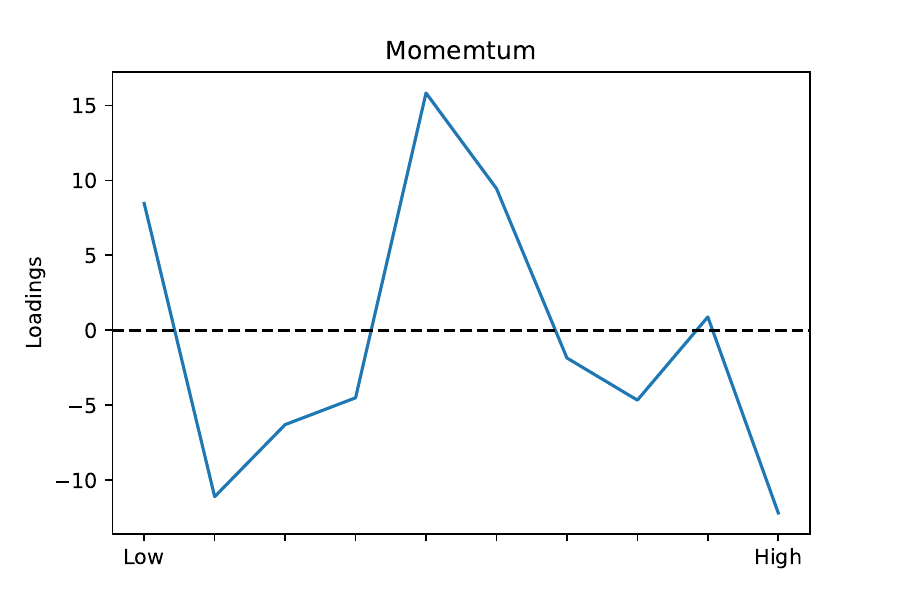}
    \includegraphics[width=0.5\textwidth]{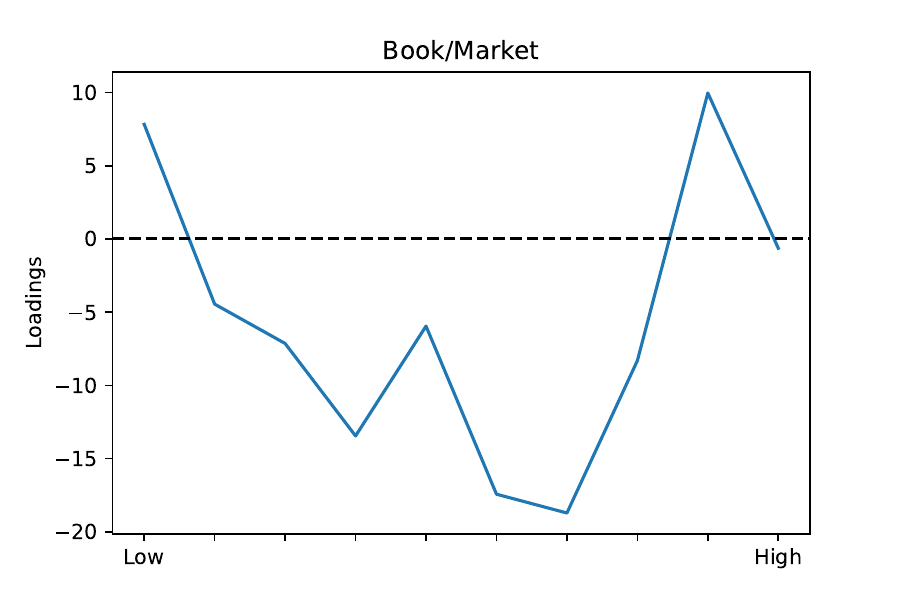}}
    
    \mbox{
    \includegraphics[width=0.5\textwidth]{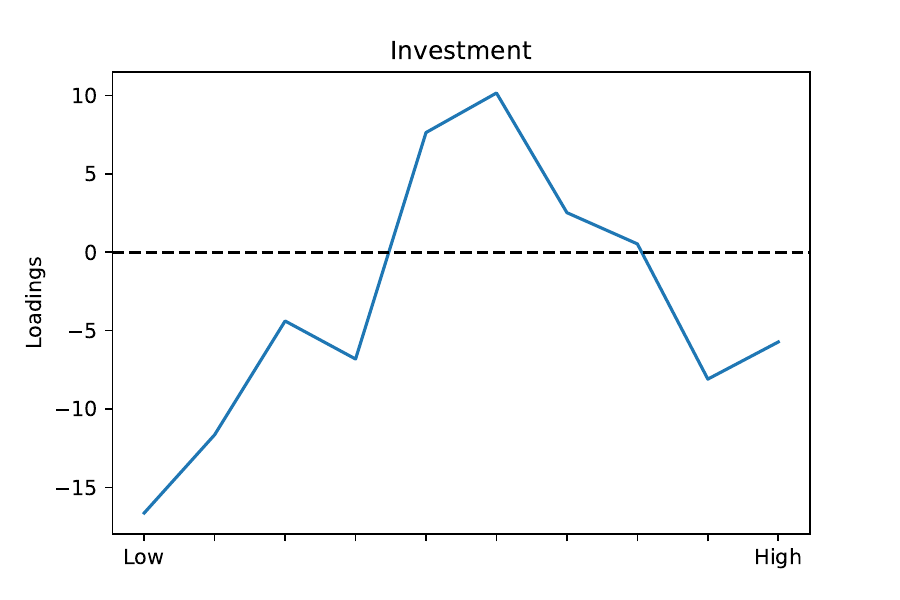}
    \includegraphics[width=0.5\textwidth]{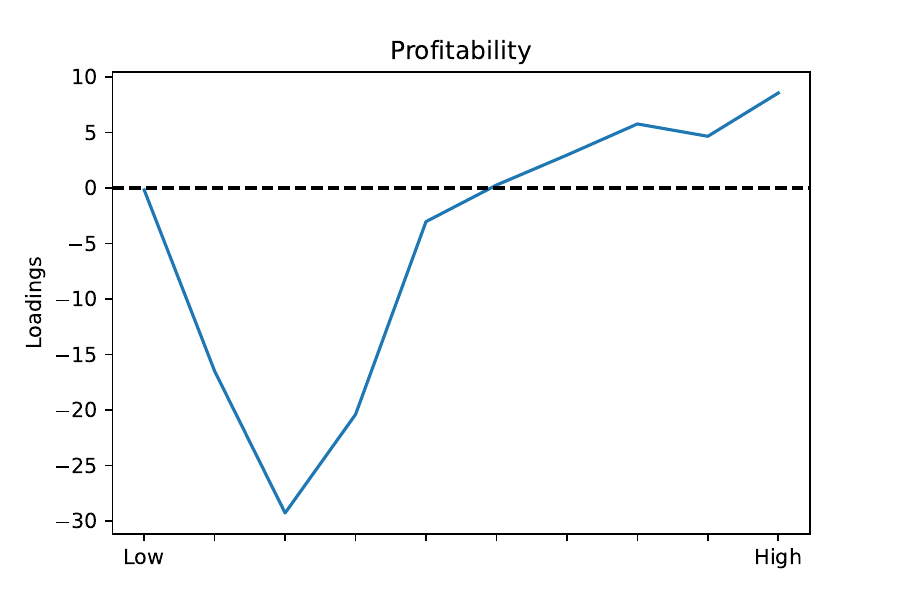}}
    
  \caption{The first panel shows loadings on $F_{\textit{ENT}}$ of 10 portfolios sorted on size. The
    other panels show loadings on $F_{\textit{ENT}}$ of portfolios sorted on momentum, book/market,
    investment, and operating profitability, respectively. For each panel, we construct
    $F_{\textit{ENT}}$ using 100 base assets: 25 portfolios formed on size-momentum,
    size-book/market, size-investment, and size-profitability. The methodology is described in
    Section \ref{s:which-hedge}.}
\label{f:FM-univariate-average}
\end{figure}

\begin{figure}[h]
  \centering
  \includegraphics[width=\textwidth]{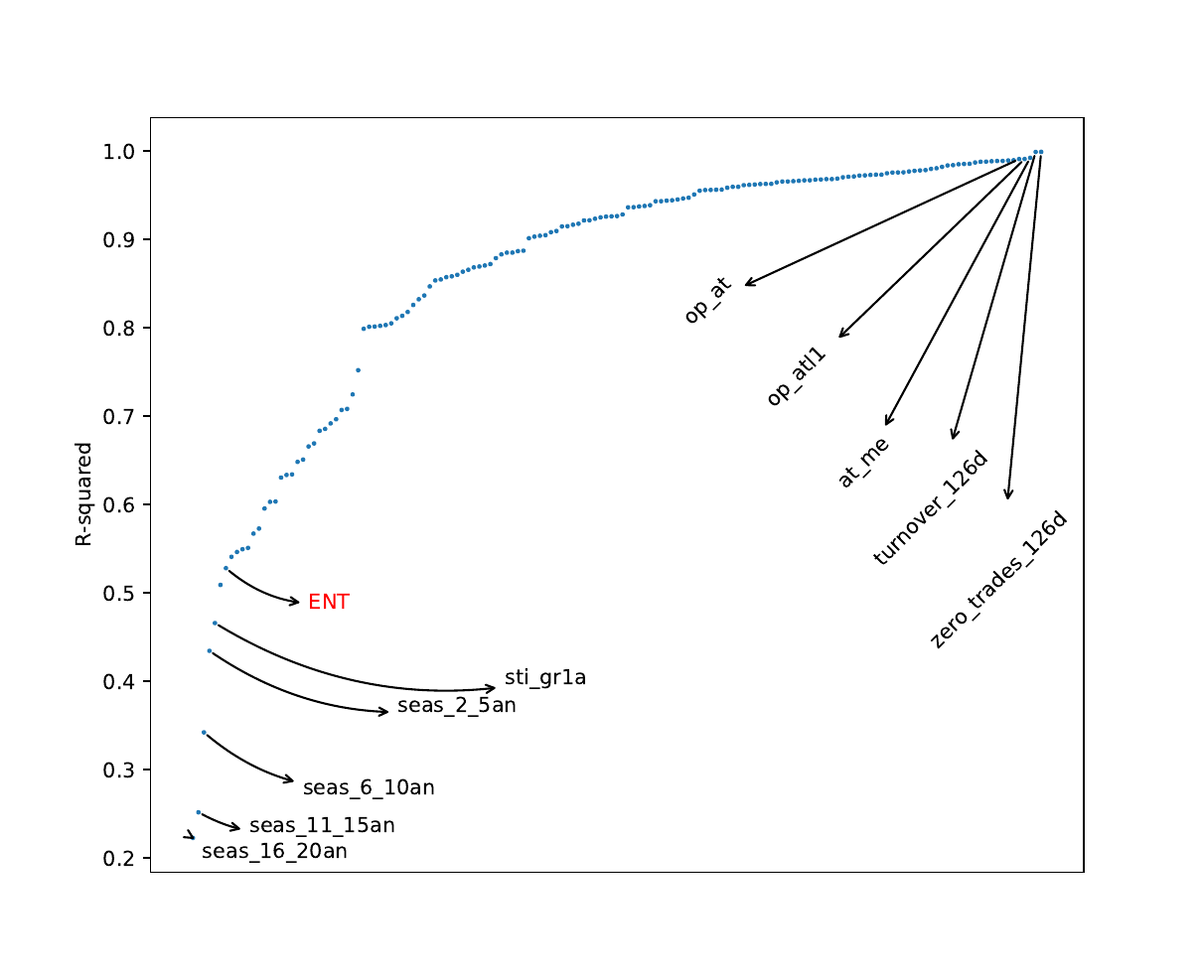}
  \caption{In-sample $R^2$, ordered from lowest to highest, of regressing each 153 value-weighted
    factors in \cite{JensenKellyPedersen2022} plus $F_{\textit{ENT}}$ and $F_{\textit{EPU}}$ on all
    other factors using the OLS model. Top and bottom five factors based on this spanning R-squared
    measure are annotated. Five factors with the highest R-squared are zero\_trades\_126d (number of
    zero trades with turnover as tiebreaker, 6 months), turnover\_126d (share turnover), at\_me
    (assets to market), ivol\_ff3\_21d (idiosyncratic volatility from the Fama-French 3-factor
    model), op\_atl1 (ball operating profit scaled by lagged assets), and op\_at (ball operating
    profit to assets). Five factors with the lowest R-squared are aseas\_2\_5an (year 2-5 lagged
    returns, annual), seas\_6\_10an (year 6-10 lagged returns, annual), sti\_gr1a (change in
    short-term investments), seas\_11\_15an (year 11-15 lagged returns, annual), seas\_16\_20an
    (year 16-20 lagged returns, annual). The entropy mimicking portfolio, $F_{\textit{ENT}}$, ranks
    as the seventh lowest R-squared among all 155 factors.}
  \label{f:channel_r2}
\end{figure}

\begin{figure}[h]
  \centering
  \includegraphics[width=\textwidth]{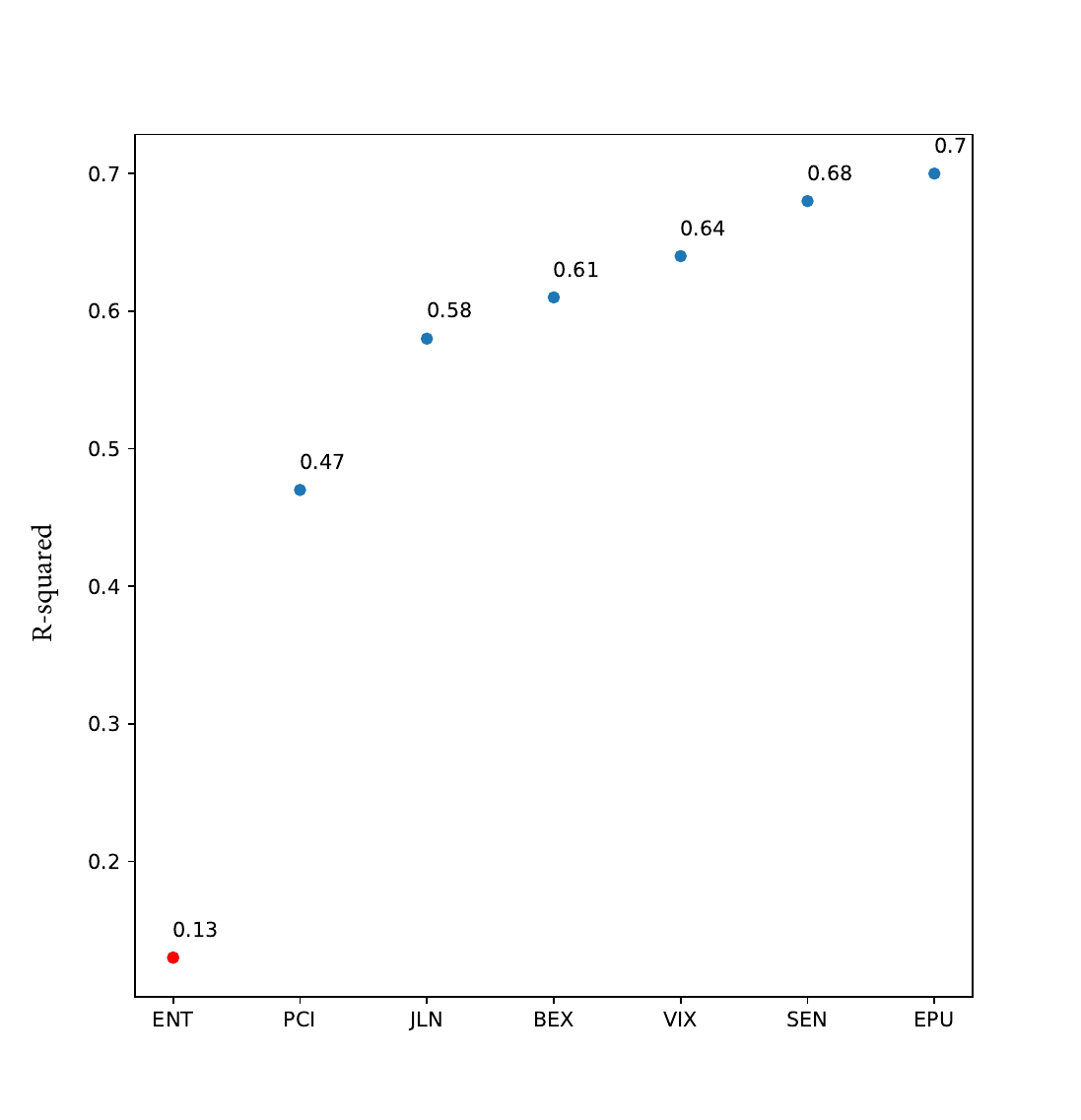}
  \caption{In-sample $R^2$, ordered from lowest to highest, of regressing each variable on all other
    variables using OLS model. Included measures of macroeconomic uncertainty, ranked from lowest to
    highest, are entropy (\textit{ENT}, marked in red), the \cite{azzimonti2018partisan} Partisan
    Conflict Index (\textit{PCI}), the \cite{jurado2015measuring} (\textit{JLN}) uncertainty
    measure, the \cite{bekaert2022time} (\textit{BEX}) uncertainty measure, the Chicago Board
    Options Exchange’s CBOE Volatility Index (\textit{VIX}), the \cite{shapiro2022measuring} news
    sentiment index (\textit{SEN}), and the \cite{baker2016measuring} Economic Policy Uncertainty
    (\textit{EPU}).  }
  \label{f:macro_r2}
\end{figure}


\begin{landscape}
\begin{table}[h]
  \caption{This table presents summary statistics of variables at the monthly frequency used in this
    study, including entropy measure (\textit{ENT}), the first part of the decomposition of entropy
    (\textit{ENT\_NEWS}), the second part of the decomposition of entropy (\textit{ENT\_MODEL}),
    economic policy uncertainty (\textit{EPU}), sentiment (\textit{SEN}), squared implied volatility
    (\textit{VIX2}), market yield on U.S. treasury securities at 10-year constant maturity
    (\textit{DGS10}), market yield on U.S. treasury securities at 10-year constant maturity minus
    2-year constant maturity (\textit{DGS10-2}), the dividend yield (\textit{DY}), the difference
    between actual consumption and the consumption level predicted by wealth and income
    (\textit{CAY}), the inverse of the cyclically adjusted price-to-earnings ratio
    (\textit{1/CAPE}), the market return of the previous month (\textit{Return1}), the cumulative
    market return of the previous 12 months excluding the most recent month (\textit{Return12}), the
    cumulative returns of the Fama-French five factors and momentum factor over the previous 60
    months (\textit{Return60}, \textit{SMB60}, \textit{HML60}, \textit{RMW60}, \textit{CMA60}, and
    \textit{UMD60}).}
    \label{tab: summary-stats-month}
    \centering
    
\begin{tabular}{@{\extracolsep{5pt}}lD{.}{.}{6} D{.}{.}{6} D{.}{.}{6} D{.}{.}{6} D{.}{.}{6} D{.}{.}{6} D{.}{.}{6} } 
\\[-1.8ex]\hline 
\hline \\[-1.8ex] 
\\[-1.8ex] & \multicolumn{1}{c}{$\text{mean}$} & \multicolumn{1}{c}{$\text{std}$} & \multicolumn{1}{c}{$\text{min}$}  & \multicolumn{1}{c}{$\text{25\%}$} & \multicolumn{1}{c}{$\text{50\%}$} & \multicolumn{1}{c}{$\text{75\%}$} & \multicolumn{1}{c}{$\text{max}$} \\ 
\hline \\[-1.8ex]

$\textit{ENT}$           &   -0.0073 &    0.0613 &   -0.1404 &   -0.0499 &   -0.0135 &    0.0322 &     0.1881 \\
$\textit{ENT{\_}NEWS}$             &    0.1542 &    0.0795 &   -0.0310 &    0.0965 &    0.1384 &    0.2023 &     0.4708 \\
$\textit{ENT{\_}MODEL}$             &   -0.1615 &    0.0413 &   -0.3006 &   -0.1830 &   -0.1563 &   -0.1343 &    -0.0733 \\
$\textit{EPU}$ &  120.1456 &   45.0389 &   57.2026 &   88.2728 &  109.7164 &  144.1681 &   350.4598 \\
$\textit{SEN}$               &    0.0225 &    0.1964 &   -0.6361 &   -0.1022 &    0.0377 &    0.1775 &     0.4019 \\
$\textit{BEX}$               &    0.4059 &    0.2150 &    0.0878 &    0.2820 &    0.3624 &    0.4875 &     1.6233 \\
$\textit{JLN}$               &    0.6704 &    0.1245 &    0.5299 &    0.5924 &    0.6310 &    0.6920 &     1.2166 \\
$\textit{PCI}$               &  116.1411 &   38.9802 &   34.7400 &   84.6175 &  107.6350 &  143.0900 &   271.2900 \\
$\textit{VIX2}$              &  492.9679 &  474.4772 &  102.5248 &  209.2536 &  375.2209 &  600.2726 &  3927.3970 \\
$\textit{DGS10}$             &    3.4407 &    1.4030 &    0.5600 &    2.3500 &    3.4405 &    4.4775 &     6.6200 \\
$\textit{DGS10-2}$           &    1.1069 &    0.8725 &   -0.7200 &    0.2725 &    1.1010 &    1.8200 &     2.8700 \\
$\textit{DY}$                &    1.8094 &    0.3790 &    1.1100 &    1.6100 &    1.8200 &    2.0000 &     3.6000 \\
$\textit{CAY}$               &   -0.0072 &    0.0188 &   -0.0444 &   -0.0232 &   -0.0049 &    0.0073 &     0.0227 \\
$\textit{1/CAPE}$              &   -0.1132 &    0.1674 &   -0.5346 &   -0.2539 &   -0.0567 &    0.0366 &     0.0659 \\
$\textit{Return1}$  &    0.6438 &    4.6863 &  -17.2300 &   -2.0200 &    1.2300 &    3.4800 &    13.6500 \\
$\textit{Return12}$ &    8.0261 &   17.0219 &  -42.8302 &    0.0748 &    9.8582 &   18.0511 &    59.6603 \\
$\textit{Return60}$ &   51.4848 &   52.2836 &  -36.2647 &    1.3241 &   55.0996 &   95.4152 &   178.0090 \\
$\textit{SMB60}$          &    9.4205 &   25.0681 &  -50.7400 &   -7.8025 &    7.7350 &   20.6425 &    78.2400 \\
$\textit{HML60}$          &    6.6328 &   27.9585 &  -53.4300 &  -12.5600 &    2.2200 &   26.3650 &    91.0700 \\
$\textit{RMW60}$          &   19.8381 &   15.8996 &  -25.4300 &    7.1425 &   20.2550 &   30.6275 &    80.4300 \\
$\textit{CMA60}$          &   12.8265 &   22.5871 &  -19.2300 &   -2.5200 &    7.1000 &   21.6575 &    76.0000 \\
$\textit{UMD60}$          &   24.4751 &   34.9969 &  -49.7600 &    2.1925 &   18.9050 &   48.9200 &   108.9900 \\

\hline 
\hline \\[-1.8ex] 
\end{tabular} 
\end{table}
\end{landscape}

\begin{landscape}
\scriptsize
  \begin{xltabular}{\linewidth}{>{\hsize=.13\hsize\linewidth=\hsize}X |
      >{\hsize=.4\hsize\linewidth=\hsize}X | X | >{\hsize=.1\hsize\linewidth=\hsize}X}
    \caption{Illustrative examples of high entropy articles. The table shows articles from three
      distinct months (first column) when \textit{ENT} were high: January 2007, November 2010, and
      October 2014. Within each month, articles are ranked from the highest to lowest based on their
      raw entropy scores (last column). The displayed headlines (second column) and first sentences
      (third column) are from selected articles, whose entropy scores are within the top 1\% for
      their respective month.} \label{t:ent-ex} \\
    \textbf{\normalsize Month} & \textbf{\normalsize Headline} & \textbf{\normalsize Body (first sentence)} & \textbf{\normalsize Score}\\

\textbf{2007-01} & Freddie sells more than half \$7 bln notes overseas & Freddie Mac says it sold 58\% of its \$4 billion 2-year notes to overseas investors, with central bankers taking 46\% of the offering. & 6.200 \\ 

\textbf{2007-01} & US House sends business issues to Senate for action & Raising the minimum wage, changing Medicare drug purchasing and other initiatives important to the business community are landing on the U.S. Senate's doorstep, having won rapid passage since Jan. 9 in the House of
 Representatives. & 6.016 \\ 

\textbf{2007-01} & UBS sees narrow GSE debt spreads through 2007 & Agency and mortgage securities yield spreads will stay narrow through most of this year, UBS strategists Laurie Goodman and Ivan Hrazdira said in an investor conference call. & 5.936 \\ 

\textbf{2007-01} & Fed's Poole says lag in GSE reform leaves crisis risk & Lack of progress in reining in mortgage lenders Fannie Mae and Freddie Mac leaves the economy at risk of possible financial crisis, St. Louis Federal Reserve Bank President William Poole said on Wednesday. & 5.754 \\

 \textbf{2010-11} & U.S. as Currency Manipulator? It's a Bit Rich & Breakingviews U.S. Editor Rob Cox says the Fed's QE policy fits its mandate, despite China's criticism over what it called an ``indirect currency manipulation.''  & 6.384 \\ 
 
 \textbf{2010-11} &  Ireland Makes Concessions to Restore Confidence & Following S\&P's downgrade of Ireland's credit rating to single A, the Ireland government said it will reduce current spending, cut minimum wage and maintain its debated corporate tax at 12.5 percent. & 6.266 \\

 \textbf{2010-11} & QE2 Critics Put Easing on Agenda at G20 Meeting in Seoul & Emerging market economies critical of the Fed's asset-purchase program have said quantitative easing threatens to flood their economies with excess liquidity. & 6.191 \\

 \textbf{2010-11} & Euro Slumps on Irish Debt Fears  & The euro falls to a 5-week low on growing concerns Ireland will be forced to default on its debts. & 5.912 \\

 \textbf{2014-10} & Wall St declines after Fed ends bond-buying program & U.S. stocks fell on Wednesday, adding to their earlier declines after the Federal Reserve ended its monthly bond purchase program, as had been expected. & 4.524 \\ 

 \textbf{2014-10} & Futures lower, investors look ahead to GDP data & U.S. stock index futures were slightly lower on Thursday as investors looked ahead to a report on economic growth and continued to digest recent comments from the Federal Reserve. & 4.187  \\

 \textbf{2014-10} & Wall St flat after GDP, but Visa lifts Dow & U.S. stocks were mostly flat on Thursday, as a strong read on third-quarter economic growth raised new questions about monetary policy, though strong results at Visa single-handedly put the Dow in positive territory. & 4.116

  \end{xltabular}
\end{landscape}

  \begin{table}[h]
    \caption{In-sample predictions of 12-month ahead cumulative market returns. Returns are measured
      in percent. \textit{Return60}, \textit{SMB60}, \dots, \textit{UMD60} convert factor returns to
      state variables as explained in Section \ref{s:data}. The pre-COVID period ends in 2019, i.e.,
      the year-ahead returns on the left-hand side of the regression in Equation~\eqref{eq: reg-is}
      do not extend past 2019. The coefficient estimates have been normalized by the standard
  deviation of the right-hand side variables. Robust t-statistics are in parentheses and are based on Newey–West
      standard errors with four lags.}
    \label{tab:IS-std}
    \centering
    \resizebox{\columnwidth}{!}{%
\begin{tabular}{@{\extracolsep{5pt}}lD{.}{.}{6} D{.}{.}{6} D{.}{.}{6} D{.}{.}{6} D{.}{.}{6} D{.}{.}{6}} \\[-1.8ex]
\hline 
\hline \\[-1.8ex] 

& \multicolumn{6}{c}{\text{12-Month Ahead Cumulative Return}} \\ 

\cline{2-7} \\[-1.8ex] 

& \multicolumn{3}{c|}{Full sample} & \multicolumn{3}{c}{Pre-COVID} \\ 

& \multicolumn{1}{c}{(1)} & \multicolumn{1}{c}{(2)} & \multicolumn{1}{c|}{(3)} & \multicolumn{1}{c}{(4)} & \multicolumn{1}{c}{(5)} & \multicolumn{1}{c}{(6)}\\ 

\hline \\[-1.8ex]

$\textit{ENT}$  &   -2.821^{***} &                &   -2.607^{***} &  -2.728^{***} &                &   -2.694^{***} \\
     &       (-3.358) &                &       (-3.055) &      (-2.999) &                &       (-2.973) \\
$\textit{EPU}$  &                &    3.841^{***} &    3.436^{***} &               &          1.239 &          0.937 \\
     &                &        (3.185) &        (2.889) &               &        (0.946) &        (0.750) \\
$\textit{SEN}$  &          0.399 &          2.518 &          1.744 &     3.274^{*} &     4.411^{**} &      3.484^{*} \\
     &        (0.231) &        (1.479) &        (1.032) &       (1.782) &        (2.365) &        (1.868) \\
$\textit{VIX2}$ &    3.856^{***} &      2.217^{*} &     2.699^{**} &         1.613 &          0.712 &          1.376 \\
     &        (2.703) &        (1.696) &        (2.011) &       (1.246) &        (0.529) &        (1.051) \\
$\textit{DGS10}$&         -1.204 &         -1.550 &         -0.557 &        -0.480 &         -1.304 &         -0.281 \\
     &       (-0.927) &       (-1.101) &       (-0.423) &      (-0.406) &       (-0.984) &       (-0.236) \\
$\textit{DGS10-2}$&   -4.269^{***} &   -4.762^{***} &   -3.682^{***} &        -1.722 &     -2.839^{*} &         -1.647 \\
     &       (-3.139) &       (-3.299) &       (-2.709) &      (-1.194) &       (-1.883) &       (-1.131) \\
$\textit{DY}$   &   16.164^{***} &   16.153^{***} &   16.894^{***} &  16.679^{***} &   16.462^{***} &   16.906^{***} \\
     &        (8.529) &        (8.288) &        (8.984) &      (10.353) &        (9.268) &        (9.981) \\
$\textit{CAY}$  &   -6.445^{***} &    -4.317^{**} &   -5.803^{***} &        -3.963 &         -2.283 &         -3.950 \\
     &       (-3.319) &       (-1.994) &       (-2.936) &      (-1.486) &       (-0.780) &       (-1.473) \\
$\textit{1/CAPE}$&         -3.402 &         -2.516 &     -3.607^{*} &        -2.462 &         -1.639 &         -2.601 \\
     &       (-1.625) &       (-1.095) &       (-1.734) &      (-1.194) &       (-0.703) &       (-1.244) \\
$\textit{Return1}$&    3.168^{***} &    3.164^{***} &    3.098^{***} &   2.396^{***} &    2.415^{***} &    2.441^{***} \\
     &        (5.700) &        (5.633) &        (5.507) &       (3.721) &        (3.735) &        (3.756) \\
$\textit{Return12}$&    9.248^{***} &    8.939^{***} &    9.046^{***} &   8.242^{***} &    8.063^{***} &    8.220^{***} \\
     &        (6.554) &        (6.220) &        (6.436) &       (6.181) &        (6.007) &        (6.149) \\
$\textit{Return60}$&     -4.773^{*} &         -4.365 &         -3.460 &        -4.381 &         -5.662 &         -3.911 \\
     &       (-1.691) &       (-1.552) &       (-1.214) &      (-1.107) &       (-1.421) &       (-0.989) \\
$\textit{SMB60}$&  -14.645^{***} &  -15.086^{***} &  -14.077^{***} & -12.970^{***} &  -14.321^{***} &  -12.898^{***} \\
     &       (-5.406) &       (-5.738) &       (-5.267) &      (-5.090) &       (-5.646) &       (-5.084) \\
$\textit{HML60}$&   -7.100^{***} &    -5.980^{**} &    -5.795^{**} &   -5.927^{**} &    -5.986^{**} &    -5.650^{**} \\
     &       (-2.688) &       (-2.216) &       (-2.229) &      (-2.090) &       (-2.088) &       (-2.001) \\
$\textit{RMW60}$&     4.574^{**} &      4.101^{*} &     4.779^{**} &   6.382^{***} &     5.579^{**} &    6.396^{***} \\
     &        (2.014) &        (1.841) &        (2.172) &       (2.849) &        (2.398) &        (2.873) \\
$\textit{CMA60}$&   19.474^{***} &   19.598^{***} &   18.537^{***} &  14.566^{***} &   15.994^{***} &   14.562^{***} \\
     &        (5.472) &        (5.848) &        (5.329) &       (4.040) &        (4.480) &        (4.027) \\
$\textit{UMD60}$&         -0.606 &          0.139 &         -0.093 &         0.112 &          0.346 &          0.217 \\
     &       (-0.388) &        (0.090) &       (-0.061) &       (0.074) &        (0.218) &        (0.144) \\

\hline 
\hline \\[-1.8ex] 
\textit{Note:}  & \multicolumn{6}{r}{$^{*}$p$<$0.1; $^{**}$p$<$0.05; $^{***}$p$<$0.01} \\

\end{tabular}%
}

  \end{table}

\begin{table}[h]
  \caption{This table shows out-of-sample R-squareds, calculated using \eqref{eq:oos-R2}, of
    12-month ahead market return forecasts using monthly rolling estimates of \eqref{eq:oos} in
    $X$-month windows. The different $X$s correspond to the columns of the table. The methodology is
    explained further in Section \ref{s:outofsample}. The data series are explained in Section
    \ref{s:data}.}
  \label{tab: OOS}
  \centering
  {\bf Out-of-Sample R-Squareds}
  
\begin{tabular}{@{\extracolsep{5pt}}lD{.}{.}{6} D{.}{.}{6} D{.}{.}{6} D{.}{.}{6} D{.}{.}{6} } \\[-1.8ex]
\hline 
\hline \\[-1.8ex] 

& \multicolumn{5}{c}{\text{Prediction Window in Months}} \\ 
\cline{2-6} \\[-1.8ex] 

& \multicolumn{1}{c}{12} & \multicolumn{1}{c}{15} & \multicolumn{1}{c}{18} & \multicolumn{1}{c}{21} & \multicolumn{1}{c}{24} \\ 

\hline \\[-1.8ex]

$\textit{ENT}$ & -0.019 &  0.024 &  0.047 &  0.051 &  0.046 \\
$\textit{ENT{\_}NEWS}$ &  0.069 &  0.087 &  0.096 &  0.091 &  0.060 \\
$\textit{ENT{\_}MODEL}$ & -0.074 & -0.061 & -0.059 & -0.057 & -0.084 \\
$\textit{EPU}$ & -0.272 & -0.322 & -0.414 & -0.417 & -0.478 \\
$\textit{SEN}$ & -0.595 & -0.642 & -0.678 & -0.668 & -0.607 \\
$\textit{VIX2}$ & -1.320 & -1.626 & -1.813 & -2.130 & -2.422 \\
$\textit{DGS10}$ & -0.214 & -0.221 & -0.284 & -0.296 & -0.312 \\
$\textit{DGS10-2}$ & -0.246 & -0.400 & -0.511 & -0.605 & -0.637 \\
$\textit{DY}$ & -2.001 & -1.982 & -1.722 & -1.580 & -1.767 \\
$\textit{CAY}$ & -0.123 & -0.141 & -0.137 & -0.136 & -0.138 \\
$\textit{1/CAPE}$ & -1.597 & -1.964 & -2.227 & -2.269 & -2.170 \\
$\textit{Return1}$ & -0.059 & -0.081 & -0.075 & -0.050 & -0.060 \\
$\textit{Return12}$ & -1.805 & -1.691 & -1.157 & -0.855 & -0.749 \\
$\textit{Return60}$ &  0.038 &  0.001 &  0.075 &  0.087 &  0.027 \\
$\textit{SMB60}$ & -0.427 & -0.532 & -0.568 & -0.573 & -0.584 \\
$\textit{HML60}$ & -0.799 & -0.797 & -0.826 & -0.935 & -1.082 \\
$\textit{RMW60}$ & -0.170 & -0.124 &  0.020 &  0.054 &  0.049 \\
$\textit{CMA60}$ & -0.913 & -0.977 & -1.023 & -1.143 & -1.232 \\
$\textit{UMD60}$ & -0.732 & -0.705 & -0.546 & -0.481 & -0.460 \\
$\textit{BEX}$ & -1.206 & -1.326 & -1.715 & -2.034 & -2.237 \\
$\textit{JLN}$ & -2.037 & -1.999 & -1.887 & -1.754 & -1.493 \\
$\textit{PCI}$ & -0.133 & -0.125 & -0.158 & -0.163 & -0.216 \\

\hline 
\hline \\[-1.8ex]

\end{tabular}

\end{table}

\begin{landscape}
\begin{table}[h]
  \caption{This table presents summary statistics of variables at the daily frequency used in this
    study, including Fama-French five factors market return minus risk-free rate (\textit{MKT}),
    small minus big (\textit{SMB}), high minus low (\textit{HML}), robust minus weak (\textit{RMW}),
    conservative minus aggressive (\textit{CMA}), and momentum factor
    (\textit{UMD}). $F_{EPU\_Momentum}$ represents the economic policy uncertainty factor mimicking
    portfolio using size-momentum returns as base asset. $F_{\textit{ENT}}$ ,
    $F_{\textit{ENT\_NEWS}}$ , $F_{\textit{ENT\_MODEL}}$ represent entropy, news updates, and model
    updates mimicking portfolios. We also use size-book to market, size-investment, and
    size-profitability returns as base assets.}
    \label{tab: summary-stats-day}
    \centering
    
\begin{tabular}{@{\extracolsep{5pt}}lD{.}{.}{6} D{.}{.}{6} D{.}{.}{6} D{.}{.}{6} D{.}{.}{6} D{.}{.}{6} D{.}{.}{6} } 
\\[-1.8ex]\hline 
\hline \\[-1.8ex] 
\\[-1.8ex] & \multicolumn{1}{c}{$\text{mean}$} & \multicolumn{1}{c}{$\text{std}$} & \multicolumn{1}{c}{$\text{min}$}  & \multicolumn{1}{c}{$\text{25\%}$} & \multicolumn{1}{c}{$\text{50\%}$} & \multicolumn{1}{c}{$\text{75\%}$} & \multicolumn{1}{c}{$\text{max}$} \\ 
\hline \\[-1.8ex]

$\textit{MKT}$           &  0.0294 & 1.2682 & -12.0000 & -0.5100 &  0.0700 & 0.6300 & 11.3500 \\
$\textit{SMB}$              &  0.0090 & 0.6425 &  -4.5500 & -0.3500 &  0.0200 & 0.3800 &  5.7100 \\
$\textit{HML}$              &  0.0062 & 0.7828 &  -5.0000 & -0.3400 & -0.0100 & 0.3200 &  6.7400 \\
$\textit{RMW}$              &  0.0173 & 0.5487 &  -3.0100 & -0.2500 &  0.0100 & 0.2800 &  4.5200 \\
$\textit{CMA}$              &  0.0136 & 0.4606 &  -5.8700 & -0.2100 &  0.0000 & 0.2200 &  2.5300 \\
$\textit{UMD}$              &  0.0171 & 1.0637 & -14.3700 & -0.4200 & 0.0700 & 0.5200 & 7.1200 \\

$F_{\textit{EPU\_Momentum}}$   & -0.0954 & 2.5804 & -22.3755 & -1.4273 & -0.1415 & 1.2136 & 20.5997 \\
$F_{\textit{ENT\_Momentum}}$   & -0.0001 & 0.0035 &  -0.0246 & -0.0018 & -0.0000 & 0.0016 &  0.0367 \\
$F_{\textit{ENT\_NEWS\_Momentum}}$ & -0.0001 & 0.0029 & -0.0218 & -0.0015 & -0.0001 & 0.0014 & 0.0325 \\
$F_{\textit{ENT\_MODEL\_Momentum}}$ & -0.0000 & 0.0018 & -0.0129 & -0.0009 &  0.0000 & 0.0010 & 0.0102 \\

$F_{\textit{EPU\_BM}}$         & -0.0547 & 2.0073 & -13.2736 & -1.0804 & -0.0717 & 0.9715 & 16.7301 \\
$F_{\textit{ENT\_BM}}$         & -0.0001 & 0.0027 &  -0.0183 & -0.0016 & -0.0001 & 0.0014 &  0.0218 \\
$F_{\textit{ENT\_NEWS\_BM}}$       &  -0.0001 & 0.0028 & -0.0164 & -0.0016 & -0.0001 & 0.0015 & 0.0187 \\
$F_{\textit{ENT\_MODEL\_BM}}$       & -0.0000 & 0.0017 & -0.0120 & -0.0010 &  0.0000 & 0.0009 & 0.0100 \\

$F_{\textit{EPU\_INV}}$     & -0.0207 & 2.2522 & -18.3677 & -1.2390 & -0.0525 & 1.1925 & 18.1130 \\
$F_{\textit{ENT\_INV}}$     & -0.0001 & 0.0029 &  -0.0287 & -0.0017 & -0.0002 & 0.0014 &  0.0199 \\
$F_{\textit{ENT\_NEWS\_INV}}$   & -0.0001 & 0.0034 & -0.0290 & -0.0020 & -0.0002 & 0.0017 & 0.0276 \\
$F_{\textit{ENT\_MODEL\_INV}}$   &  0.0000 & 0.0019 & -0.0114 & -0.0011 & -0.0000 & 0.0011 & 0.0093 \\

$F_{\textit{EPU\_OP}}$     & -0.0039 & 1.9306 & -15.5703 & -0.9966 & -0.0204 & 0.9854 & 14.9994 \\
$F_{\textit{ENT\_OP}}$     & -0.0001 & 0.0032 &  -0.0236 & -0.0018 & -0.0001 & 0.0016 &  0.0181 \\
$F_{\textit{ENT\_NEWS\_OP}}$   & -0.0001 & 0.0032 & -0.0232 & -0.0018 & -0.0001 & 0.0016 & 0.0207 \\
$F_{\textit{ENT\_MODEL\_OP}}$   & -0.0000 & 0.0015 & -0.0100 & -0.0009 & -0.0000 & 0.0008 & 0.0087 \\

\hline 
\hline \\[-1.8ex] 
\end{tabular} 
\end{table}
\end{landscape}

  \begin{table}[h]
    \caption{Fama-MacBeth regressions using \textit{MKT} (market minus risk-free rate), \textit{SMB}
      (small minus big), \textit{HML} (high minus low), \textit{RMW} (robust minus weak),
      \textit{CMA} (conservative minus aggressive), \textit{UMD} (momentum factor), all of which are
      obtained from French's website, along with factor mimicking portfolios with base assets
      corresponding to the test assets. The first four columns represent different base assets:
      size--momentum, size--book/market, size--investment, and size--profitability. The last column
      represents the sign of the corresponding coefficient in Table \ref{tab:IS-std} for forecasting
      12-month ahead market returns if that coefficient is significant at 5\% level. Risk premia
      coefficients are in percent per month. \scaledCoeffDesc General method of moments t-statistics are in
      parentheses.} \label{t:fm-same-ent-std}
    \centering {\bf Fama-MacBeth Factor Risk Premia (scaled coefficients)}
    
\begin{tabular}{@{\extracolsep{5pt}}lD{.}{.}{6} D{.}{.}{6} D{.}{.}{6} D{.}{.}{6} c }
\\[-1.8ex]\hline 
\hline \\[-1.8ex] 
 & \multicolumn{4}{c}{\text{Base Assets the Same as Test Assets: Size -}} \\ 
\cline{2-5} 
\\[-1.8ex] & \multicolumn{1}{c}{$\text{Momentum}$} & \multicolumn{1}{c}{$\text{Book/Market}$} & \multicolumn{1}{c}{$\text{Investment}$} & \multicolumn{1}{c}{$\text{Profitability}$} & \multicolumn{1}{c}{$\text{Mkt-RF}$} \\ 
\\[-1.8ex] & \multicolumn{1}{c}{(1)} & \multicolumn{1}{c}{(2)} & \multicolumn{1}{c}{(3)} & \multicolumn{1}{c}{(4)} & \multicolumn{1}{c}{(5)} \\ 
\hline \\[-1.8ex]

$\textit{MKT}$ &   0.048^{*} &  0.083^{**} &    0.064^{**} &   0.056^{**} &  \\
&     (1.681) &      (2.189) &       (2.023) &      (1.997) \\
$\textit{SMB}$ &         0.100 &  0.067 &         0.008 &        0.079 & \text{--} \\
&     (1.371) &     (0.907) &       (0.108) &      (1.051) \\
$\textit{HML}$ &   0.078^{*} &  0.048 &   0.136^{***} &        0.041 & \text{--}  \\
&     (1.754) &      (0.601) &       (2.879) &      (0.905) \\
$\textit{RMW}$ &      -0.066 & 0.062 &        -0.018 &    0.119^{*} & \text{+}  \\
&     (-1.050) &      (1.283) &      (-0.337) &      (1.953) \\
$\textit{CMA}$ &  0.078^{**} & -0.007 &         0.065 &        0.018 & \text{+} \\
&     (2.142) &      (-0.166) &       (1.498) &      (0.438) \\
$\textit{UMD}$ &       0.149 &  0.083^{***} &    0.064^{**} &        0.029 \\
&     (1.342) &      (2.896) &       (2.346) &      (0.928) \\
$F_{\textit{EPU}}$ &  -0.063^{*} & -0.083^{**} &        -0.008 &       -0.013 & \text{+} \\
&    (-1.746) &     (-2.342) &      (-0.326) &     (-0.432) \\
$F_{\textit{ENT}}$ &   -0.060^{*} & -0.066^{**} &  -0.091^{***} &  -0.082^{**} & \text{--} \\
&    (-1.757) &     (-2.398) &      (-2.669) &     (-2.175) \\

\hline 
\hline \\[-1.8ex] 
\textit{Note:}  & \multicolumn{5}{r}{$^{*}$p$<$0.1; $^{**}$p$<$0.05; $^{***}$p$<$0.01} \\

\end{tabular}

  \end{table}

  \begin{table}[h]
    \caption{Fama-MacBeth regressions using \textit{MKT} (market minus risk-free rate), \textit{SMB}
      (small minus big), \textit{HML} (high minus low), \textit{RMW} (robust minus weak),
      \textit{CMA} (conservative minus aggressive), \textit{UMD} (momentum factor), all of which are
      obtained from French's website, along with factor mimicking portfolios with base assets
      corresponding to the test assets. The first four columns represent different base assets:
      size--momentum, size--book/market, size--investment, and size--profitability. The last column
      represents the sign of the corresponding coefficient in Table \ref{tab:IS-std-news-model} for forecasting
      12-month ahead market returns if that coefficient is significant at 5\% level. Risk premia
      coefficients are in percent per month. \scaledCoeffDesc General method of moments t-statistics are in
      parentheses.} \label{t:fm-same-part1-std}
    \centering {\bf Fama-MacBeth Factor Risk Premia (scaled coefficients)}
    
\begin{tabular}{@{\extracolsep{5pt}}lD{.}{.}{6} D{.}{.}{6} D{.}{.}{6} D{.}{.}{6} c }
\\[-1.8ex]\hline 
\hline \\[-1.8ex] 
 & \multicolumn{4}{c}{\text{Base Assets the Same as Test Assets: Size -}} \\ 
\cline{2-5} 
\\[-1.8ex] & \multicolumn{1}{c}{$\text{Momentum}$} & \multicolumn{1}{c}{$\text{Book/Market}$} & \multicolumn{1}{c}{$\text{Investment}$} & \multicolumn{1}{c}{$\text{Profitability}$} & \multicolumn{1}{c}{$\text{Mkt-RF}$} \\ 
\\[-1.8ex] & \multicolumn{1}{c}{(1)} & \multicolumn{1}{c}{(2)} & \multicolumn{1}{c}{(3)} & \multicolumn{1}{c}{(4)} & \multicolumn{1}{c}{(5)} \\ 
\hline \\[-1.8ex]

$\textit{MKT}$ &    0.050^{*} &  0.077^{**} &   0.053^{**} &  0.051^{**} &  \\
&      (1.720) &      (2.157) &      (2.022) &       (2.000) \\
$\textit{SMB}$ &       0.093 &  0.071 &         0.010 &       0.082 & \text{--} \\
&     (1.287) &     (0.966) &      (0.131) &     (1.095) \\
$\textit{HML}$ &   0.081^{*} &   0.054 &  0.133^{***} &       0.035 &  \\
&      (1.780) &      (0.684) &      (2.766) &     (0.758) \\
$\textit{RMW}$ &      -0.065 &        0.067 &       -0.014 &  0.122^{**} &  \\
&    (-1.038) &      (1.316) &     (-0.265) &     (1.978) \\
$\textit{CMA}$ &  0.071^{**} &        -0.013 &        0.063 &       0.009 & + \\
&     (2.055) &      (-0.326) &      (1.457) &     (0.216) \\
$\textit{UMD}$ &       0.148 &        0.086^{***} &   0.066^{**} &        0.030 \\
&     (1.335) &      (3.032) &      (2.332) &     (0.937) \\
$F_{\textit{EPU}}$ &  -0.063^{*} &       -0.083^{**} &       -0.013 &      -0.009 & + \\
&    (-1.764) &     (-2.297) &     (-0.484) &    (-0.311) \\
$F_{\textit{ENT\_NEWS}}$ &      -0.044 &   -0.060^{**} &  -0.055^{**} &  -0.059^{*} \\
&    (-1.426) &     (-2.237) &       (-2.000) &    (-1.771) \\

\hline 
\hline \\[-1.8ex] 
\textit{Note:}  & \multicolumn{5}{r}{$^{*}$p$<$0.1; $^{**}$p$<$0.05; $^{***}$p$<$0.01} \\

\end{tabular}

  \end{table}


\begin{landscape}
  \begin{table}[h]
    \caption{In-sample predictions of 12-month ahead economic policy uncertainty (\textit{EPU}) and
      fundamental variables including unemployment rate (\textit{UNRATE}), year-over-year change of
      industrial production (\textit{INDPRO{\_}YOY}), year-over-year change of the consumer price
      index (\textit{CPI{\_}YOY}), interest rates (\textit{DGS10}, \textit{DGS2}, and
      \textit{DGS10-2}), the Chicago Board Options Exchange’s CBOE Volatility Index (\textit{VIX}),
      and the S\&P500 earnings per share (\textit{EPS}). The columns correspond to different
      dependent variables in \eqref{eq:fund}. Each specification includes the lagged value of the
      dependent variable in question and the lagged value of all other macro variables as
      controls. The lagged value of one entropy measure
  (\textit{ENT}, \textit{ENT\_NEWS}, or \textit{ENT\_MODEL}) is also used as a control,
  corresponding to each row. The coefficient estimates have been normalized by the standard
  deviation of entropy, \textit{ENT\_NEWS}, or \textit{ENT\_MODEL} respectively. This table
  summarizes results in Online Appendix Tables \ref{tab: macro-std}, \ref{tab: macro-part1-std}, and
  \ref{tab: macro-part2-std}. Robust t-statistics are in parentheses and are based on Newey–West standard
      errors with four lags.}
    \label{tab: macro-std-short}
    \centering
    {\bf Macro Forecasting Using Entropy and Other Macro Controls} \\[5pt]
    \resizebox{\columnwidth}{!}{%
\begin{tabular}{@{\extracolsep{5pt}}lD{.}{.}{6} D{.}{.}{6} D{.}{.}{6} D{.}{.}{6} D{.}{.}{6} D{.}{.}{6} D{.}{.}{6} D{.}{.}{6} D{.}{.}{6} } \\[-1.8ex]
\hline 
\hline \\[-1.8ex]

& \multicolumn{9}{c}{\text{12-Month Ahead Macro Variables}} \\ 
\cline{2-10} \\[-1.8ex] 
& \multicolumn{1}{c}{EPU} & \multicolumn{1}{c}{UNRATE} & \multicolumn{1}{c}{INDPRO{\_}YOY} & \multicolumn{1}{c}{CPI{\_}YOY} & \multicolumn{1}{c}{DGS10} & \multicolumn{1}{c}{DGS2} & \multicolumn{1}{c}{DGS10-2} & \multicolumn{1}{c}{VIX} & \multicolumn{1}{c}{EPS} \\ 
\hline \\[-1.8ex]

$\textit{ENT}$ &       11.936 &  0.723^{***} &  -1.348^{***} &  -0.371^{**} &  -0.261^{***} &  -0.371^{***} &         0.049 &  1.571^{***} &    -4.890^{**} \\
&      (1.608) &       (4.050) &      (-3.248) &     (-2.183) &      (-3.569) &      (-3.552) &       (0.807) &       (3.360) &      (-2.239) \\
\hline \\[-0.8ex] 
$\textit{ENT{\_}NEWS}$ &        7.779 &  0.642^{***} &  -1.213^{***} &   -0.356^{*} &   -0.182^{**} &  -0.351^{***} &         0.065 &  2.059^{***} &   -5.242^{**} \\
&      (1.333) &      (4.441) &      (-2.809) &     (-1.906) &      (-2.509) &      (-3.133) &       (0.999) &      (3.786) &      (-2.068) \\
\hline \\[-0.8ex] 
$\textit{ENT{\_}MODEL}$ &        3.892 &      -0.079 &        0.178 &        0.090 &        -0.063 &       0.080 &        -0.045 &  -1.397^{**} &         2.197 \\
&      (0.530) &    (-0.486) &      (0.441) &      (0.606) &      (-1.009) &     (0.762) &      (-0.791) &     (-1.994) &       (0.909) \\

\hline 
\hline \\[-1.8ex] 
\textit{Note:}  & \multicolumn{9}{r}{$^{*}$p$<$0.1; $^{**}$p$<$0.05; $^{***}$p$<$0.01} \\

\end{tabular}%
}
  \end{table}
\end{landscape}


  \begin{table}[h]
    \caption{In-sample predictions of 12-month ahead market returns, measured in percent, using the
      specification in \eqref{eq:reg-is-robust}. Uncertainty measures include \cite{bekaert2022time}
      (\textit{BEX}), \cite{jurado2015measuring} (\textit{JLN}), the \cite{azzimonti2018partisan}
      Partisan Conflict Index (\textit{PCI}), and squared implied volatility (\textit{VIX2}). The
      other control variables are explained in Section \ref{s:data}. Full results are shown in Online
  Appendix Table~\ref{tab:IS-std-robust}. The coefficient estimates have been normalized by the
  standard deviation of the right-hand side variables. Unscaled full results are shown in Online
  Appendix Table~\ref{tab:IS-robust}. Robust t-statistics are in
      parentheses and are based on Newey–West standard errors with four lags.}
    \label{tab:IS-std-short} \centering
    {\bf In-Sample Return Forecasting with Uncertainty Controls} \\[1pt]
    \scalebox{1}{\resizebox{\columnwidth}{!}{%
\begin{tabular}{@{\extracolsep{5pt}}lD{.}{.}{6} D{.}{.}{6} D{.}{.}{6} D{.}{.}{6} D{.}{.}{6} D{.}{.}{6}} \\[-1.8ex]
\hline 
\hline \\[-1.8ex] 

& \multicolumn{6}{c}{\text{12-Month Ahead Cumulative Return}} \\ 

\cline{2-7} \\[-1.8ex]

& \multicolumn{1}{c}{(1)} & \multicolumn{1}{c}{(2)} & \multicolumn{1}{c}{(3)} & \multicolumn{1}{c}{(4)} & \multicolumn{1}{c}{(5)} & \multicolumn{1}{c}{(6)}\\ 

\hline \\[-1.8ex]

$\textit{ENT}$  &    -2.266^{**} &  -2.530^{***} &                &                &    -2.306^{**} &   -2.413^{***} \\
     &       (-2.545) &      (-3.040) &                &                &       (-2.522) &       (-2.709) \\
$\textit{EPU}$  &                &               &    4.839^{***} &     3.674^{**} &    4.686^{***} &     3.531^{**} \\
     &                &               &        (3.928) &        (2.502) &        (3.924) &        (2.385) \\
$\textit{SEN}$  &                &               &          2.091 &          2.319 &          1.319 &          1.821 \\
     &                &               &        (1.212) &        (1.465) &        (0.768) &        (1.158) \\

\hline \\[-1.8ex]

Other Uncertainty&	\multicolumn{1}{c}{No}	&	\multicolumn{1}{c}{Yes}	&	\multicolumn{1}{c}{No}	&	\multicolumn{1}{c}{Yes}	&	\multicolumn{1}{c}{No}	&	\multicolumn{1}{c}{Yes} \\
Controls	&	\multicolumn{1}{c}{Yes}	&	\multicolumn{1}{c}{Yes}	&	\multicolumn{1}{c}{Yes}	&	\multicolumn{1}{c}{Yes}	&	\multicolumn{1}{c}{Yes}	&	\multicolumn{1}{c}{Yes} \\
Adj. $R^{2}$	&	0.648	&	0.671	&	0.657	&	0.667	&	0.666	&	0.677 \\

\hline 
\hline \\[-1.8ex] 
\textit{Note:}  & \multicolumn{6}{r}{$^{*}$p$<$0.1; $^{**}$p$<$0.05; $^{***}$p$<$0.01} \\ 

\end{tabular}%
}

}
  \end{table}

\makeatletter\@input{yy.tex}\makeatother
\end{document}


\title{Online Appendix for: \paperTitle}
\authorLine

\date{\today}
\maketitle

\section{List of results}

\begin{figure}[ht]
  \begin{center}
    \includegraphics[width=16cm]{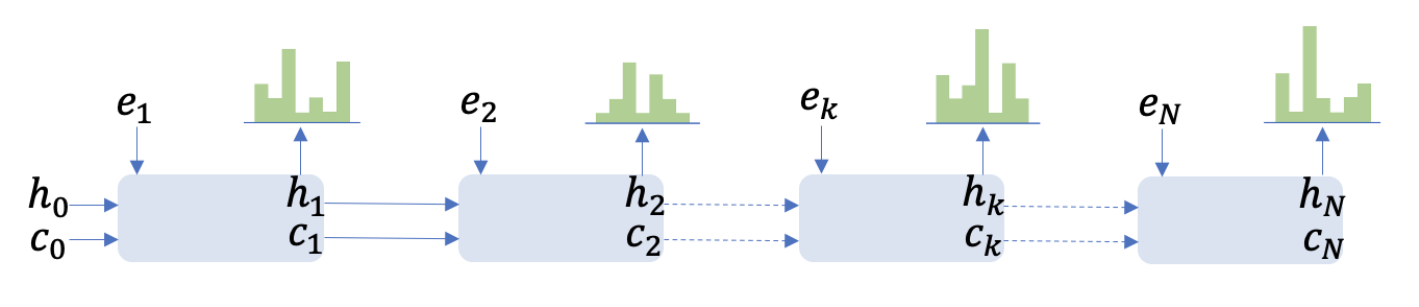}
  \end{center}
  \vskip -10pt
  \caption{Illustration of RNN model with LSTM architecture.} \label{f:rnn-architecture}
\end{figure}

\begin{figure}[ht]
  \begin{center}
    \includegraphics[width=17cm]{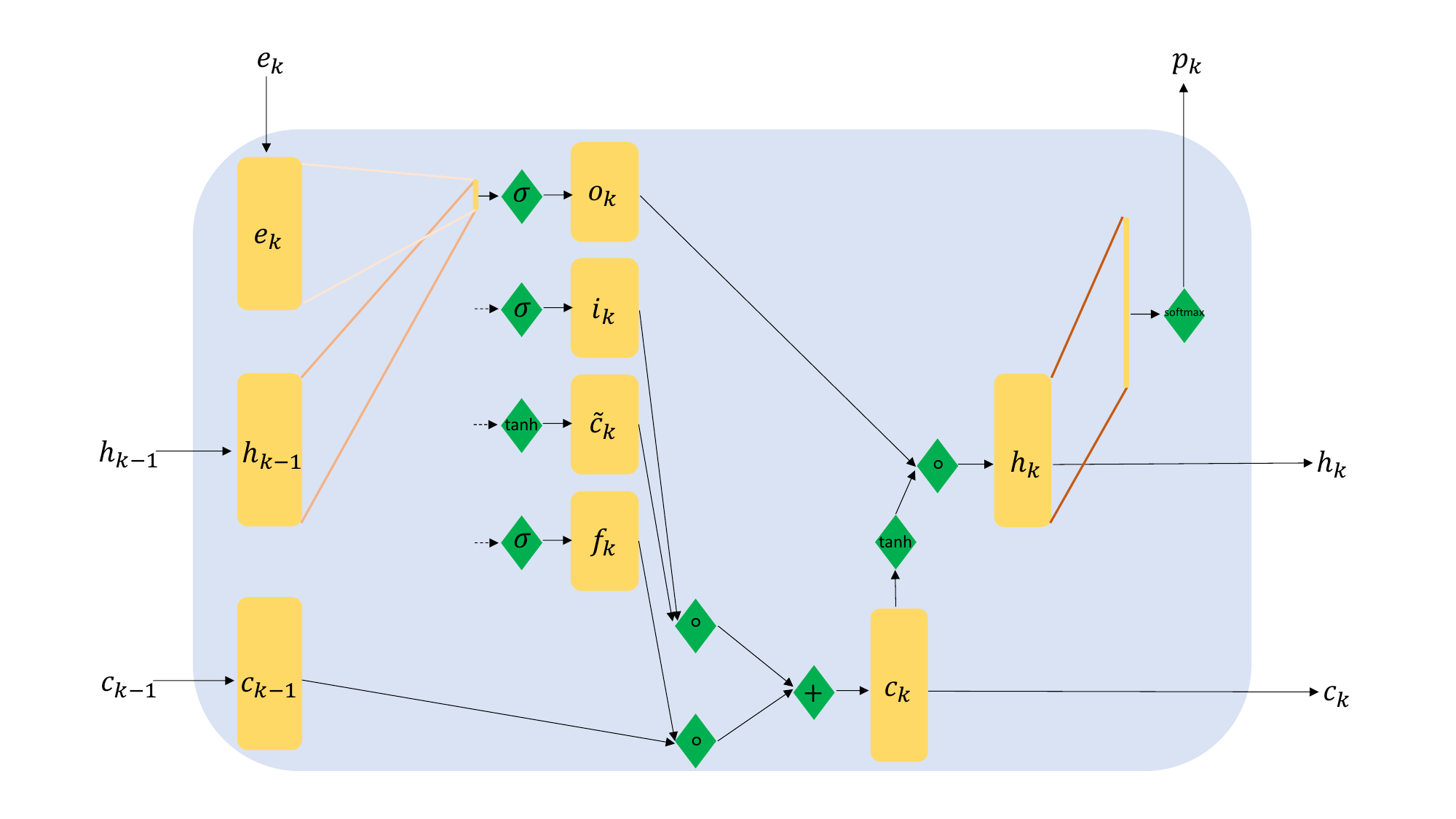}
  \end{center}
  \caption{Illustration of a single LSTM cell.} \label{f:lstm-gate}
\end{figure}

\begin{landscape}
\begin{figure}[ht]
  \centering
  \includegraphics[width=1.75\textwidth,trim={3.25cm 0 0 0},clip]{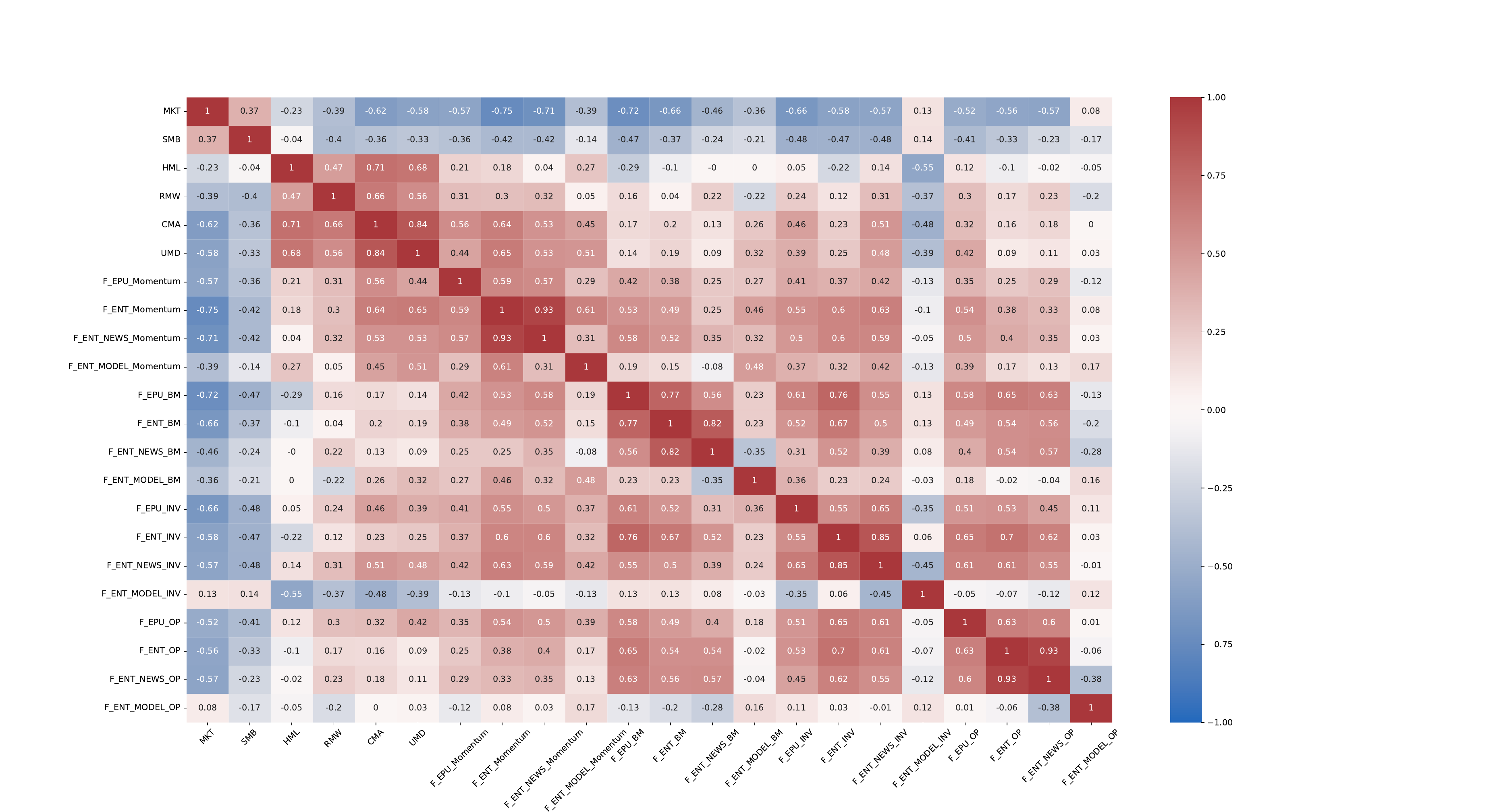}
  \caption{Correlations of factors and replicating portfolios for EPU, entropy, and news updates and
    model updates of entropy, using daily observations.} \label{f:daily-corrs}
\end{figure}
\end{landscape}

\begin{figure}[ht]
  \centering
  \mbox{\includegraphics[width=0.5\textwidth]{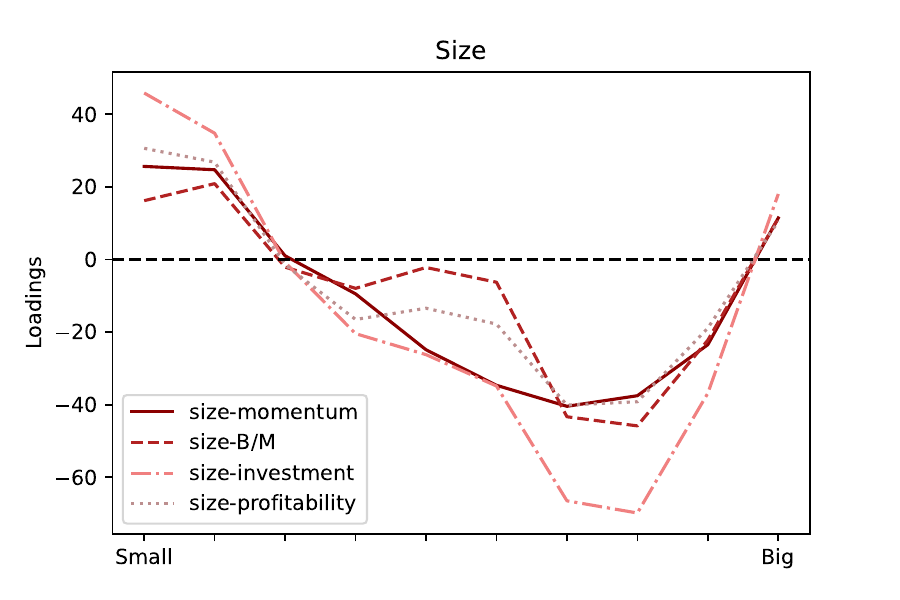}}
  \mbox{
  \includegraphics[width=0.5\textwidth]{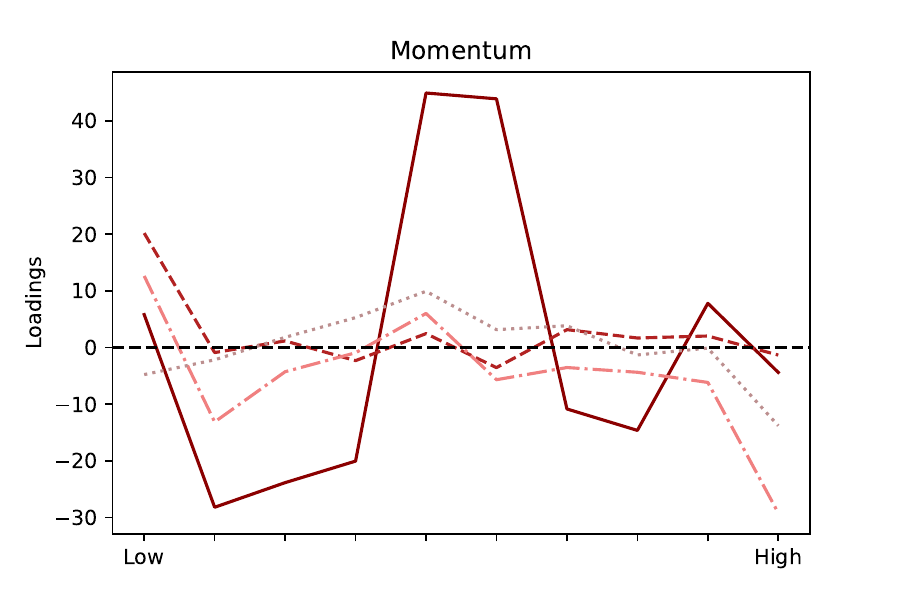}
  \includegraphics[width=0.5\textwidth]{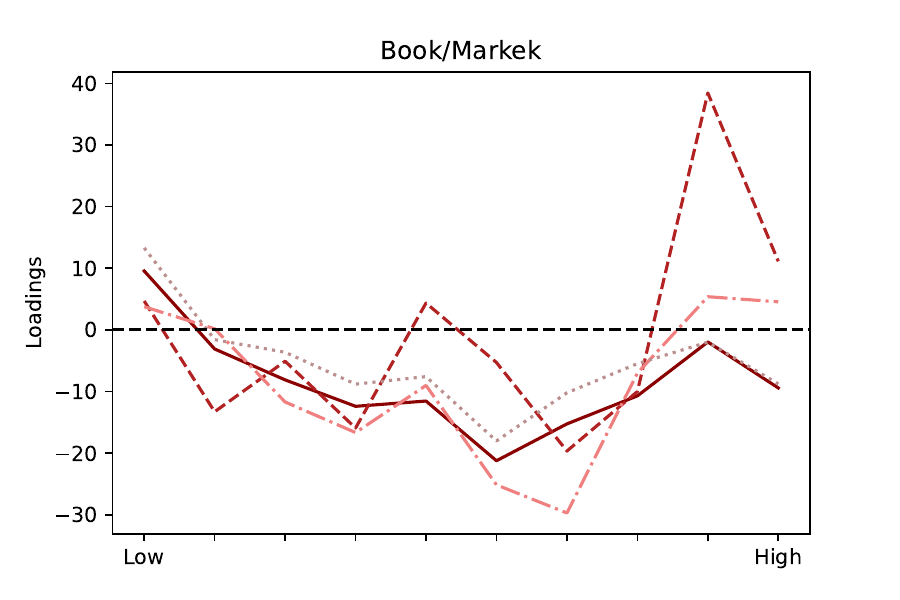}}
  \mbox{
  \includegraphics[width=0.5\textwidth]{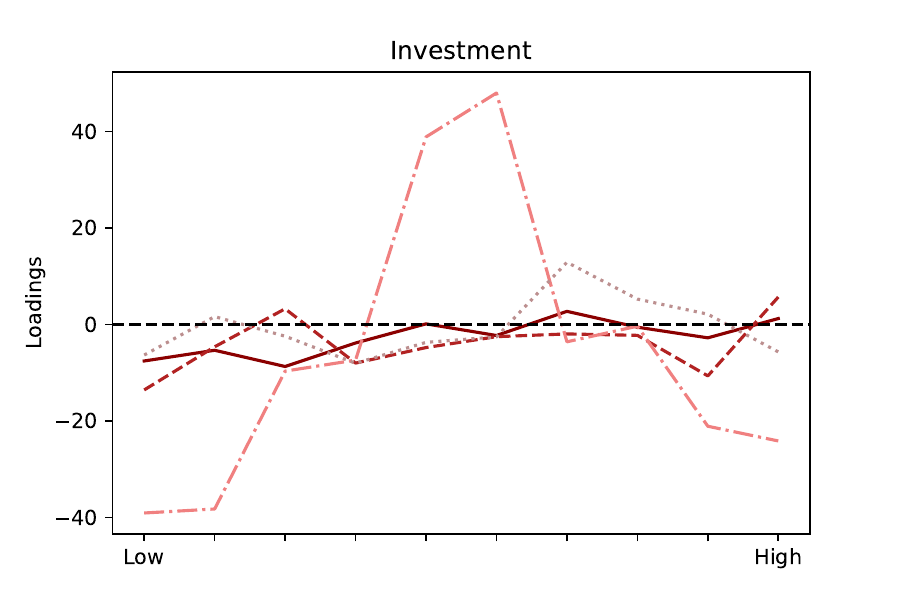}
  \includegraphics[width=0.5\textwidth]{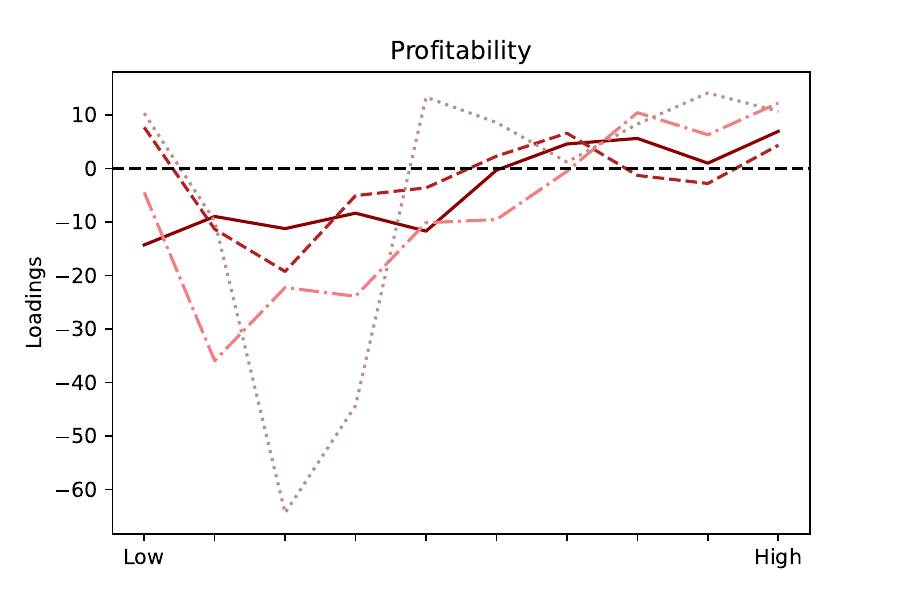}}
  \caption{Loadings on $F_{\textit{ENT}}$ for various test portfolios. This figure repeats the
    analysis of Section \ref{s:which-hedge} in the main paper, but with different sets of base
    assets in the construction of $F_{\textit{ENT}}$. In each panel, size-momentum, size-book/market,
    size-investment, and size-profitability  sorted portfolios are used as the base assets
    respectively.}
  \label{f:FM-univariate}
\end{figure}

\underline{\it Figures}
\begin{itemize}
\item Figures \ref{f:rnn-architecture} and \ref{f:lstm-gate} show the architecture of our recurrent
  neural network (RNN) and a closeup of the LSTM (long short-term memory) gate.
\item Figure \ref{f:daily-corrs} shows the correlations of daily returns of factors
  (\citealt{fama2015five}) and replicating portfolios for \textit{EPU}, \textit{ENT},
  \textit{ENT\_NEWS} and \textit{ENT\_MODEL}.
\item Figure \ref{f:FM-univariate} repeats the analysis of Section \ref{s:which-hedge} in the main
  paper, but with different sets of base assets in the construction of $F_{\textit{ENT}}$.
\end{itemize}

\noindent
\underline{\it Tables}
\begin{itemize}
\item Table~\ref{tab: IS} shows the unscaled coefficients corresponding to Equation~\eqref{eq:
    reg-is} and Table~\ref{tab:IS-std} in the main paper.
\item Tables~\ref{tab:IS-std-news-model}--\ref{tab:IS-news-model} show the scaled and unscaled
  coefficients when replacing \textit{ENT} in Table~\ref{tab:IS-std} columns (3) and (6) in the main
  paper with \textit{ENT\_NEWS} and \textit{ENT\_MODEL}.
\item Table~\ref{tab: OOS-coeff} shows the average out-of-sample forecasting coefficients and
  associated t-statistics corresponding to rolling estimates of \eqref{eq:oos} and the out-of-sample
  R-squareds shown in Table~\ref{tab: OOS} in the main paper.
\item Tables~\ref{t:fm-same-ent}--\ref{t:fm-same-part1} show the unscaled Fama-MacBeth risk
  premia corresponding to Tables~\ref{t:fm-same-ent-std} and \ref{t:fm-same-part1-std} (\textit{ENT} and
  \textit{ENT\_NEWS}) in the main paper.
\item Tables~\ref{t:fm-same-part2-std}--\ref{t:fm-same-part2} show the scaled and unscaled
  Fama-MacBeth coefficients associated with the model innovation, \textit{ENT\_MODEL}, part of entropy.
\item Tables~\ref{tab: macro-std}--\ref{tab: macro-part2-std} show the full results of the scaled
  coefficients corresponding to the macro forecasting regressions of equation~\eqref{eq:fund} and
  Table~\ref{tab: macro-std-short} of the main paper.
\item Tables~\ref{tab: macro}--\ref{tab: macro-part2} show the full results of the raw coefficients
  corresponding to the macro forecasting regressions in equation~\eqref{eq:fund} and Table~\ref{tab:
    macro-std-short} of the main paper.
\item Tables~\ref{t:fm-same-std-incSEN}--\ref{t:fm-same-incSEN} show the scaled and unscaled
  coefficients augmenting the Fama-MacBeth analysis in the main paper
  (Table~\ref{t:fm-same-ent-std}) with mimicking portfolios for the \textit{SEN} sentiment variable.
\item Tables~\ref{tab:IS-std-robust}--\ref{tab:IS-robust} show the full results of the unscaled and
  raw coefficients respectively corresponding to Equation~\eqref{eq:reg-is-robust} and
  Table~\ref{tab:IS-std-short} in the main paper.
\end{itemize}

\section{LSTM Network} \label{s:lstm}

LSTM is a particular architecture of RNN designed to address long-term dependencies in sequences by
effectively utilizing distant information. This architecture tackles two main challenges: discarding
information that is no longer relevant to the context and incorporating new information for future
use. LSTM utilizes a context or memory state ($c$), in addition to the hidden state ($h$).

Each component of the LSTM network is modeled as follows. It utilizes ``gates'' to control the flow
of information into and out of the units. All gates process the linear combinations of previous
hidden state and current input vector through a sigmoid function ($\sigma$). Since the sigmoid
function has an output range of (0, 1), it allows the unit to either fully open, partially open, or
close the gates based on the probabilities. When we extract information, we use tanh as the
activation function, as it allows complex relationships and learns intricate patterns in the data.

A new memory state ($\Tilde{c}_k$) is computed from linear combinations of previous hidden state
($h_{k-1}$) and current word embedding ($e_k$). A forget gate $f_k$ deletes information in the
previous context that is no longer needed. Input gate $i_k$ adds information from the new memory to
the final memory passed to the next step. Thus, the final memory passed to the next step comes from
the combination of the filtered old memory ($f_k \circ c_{k-1}$) and the filtered new memory
($i_k \circ \Tilde{c}_k$), where $\circ$ denotes element-wise multiplication of two vectors. The
output gate is used to decide what information is needed for the current hidden state $h$. Note that
this is different from the information preserved for future use $c$. The current hidden state is
thus computed as the tanh activated context filtered by the output gate ($o_k \circ \tanh(c_k)$).

The model is parameterized by matrices $W_i, W_f, W_o, W_c, U_i, U_f, U_o, U_c, U_s$, which are
estimated using the reference text. The detailed training process will be introduced in
Section~\ref{s:rnn-train}.  Once the parameters are chosen, the model can be applied to evaluate
next-word probabilities in an evaluation text. Given a sequence of words $w_{1} \cdots w_{k-1}$, the
RNN evaluates the probability $p_k^x$ that the next word will be $x$, for each $x$ in the
vocabulary. The probability vector $p_k$ is calculated using
\begin{align*}
   i_k &= \sigma(W_i e_k + U_i h_{k-1}+b_i)  ~\mbox{  }~ (\text{Input gate}) \\
   f_k &= \sigma(W_f e_k + U_f h_{k-1}+b_f)  ~\mbox{  }~ (\text{Forget gate}) \\
   o_k &= \sigma(W_o e_k + U_o h_{k-1}+b_o)  ~\mbox{  }~ (\text{Output gate}) \\
   \Tilde{c}_k &= \tanh (W_c e_k + U_c h_{k-1}+b_c)  ~\mbox{  }~ (\text{New memory}) \\
   c_k &= f_k \circ c_{k-1} + i_k \circ \Tilde{c}_k  ~\mbox{  }~ (\text{Final memory}) \\
   h_k &= o_k \circ \tanh(c_k) \\
   p_k &= \text{softmax} (U_s h_k+b_s)
\end{align*} 
where $\circ$ denotes element-wise multiplication of two vectors. We refer to the input, hidden, and
output layer dimensions as $d_{in}$, $d_{h}$, and $d_{out}$ respectively. In our case, $d_{in}=100$,
$d_{h}=16$, and $d_{out}=10,000$. The parameter matrices
$W_i, W_f,W_o, W_c \in \mathbb{R}^{d_h \times d_{in}}$,
$U_i, U_f, U_o, U_c \in \mathbb{R}^{d_h \times d_h}$, $U_s \in \mathbb{R}^{d_{out} \times d_h}$,
$b_i,b_f,b_o,b_c \in \mathbb{R}^{d_h}$, and $b_s \in \mathbb{R}^{d_{out}}$.

\section{Economic Policy Uncertainty and Sentiment} \label{s:epu-and-sen}

\cite{baker2016measuring} develop a monthly Economic Policy Uncertainty (\textit{EPU}) index, which
consists of three underlying components. The first component is the newspaper-based component
derived from 10 large newspapers. A normalized index of the volume of news articles discussing
economic policy uncertainty is constructed from these newspapers. The second component is derived
from reports by the Congressional Budget Office (CBO) that compile lists of temporary federal tax
code provisions. The third component utilizes the dispersion between individual forecasters'
predictions about future levels of the Consumer Price Index, Federal Expenditures, and State and
Local Expenditures to construct indices of uncertainty about policy-related macroeconomic
variables. We download the data from the authors' website \url{https://www.policyuncertainty.com/}.

The \textit{SEN} index of \cite{shapiro2022measuring} is based on lexical analysis of
economics-related news articles. The sentiment scores are calculated from economics-related news
articles published in 24 major U.S. newspapers. It combines publicly available lexicons with a
news-specific lexicon created by the authors and is constructed as a trailing weighted average of
time series, with weights that decline geometrically with the length of time since article
publication. We download this index from
\url{https://www.frbsf.org/economic-research/indicators-data/daily-news-sentiment-index/}.

\bibliography{ref.bib}


\isForecasts{IS.txt}{tab: IS}{The table shows the raw coefficient estimates, which correspond to the
  standardized coefficient estimates of Table \ref{tab:IS-std}.}

\isForecasts{IS-std_part12.txt}{tab:IS-std-news-model}{The coefficient estimates have been
  normalized by the standard deviation of the right-hand side variables.}

\isForecasts{IS_part12.txt}{tab:IS-news-model}{This table shows the raw (unscaled) coefficient
  estimates.}

\begin{table}[h]
  \caption{This table shows the average forecasting coefficients ($\beta_t$'s from \eqref{eq:oos})
    and t-statistics of the out-of-sample prediction analysis of 12-month ahead market returns using
    monthly rolling estimates of \eqref{eq:oos}. The t-statistic (shown in parentheses) uses
    Newey-West with a 4-month lag.} \label{tab: OOS-coeff}
  \centering
  \scalebox{0.8}{
\begin{tabular}{@{\extracolsep{5pt}}lD{.}{.}{6} D{.}{.}{6} D{.}{.}{6} D{.}{.}{6} D{.}{.}{6} } \\[-1.8ex]
\hline 
\hline \\[-1.8ex] 

& \multicolumn{5}{c}{\text{Prediction Window in Months}} \\ 
\cline{2-6} \\[-1.8ex] 

& \multicolumn{1}{c}{12} & \multicolumn{1}{c}{15} & \multicolumn{1}{c}{18} & \multicolumn{1}{c}{21} & \multicolumn{1}{c}{24} \\ 

\hline \\[-1.8ex]

$\textit{ENT}$          &           -15.659 &      -21.943^{**} &      -25.452^{**} &     -28.433^{***} &     -30.662^{***} \\
&          (-1.593) &          (-2.163) &          (-2.513) &          (-2.844) &          (-3.110) \\
$\textit{ENT{\_}NEWS}$  &           -12.291 &           -15.715 &           -16.372 &           -16.435 &           -16.915 \\
&          (-1.301) &          (-1.615) &          (-1.579) &          (-1.493) &          (-1.467) \\
$\textit{ENT{\_}MODEL}$ &      -54.636^{**} &     -64.012^{***} &     -70.567^{***} &     -72.046^{***} &     -73.283^{***} \\
&          (-2.425) &          (-2.806) &          (-3.118) &          (-3.155) &          (-3.162) \\
$\textit{EPU}$          &       0.099^{***} &       0.107^{***} &       0.116^{***} &       0.125^{***} &       0.128^{***} \\
&           (3.961) &           (4.485) &           (5.024) &           (5.703) &           (5.854) \\
$\textit{SEN}$          &     -35.470^{***} &     -34.705^{***} &     -32.388^{***} &     -29.996^{***} &     -27.746^{***} \\
&          (-6.971) &          (-6.439) &          (-6.231) &          (-6.332) &          (-6.603) \\
$\textit{VIX2}$         &       0.025^{***} &       0.023^{***} &       0.022^{***} &       0.020^{***} &       0.018^{***} \\
&           (9.231) &           (8.828) &           (8.320) &           (7.522) &           (6.935) \\
$\textit{DGS10}$        &      -3.256^{***} &      -4.007^{***} &      -4.635^{***} &      -5.143^{***} &      -5.631^{***} \\
&          (-3.278) &          (-3.855) &          (-4.366) &          (-4.848) &          (-5.350) \\
$\textit{DGS10-2}$      &             0.727 &             0.723 &             1.056 &             1.261 &             1.390 \\
&           (0.634) &           (0.684) &           (0.963) &           (1.136) &           (1.260) \\
$\textit{DY}$           &      61.360^{***} &      62.551^{***} &      62.912^{***} &      62.016^{***} &      60.283^{***} \\
&          (12.173) &          (12.687) &          (13.475) &          (13.975) &          (14.269) \\
$\textit{CAY}$          &     755.599^{***} &     850.803^{***} &     881.254^{***} &     585.782^{***} &     514.538^{***} \\
&           (3.647) &           (3.801) &           (3.861) &           (8.077) &           (8.411) \\
$\textit{1/CAPE}$       &     370.914^{***} &     329.612^{***} &     293.478^{***} &     262.436^{***} &     233.926^{***} \\
&           (3.867) &           (3.486) &           (3.208) &           (3.063) &           (3.027) \\
$\textit{Return1}$      &      -0.454^{***} &      -0.432^{***} &      -0.389^{***} &      -0.347^{***} &      -0.312^{***} \\
&          (-8.746) &          (-8.362) &          (-7.202) &          (-6.653) &          (-6.109) \\
$\textit{Return12}$     &      -0.177^{***} &      -0.165^{***} &      -0.157^{***} &      -0.147^{***} &      -0.153^{***} \\
&          (-2.874) &          (-2.806) &          (-3.060) &          (-3.302) &          (-3.832) \\\
$\textit{Return60}$     &      -0.363^{***} &      -0.398^{***} &      -0.421^{***} &      -0.428^{***} &      -0.433^{***} \\
&          (-8.738) &         (-10.091) &         (-12.048) &         (-13.030) &         (-13.155) \\
$\textit{SMB60}$        &      -0.330^{***} &      -0.347^{***} &      -0.346^{***} &      -0.342^{***} &      -0.337^{***} \\
&          (-3.111) &          (-3.213) &          (-3.254) &          (-3.271) &          (-3.404) \\
$\textit{HML60}$        &        0.166^{**} &             0.096 &             0.027 &            -0.029 &            -0.051 \\
&           (2.221) &           (1.300) &           (0.373) &          (-0.393) &          (-0.700) \\
$\textit{RMW60}$        &            -0.039 &             0.096 &         0.231^{*} &       0.330^{***} &       0.414^{***} \\
&          (-0.270) &           (0.722) &           (1.896) &           (2.972) &           (4.417) \\
$\textit{CMA60}$        &            -0.130 &            -0.055 &             0.012 &             0.038 &             0.059 \\
&          (-0.878) &          (-0.389) &           (0.091) &           (0.299) &           (0.496) \\
$\textit{UMD60}$       &       0.283^{***} &       0.277^{***} &       0.260^{***} &       0.231^{***} &       0.193^{***} \\
&           (4.981) &           (4.866) &           (4.671) &           (4.366) &           (3.890) \\
$\textit{BEX}$          &  3.970^{***} &  3.634^{***} &  3.396^{***} &  3.238^{***} &  3.044^{***} \\
&           (8.564) &           (8.009) &           (7.354) &           (7.056) &           (6.727) \\
$\textit{JLN}$         &      72.916^{***} &      62.228^{***} &       47.377^{**} &            29.101 &            11.726 \\
&           (3.702) &           (3.237) &           (2.534) &           (1.584) &           (0.654) \\
$\textit{PCI}$          &            -0.012 &             0.001 &             0.011 &             0.021 &             0.031 \\
&          (-0.497) &           (0.059) &           (0.461) &           (0.791) &           (1.156) \\

\hline 
\hline \\[-1.8ex]

\end{tabular}

}
\end{table}


\famaMacBeth{FM-Same.txt}{t:fm-same-ent}{tab:IS-std}{\rawCoeffDesc}{raw}

\famaMacBeth{FM-Same-part1.txt}{t:fm-same-part1}{tab:IS-std-news-model}{\rawCoeffDesc}{raw}

\famaMacBeth{FM-Same-part2-std.txt}{t:fm-same-part2-std}{tab:IS-std-news-model}{\scaledCoeffDesc}{scaled}

\famaMacBeth{FM-Same-part2.txt}{t:fm-same-part2}{tab:IS-std-news-model}{\rawCoeffDesc}{raw}


\macroForecasts{macro-std.txt}{tab: macro-std}{Entropy $ENT$ is also used as a control. The coefficient estimates have been normalized by the standard deviation of the right-hand side variables.}

\macroForecasts{macro-part1-std.txt}{tab: macro-part1-std}{\textit{News Updates} (\textit{ENT\_NEWS}) is also used as a control. The coefficient estimates have been normalized by the standard deviation of the right-hand side variables.}

\macroForecasts{macro-part2-std.txt}{tab: macro-part2-std}{\textit{Model Updates} (\textit{ENT\_MODEL}) is also used as a control. The coefficient estimates have been normalized by the standard deviation of the right-hand side variables.}

\macroForecasts{macro.txt}{tab: macro}{Entropy $ENT$ is also used as a control. This table reports the raw (unscaled) regression coefficients.}

\macroForecasts{macro-part1.txt}{tab: macro-part1}{\textit{News Updates} (\textit{ENT\_NEWS}) is also used as a control. This table reports the raw (unscaled) regression coefficients.}

\macroForecasts{macro-part2.txt}{tab: macro-part2}{\textit{Model Updates} (\textit{ENT\_MODEL}) is also used as a control. This table reports the raw (unscaled) regression coefficients.}


\famaMacBeth{FM-Same-std-incSEN.txt}{t:fm-same-std-incSEN}{tab:IS-std}{\scaledCoeffDesc}{scaled}

\famaMacBeth{FM-Same-incSEN.txt}{t:fm-same-incSEN}{tab:IS-std}{\rawCoeffDesc}{raw}


\isForecastsRobust{IS-std_robust.txt}{tab:IS-std-robust}{The coefficient estimates have been
  normalized by the standard deviation of the right-hand side variables.}{0.85}

\isForecastsRobust{IS_robust.txt}{tab:IS-robust}{This table shows the raw (unscaled) coefficient
  estimates.}{0.85}

\makeatletter\@input{xx.tex}\makeatother